\documentclass[fleqn,usenatbib]{mnras}

\usepackage[T1]{fontenc}
\usepackage{ae,aecompl}

\usepackage{graphicx}	
\usepackage{amsmath}	
\usepackage{amssymb}	
\usepackage{times}
\usepackage{booktabs}
\usepackage{multirow}
\usepackage{mathtools}
\usepackage{fix-cm}
\usepackage{xcolor}

\makeatletter
\newlength{\abovecaptionskip}%
\makeatother
\usepackage{threeparttable}
\usepackage{tabularx}
\makeatletter
\g@addto@macro{\UrlBreaks}{\UrlOrds}
\makeatother

\usepackage{newtxtext,newtxmath}

\newcommand{\ez}{EZmock}

\title[EZmock catalogues for eBOSS DR16]{The completed SDSS-\uppercase\expandafter{\romannumeral 4} extended Baryon Oscillation Spectroscopic Survey: 1000 multi-tracer mock catalogues with redshift evolution and systematics for galaxies and quasars of the final data release}

\author[C. Zhao et al.]{
\parbox{\textwidth}{
Cheng Zhao$^{1,}$\thanks{E-mail: \texttt{\href{mailto:cheng.zhao@epfl.ch}{cheng.zhao@epfl.ch}}},
Chia-Hsun Chuang$^{2}$,
Julian Bautista$^{3}$,
Arnaud de Mattia$^{4}$,
Anand Raichoor$^{1}$,
Ashley J. Ross$^{5}$,
Jiamin Hou$^{6}$,
Richard Neveux$^{4}$,
Charling Tao$^{7}$,
Etienne Burtin$^{4}$,
Kyle S. Dawson$^{8}$,
Sylvain de la Torre$^{9}$,
H\'ector Gil-Mar\'in$^{10,11}$,
Jean-Paul Kneib$^{1,9}$,
Will J. Percival$^{12,13,14}$,
Graziano Rossi$^{15}$,
Am\'elie Tamone$^{1}$,
Jeremy L. Tinker$^{16}$,
Gong-Bo Zhao$^{17,18}$,
Shadab Alam$^{19}$
and Eva-Maria Mueller$^{20}$
}
\\
\vspace*{4pt} \\
\small $^{1}$Institute of Physics, Laboratory of Astrophysics, \'Ecole Polytechnique F\'ed\'erale de Lausanne (EPFL), Observatoire de Sauverny, CH-1290 Versoix, Switzerland\vspace*{-2pt} \\
\small $^{2}$Kavli Institute for Particle Astrophysics and Cosmology, Stanford University, 452 Lomita Mall, Stanford, CA 94305, USA\vspace*{-2pt} \\
\small $^{3}$Institute of Cosmology \& Gravitation, Dennis Sciama Building, University of Portsmouth, Portsmouth PO1 3FX, UK\vspace*{-2pt} \\
\small $^{4}$IRFU,CEA, Universit\'e Paris-Saclay, F-91191 Gif-sur-Yvette, France\vspace*{-2pt} \\
\small $^{5}$Center for Cosmology and Astro-Particle Physics, Ohio State University, Columbus, OH 43210, USA\vspace*{-2pt} \\
\small $^{6}$Max-Planck-Institut f\"ur Extraterrestrische Physik, Postfach 1312, Giessenbachstr., 85748 Garching bei M\"unchen, Germany\vspace*{-2pt} \\
\small $^{7}$CPPM, Aix-Marseille Universit\'e, CNRS/IN2P3, CPPM UMR 7346, F13288 Marseille, France\vspace*{-2pt} \\
\small $^{8}$Department Physics and Astronomy, University of Utah, 115 S 1400 E, Salt Lake City, UT 84112, USA\vspace*{-2pt} \\
\small $^{9}$Aix Marseille Univ, CNRS, CNES, LAM, F13388 Marseille, France\vspace*{-2pt} \\
\small $^{10}$Institut de Ci\`encies del Cosmos, Universitat de Barcelona, ICCUB, Mart\'i i Franqu\`es 1, E08028 Barcelona, Spain\vspace*{-2pt} \\
\small $^{11}$Institut  d'Estudis  Espacials  de  Catalunya  (IEEC),  E08034  Barcelona,  Spain\vspace*{-2pt} \\
\small $^{12}$Waterloo Centre for Astrophysics, University of Waterloo, Waterloo, ON N2L 3G1, Canada\vspace*{-2pt} \\
\small $^{13}$Department of Physics and Astronomy, University of Waterloo, Waterloo, ON N2L 3G1, Canada\vspace*{-2pt} \\
\small $^{14}$Perimeter Institute for Theoretical Physics, 31 Caroline St. North, Waterloo, ON N2L 2Y5, Canada\vspace*{-2pt} \\
\small $^{15}$Department of Physics and Astronomy, Sejong University, Seoul 143-747, Korea\vspace*{-2pt} \\
\small $^{16}$Center for Cosmology and Particle Physics, Department of Physics, New York University, New York, NY 10003, USA\vspace*{-2pt} \\
\small $^{17}$National Astronomy Observatories, Chinese Academy of Science, Beijing 100101, P.R. China\vspace*{-2pt} \\
\small $^{18}$School of Astronomy and Space Science, University of Chinese Academy of Sciences, Beijing 100049, P.R.China\vspace*{-2pt} \\
\small $^{19}$Institute for Astronomy, University of Edinburgh, Royal Observatory, Edinburgh EH9 3HJ, UK\vspace*{-2pt}\\
\small $^{20}$Department of Astrophysics, Department of Physics, University of Oxford, Denys Wilkinson Building, Keble Road, Oxford OX1 3RH\vspace*{-2pt}
}

\date{Accepted XXX. Received YYY; in original form ZZZ}

\pubyear{2020}

\begin{document}
\label{firstpage}
\pagerange{\pageref{firstpage}--\pageref{lastpage}}
\maketitle

\begin{abstract}
We produce 1000 realizations of synthetic clustering catalogues for each type of the tracers used for the baryon acoustic oscillation and redshift space distortion analysis of the Sloan Digital Sky Surveys-\uppercase\expandafter{\romannumeral 4} extended Baryon Oscillation Spectroscopic Survey final data release (eBOSS DR16), covering the redshift range from 0.6 to 2.2, to provide reliable estimates of covariance matrices and test the robustness of the analysis pipeline with respect to observational systematics. By extending the Zel'dovich approximation density field with an effective tracer bias model calibrated with the clustering measurements from the observational data, we accurately reproduce the two- and three-point clustering statistics of the eBOSS DR16 tracers, including their cross-correlations in redshift space with very low computational costs. In addition, we include the gravitational evolution of structures and sample selection biases at different redshifts, as well as various photometric and spectroscopic systematic effects. The agreements on the auto-clustering statistics between the data and mocks are generally within $1\,\sigma$ variances inferred from the mocks, for scales down to a few $h^{-1}\,{\rm Mpc}$ in configuration space, and up to $0.3\,h\,{\rm Mpc}^{-1}$ in Fourier space. For the cross correlations between different tracers, the same level of consistency presents in configuration space, while there are only discrepancies in Fourier space for scales above $0.15\,h\,{\rm Mpc}^{-1}$. The accurate reproduction of the data clustering statistics permits reliable covariances for multi-tracer analysis.
\end{abstract}

\begin{keywords}
methods: numerical -- catalogues -- cosmology: large-scale structure of Universe
\end{keywords}



\section{Introduction}

The spatial clustering of large-scale structures (LSS) offers insights into the expansion history of the Universe and the growth of structures. In particular, the baryon acoustic oscillation \citep[BAO;][]{Eisenstein1998} feature is known as a standard ruler for geometrical measurements and provides constraints on the nature of dark energy \citep[][]{Eisenstein2005}. Redshift-space distortions \citep[RSD;][]{Kaiser1987} of the clustering statistics can be used to estimate the structure formation rate and test gravity theories \citep[][]{Percival2009,Raccanelli2013}. Precise cosmological constraints with clustering measurements require the 3D positions -- 2D angular position and redshift -- of tracers of the dark matter density field over a large volume, and possibly several different types of tracers to probe different cosmic epochs.
 
Recent large-scale galaxy spectroscopic surveys, such as the Baryon Oscillation Spectroscopic Survey \citep[BOSS;][]{Dawson2013} -- which belongs to the phase \uppercase\expandafter{\romannumeral 3} of the Sloan Digital Sky Surveys (SDSS) -- have measured the redshifts of over one million luminous red galaxies (LRG) with redshifts up to 0.75, covering more than 9000\,${\rm deg}^{2}$, and achieved percent-level measurements of both distance scales and growth rate of structures \citep[][]{Alam2017}. In addition, the extended BOSS \citep[eBOSS;][]{Dawson2016}, as part of SDSS-\uppercase\expandafter{\romannumeral 4} \citep[][]{Blanton2017} and a complement to BOSS, has probed $\sim 0.8$ million LRGs, star-forming emission line galaxies (ELG), and quasi stellar objects (QSO) in total, with the redshift range $0.6 < z < 2.2$, for the LSS analysis of its final data release \citep[DR16, see Section~\ref{sec:data_catalogue};][]{Ross2020, Raichoor2021}.
In addition, $\sim 0.2$ million BOSS/eBOSS QSOs at $z > 2.1$ are used for Lyman-$\alpha$ absorption measurements \citep[][]{duMas2020, Lyke2020}, which extend the clustering analysis to higher redshift.

Apart from the sample size, accurate estimates of the uncertainties in the clustering statistics are also essential for LSS analysis. One can obtain the covariance matrices directly from the observational catalogues, by sampling the data in subvolumes with jackknife or bootstrap estimations. However, variances on scales larger than the size of the subvolumes cannot be sampled, and systematic errors that apply to all subsamples are not accounted for. An alternative way is to rely on the theoretical model for clustering statistics, and derive Gaussian covariances \citep[e.g.][]{Grieb2016, Wadekar2019}. Further improvements can be achieved by rescaling the shot noise power to include non-Gaussianity \citep[][]{Philcox2020}.
Nevertheless, the robustness of analytical approaches depend on the accuracies of the models in nonlinear regimes of the cosmic evolution, and it is challenging for them to include observational systematic errors.

In principle, these issues can be solved with catalogues generated by $N$-body simulations: they encode the full nonlinear gravitational evolution, and can be applied known observational effects to sample systematic errors. However, the estimate of covariance matrices requires a large number of realizations, and this is generally too computational expensive to be practical for $N$-body simulations with sufficient mass resolution and volume for current large-scale galaxy surveys. To circumvent this problem, some more efficient but less accurate methods for constructing mock catalogues are proposed, such as the bias assignment method \citep[BAM;][]{Balaguera2019}, COmoving Lagrangian Acceleration \citep[COLA;][]{Tassev2013, Izard2016, Koda2016}, effective Zel'dovich approximation mock \citep[\ez{};][]{Chuang2015EZ}, FastPM \citep[][]{Feng2016}, GaLAxy Mocks \citep[GLAM;][]{Klypin2018}, log-normal \citep[][]{Coles1991, Agrawal2017}, peak patch \citep[][]{Bond1996, Stein2019}, PerturbAtion Theory Catalog generator of Halo and galaxY distributions \citep[PATCHY;][]{Kitaura2014}, and quick particle mesh \citep[QPM;][]{White2014}.

These fast mock generation methods can be classified into three general categories. COLA, FastPM, GLAM, peak patch, and QPM are predictive algorithms that solve the dynamic evolution of structures approximately. BAM, \ez{}, and PATCHY generate the dark matter density field using perturbation theories, and then populate tracers with effective descriptions of their biases. While the log-normal method models halo distributions through modifications of the matter density field.
In particular, comparisons of some of the mock construction techniques with $N$-body simulations have shown that methods with bias models, including \ez{} and PATCHY, are not only among the most accurate ones, but also significantly faster than methods with comparable precisions \citep[][]{Chuang2015NIFTY, Blot2019, Colavincenzo2019, Lippich2019}. Actually, PATCHY has been used for the BOSS DR12 analyses \citep[e.g.][]{Kitaura2016, Alam2017}. We choose \ez{} for this work, due to its higher efficiency, and fewer free parameters of the bias model, which makes it easier to be calibrated.

The \ez{} algorithm uses Zel'dovich approximation \citep[][]{Zeldovich1970} to construct the density field at a given redshift, and populate matter tracers (haloes/galaxies/quasars) in the field with a parameterized modelling of tracer bias. This effective bias description includes linear, nonlinear, deterministic, and stochastic effects, which have to be calibrated with clustering statistics from observations or $N$-body simulations, including typically the two-point correlation function (2PCF), power spectrum, and bispectrum.
\ez{} is able to reproduce both two- and three-point statistics of a reference $N$-body simulation precisely down to mildly nonlinear scales. For instance, the discrepancies of redshift space power spectrum produced by \ez{} are less than 5 per cent for $k \lesssim 0.3\,h\,{\rm Mpc}^{-1}$ \citep[][]{Chuang2015NIFTY}. Moreover, thanks to the incomparable efficiency of ZA, the remarkably low computational cost makes \ez{} extremely suitable for estimating covariances for large-scale analysis.

In this work, we use the revised \ez{} method to construct mock catalogues for all eBOSS direct LSS tracers, including LRGs, ELGs, and QSOs. For the estimates of the covariance matrices, we produce 1000 realizations of mock catalogues for each type of the tracers. They are constructed from 46\,000 simulation boxes with the side length of $5\,h^{-1}\,{\rm Gpc}$, at several different redshifts, to account for the redshift evolution of structures. Furthermore, the mock tracers are populated from shared density fields, to ensure reliable estimates of the cross covariances. 
Besides, two sets of mocks are generated, {\it complete} and {\it realistic}, i.e., without and with applying observational systematic effects.
They are used for the analysis of the eBOSS LRG samples \citep[][]{GilMarin2020, Bautista2021}, ELG samples \citep[][]{Tamone2020, deMattia2021, Raichoor2021}, QSO samples \citep[][]{Neveux2020, Hou2021}, and the final cosmological constraints \citep[][]{eBOSS2020}, with the systematic errors assessed using $N$-body simulations \citep[][]{Avila2020, Alam2020, Rossi2020, Smith2020}. Moreover, \citet[][]{Lin2020} use the GLAM method to construct the density field and adopt the bias model of the QPM method to generate mock catalogues for eBOSS ELGs. The eBOSS DR16 \ez{} catalogues presented in this work will be publicly available\footnote{\url{https://data.sdss.org/sas/dr16/eboss/lss/catalogs/EZmocks}}. In addition, all SDSS BAO and RSD measurements and the cosmological interpretations can be found on the SDSS website\footnote{The BAO and RSD measurements are available at \url{https://sdss.org/science/final-bao-and-rsd-measurements}, and see \url{https://sdss.org/science/cosmology-results-from-eboss} for the cosmological results.}.

This paper is organized as follows. In Section~\ref{sec:method} we describe the methodology for constructing the mock catalogues. The clustering statistics of the mock catalogues are shown in Section~\ref{sec:result}. We perform the cross correlation analysis between different tracers in Section~\ref{sec:cross}. Finally, in Section~\ref{sec:conclusion}, we present the conclusions.

\section{Methodology}
\label{sec:method}

We present in this section the improved version of the \ez{} method, compared to the algorithm introduced in \citet[][]{Chuang2015EZ}. In particular, the method used for this work does not require the enhancement of the BAO signal for the initial conditions, and relies on less bias parameters to be calibrated. Moreover, the calibration is done directly with the observed clustering measurements of the BOSS and eBOSS catalogues, without taking $N$-body simulations as references.
This is because no reliable $N$-body simulation multi-tracer catalogue is available when the mocks are constructed, and the accuracy of the \ez{} method has been validated using the BigMultiDark simulation \citep[][]{Chuang2015NIFTY}.
As the result, the effective bias model of \ez{} further accounts for the halo occupation distributions \citep[HOD; e.g.][]{Berlind2002} of different matter tracers.
 We have made the python interface for constructing and calibrating \ez{} catalogues publicly available\footnote{\url{https://github.com/cheng-zhao/pyEZmock}}.

\subsection{Reference data catalogues}
\label{sec:data_catalogue}

The catalogues for LSS analysis in eBOSS DR16 consist of $\sim 0.20$ million LRGs, $\sim 0.27$ million ELGs, and $\sim 0.34$ million QSOs, with the redshift ranges of
\begin{align}
0.6 < z_{\,\rm LRG} < 1.0 , \label{eq:lrg_zrange} \\
0.6 < z_{\,\rm ELG} < 1.1 , \label{eq:elg_zrange} \\
0.8 < z_{\,\rm QSO} < 2.2 . \label{eq:qso_zrange}
\end{align}
Moreover, a subsample of the BOSS DR12 complete-mass (CMASS) LRGs with the same redshift range as Eq.~\eqref{eq:lrg_zrange} is also included for the cosmological analysis. As the result, the combined LRG sample contains $\sim 0.38$ million galaxies. For each of the sample, regions with low spectra completeness and qualities are masked to ensure reliable clustering measurements. Besides, various weights are applied to correct for known observational systematics, and minimize the bias of the clustering statistics \citep[see][for details]{Ross2020, Raichoor2021}.

\begin{figure*}
\centering
\includegraphics[width=.98\textwidth]{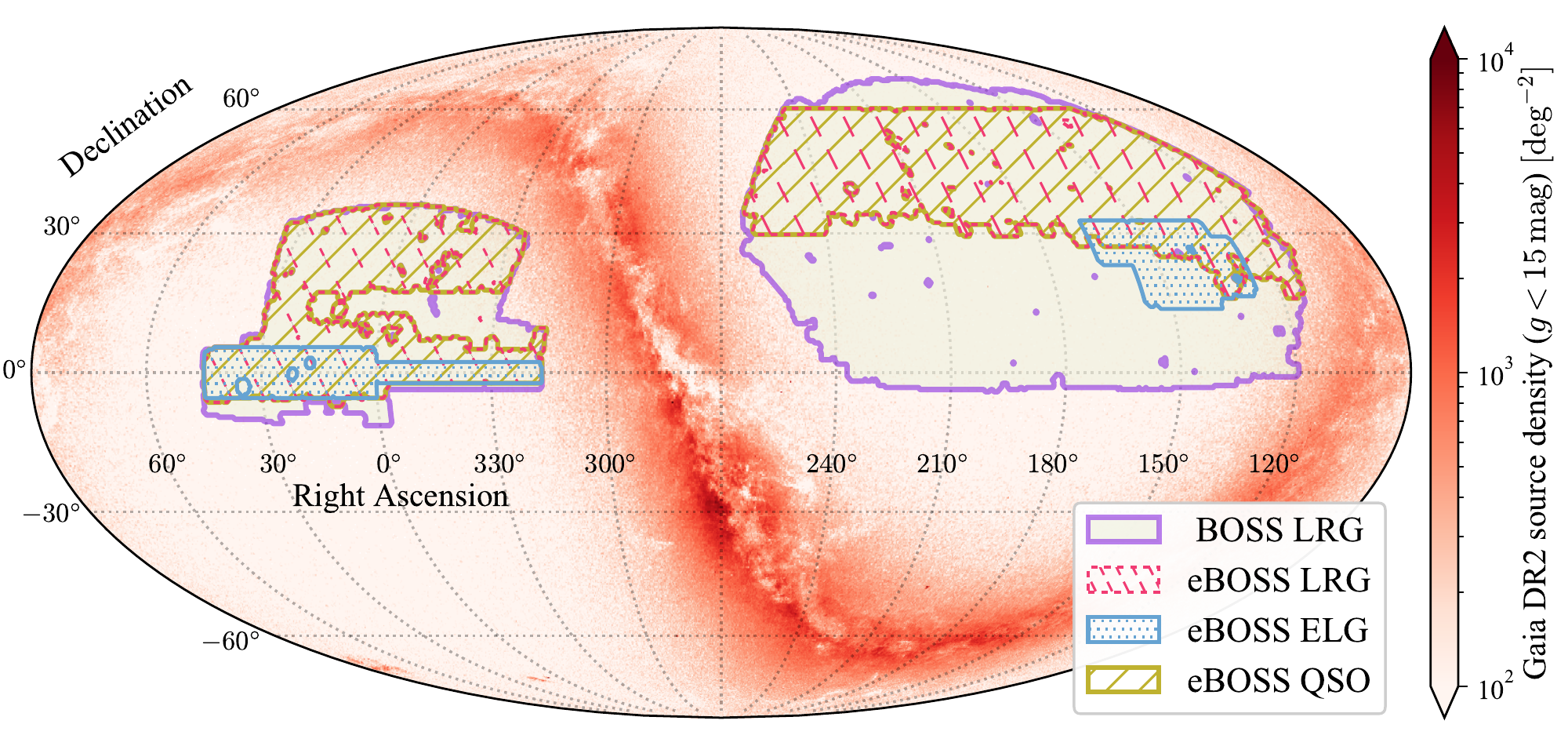}
\caption{The sky coverage of eBOSS DR16 tracers and BOSS DR12 LRGs, as well as the density map of Gaia DR2 sources with $g < 15\,{\rm mag}$.}
\label{fig:eboss_foot}
\end{figure*}

The sky coverage of the BOSS DR12 and eBOSS DR16 data, with various masks applied, are illustrated in Fig.~\ref{fig:eboss_foot}\footnote{See \url{https://skfb.ly/6TPBH} and \url{https://skfb.ly/6TPBI} for 3D illustrations.}, where the background colour map indicates the angular source density of the Gaia DR2 public data \citep[][]{Gaia2018}, with a selection of the $g$ band magnitude (\texttt{phot\char`_g\char`_mean\char`_mag} < 15). In particular, the left and right patches of the BOSS/eBOSS footprints are dubbed northern and southern Galactic caps (NGC and SGC) respectively. Since the two Galactic caps are spatially far away from each other, we construct \ez{} catalogues for NGC and SGC independently, but with the same input parameters.
Therefore, the expected clustering statistics of \ez{} catalogues in both Galactic caps are identical, if no radial selection (see Section~\ref{sec:ezmock_nbar}) is applied. 

The total effective area of the CMASS LRG, eBOSS LRG, eBOSS ELG, and eBOSS QSO samples is 9376, 4103, 727, and 4702\,deg$^2$, respectively \citep[][]{Reid2016, Ross2020, Raichoor2021}. The effective overlapped area between the eBOSS LRG and ELG samples is 458\,deg$^2$, and it is 509\,deg$^2$ for the overlapping region between eBOSS ELG and QSO samples.

\begin{figure}
\centering
\includegraphics[width=.95\columnwidth]{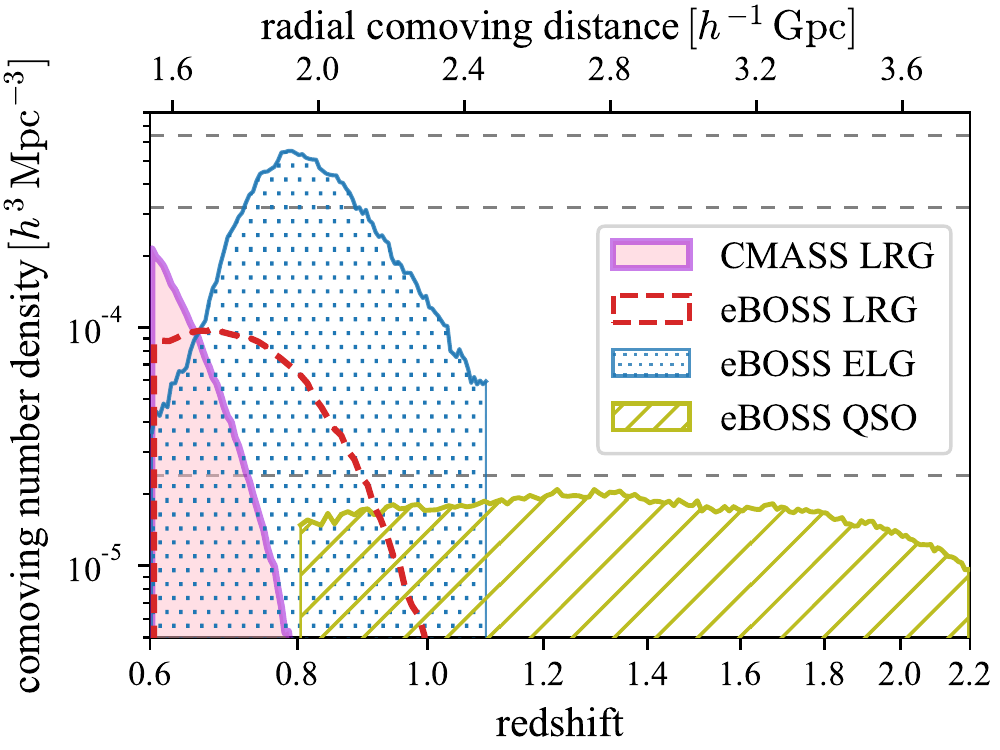}
\caption{The weighted comoving number densities of eBOSS DR16 tracers and BOSS DR12 CMASS LRGs, with all the photometric and spectroscopic systematic weights included. The comoving distances and volumes are evaluated in the flat $\Lambda$CDM cosmology with $\Omega_{\rm m} = 0.31$. The three horizontal dashed lines show the number densities of the cubic LRG, ELG, and QSO \ez{} catalogues, i.e. $3.2 \times 10^{-4}$, $6.4 \times 10^{-4}$, and $2.4 \times 10^{-5}\,h^3\,{\rm Mpc}^{-3}$, respectively.}
\label{fig:eboss_nbar}
\end{figure}

Fig.~\ref{fig:eboss_nbar} shows the effective radial comoving number densities of the tracers, evaluated in the framework of flat $\Lambda$CDM cosmology, with $\Omega_{\rm m} = 0.31$. To estimate the statistical uncertainty of the observed data, the mock catalogues should be constructed with at least the peak number densities of different tracers. Meanwhile, we would like to avoid generating much more tracers than necessary to reduce the computational costs. Consequently, the number densities of LRGs, ELGs, and QSOs that we set for the generation of the mock catalogues are
\begin{align}
n_{\rm LRG}^{\rm box} = 3.2 \times 10^{-4}\,h^3\,{\rm Mpc}^{-3},
\label{eq:lrg_dens} \\
n_{\rm ELG}^{\rm box} = 6.4 \times 10^{-4}\,h^3\,{\rm Mpc}^{-3},
\label{eq:elg_dens} \\
n_{\rm QSO}^{\rm box} = 2.4 \times 10^{-5}\,h^3\,{\rm Mpc}^{-3},
\label{eq:qso_dens}
\end{align}
respectively.

Since all BOSS and eBOSS tracers share the sky area and redshift range to some extent, in order to combine their results for final cosmological analysis, it is crucial to account for the cross covariance between different tracers.
To this end, we construct the mock catalogues for different tracers -- including BOSS CMASS LRGs, and eBOSS LRGs/ELGs/QSOs -- in the same comoving volume, and with identical initial conditions, to ensure the same underlying dark matter density field for all of them.

\subsection{Cubic mock catalogue generation}

The starting point of our mock generation process is a Gaussian random field in a periodic cubic volume, with a given initial power spectrum. The side length of the Gaussian random field in this work is $5\,h^{-1}\,{\rm Gpc}$, which is large enough to cover the survey volume of all tracers for clustering analysis. The same white noises are used for the construction of the Gaussian random field for different tracers.

The fiducial cosmological model for constructing the mocks is flat $\Lambda$CDM, with $\Omega_{\rm m} = 0.307115$, $\Omega_{\rm b}=0.048206$, $h=0.6777$, $\sigma_8 = 0.8225$, and $n_{\rm s} = 0.9611$, which are the best-fitting values from the Planck 2013 results \citep[][]{Planck2014}.
This is the same cosmological model used by the \textsc{patchy} mock catalogues for the final BOSS data release \citep[][]{Kitaura2016} which is calibrated based on the MultiDark simulations \citep[][]{Klypin2016}.
The linear matter power spectrum we use is generated by the \textsc{camb}\footnote{\url{https://camb.info/}} software \citep[][]{Lewis2000}. It has been shown that the covariance matrix of two-point clustering measurements are insensitive to the input power spectrum, if the two- and three-point statistics of the mocks are consistent with the observed measurements \citep[][]{Baumgarten2018}.

\subsubsection{Zel'dovich approximation}

To generate the dark matter field at the desired redshift, we rely on the Zel'dovich approximation \citep[ZA;][]{Zeldovich1970}, which is the linear solution of the Lagrangian Perturbation Theory \citep[LPT; see e.g.][]{Bernardeau2002}. In the Lagrangian description, the Eulerian position $\boldsymbol{x}$ of a particle at time $t$ is expressed by its initial comoving position $\boldsymbol{q}$ (i.e., the position in the Gaussian random field) and a displacement $\boldsymbol{\Psi}$:
\begin{equation}
\boldsymbol{x} (\boldsymbol{q}, t) = \boldsymbol{q} + \boldsymbol{\Psi} (\boldsymbol{q}, t) .
\label{eq:lagrangian_disp}
\end{equation}
And the linear solution to the equation of motion yields
\begin{equation}
\nabla_{\boldsymbol{q}} \cdot \boldsymbol{\Psi}_{\rm ZA} (\boldsymbol{q}, t) = - D_1 (t) \delta (\boldsymbol{q}).
\label{eq:za_disp}
\end{equation}
Here $\boldsymbol{\Psi}_{\rm ZA}$ stands for the displacement field in the Zel'dovich approximation, $D_1 (t)$ denotes the linear growth factor, and $\delta (\boldsymbol{q})$ indicates the initial density contrast in Lagrangian coordinates, which is sampled in Fourier space with random phases, with the amplitude being defined by the linear matter power spectrum $P_{\rm lin} (k)$:
\begin{equation}
P_{\rm lin} (k) = \left\langle \left| \delta (\boldsymbol{k}) \right| \right\rangle^2 .
\end{equation}

In the framework of $\Lambda$CDM, the linear growth factor can be evaluated numerically through the integral representation \citep[][]{Heath1977, Carroll1992}:
\begin{equation}
D_1 (a) = a^3 H(a) \frac{5 \Omega_{\rm m}}{2} \int_0^a \frac{{\rm d}\Tilde{a}}{\Tilde{a}^3 H^3 (\Tilde{a})},
\end{equation}
where $a$ indicates the scale factor, and $H(a)$ is the Hubble function.

The displacement field in the ZA can be obtained through the Fourier transform of Eq.~\eqref{eq:za_disp}:
\begin{equation}
\boldsymbol{\Psi}_{\rm ZA} (\boldsymbol{q}, a) = D_1 (a) \int \frac{{\rm d}^3 k}{(2 \uppi)^3}  {\rm e}^{i \boldsymbol{k} \cdot \boldsymbol{q}} \frac{i \boldsymbol{k}}{k^2} \hat{\delta} (\boldsymbol{k}),
\end{equation}
where $\hat{\delta} (\boldsymbol{k})$ denotes the density contrast in Fourier space. Thus, the ZA density field $\rho_{\rm m}$ can be efficiently computed with Fast Fourier Transforms (FFT). In this work, FFTs are performed with the grid size of $1024^3$, in the $5^3\,h^{-3}\,{\rm Gpc}^3$ cubic volume.

Once the displacement field is computed on the grid points, dark matter particles are moved from their initial Lagrangian positions -- Cartesian grid points -- to the final Eulerian positions following Eq.~\eqref{eq:lagrangian_disp}. The dark matter density field is evaluated on the same grids, using the Cloud-in-Cell \citep[CIC;][]{Hockney1981} particle assignment scheme.
Consequently, the number density fields of the observational tracers described hereafter, are all based on this grid size. In general, the \ez{} parameters have to be re-calibrated, with a different number of grids in the same comoving volume.

\subsubsection{Deterministic bias relations}

To populate tracers in the simulation box, we need to introduce a bias model describing the relationship between tracers and dark matter, or in other words, to construct the tracer number density field $\rho_{\rm t}$ based on the dark matter density field $\rho_{\rm m}$. This process can be expressed by a general bias function $B$:
\begin{equation}
\rho_{\rm t} = B( \rho_{\rm m} ) .
\label{eq:bias_model}
\end{equation}
In particular, the density $\rho$ is defined on Cartesian grid points in the comoving volume, as
\begin{equation}
\rho \equiv \tilde{\rho} / \langle \tilde{\rho} \rangle  ,
\end{equation}
where $\tilde{\rho}$ denotes the ratio between the number of objects and comoving volume for each grid cell, and $\langle \, \cdot \, \rangle$ indicates the ensemble average over all the grids.

To implement Eq.~\eqref{eq:bias_model} for the mock tracers, we begin with some analytical bias relations that have been confirmed with $N$-body simulations. However, due to the inaccuracy of ZA in the nonlinear regime, the analytical form of $B$ is not enough for a precise bias model. The actual $\rho_{\rm t}$ is generated with a rank ordering process detailed in Section~\ref{sec:ezmock_pdfmap}, in a numerical manner.

In this section, we focus on the deterministic part of the analytical bias description. To form gravitational bound systems, such as dark matter haloes, a minimum local density is required to overcome the background expansion \citep[e.g.][]{Percival2005}.
This density threshold is crucial for the correct modelling of the three-point statistics of dark matter haloes \citep[][]{Kitaura2015}.
Thus, we introduce a critical density $\rho_{\rm c}$, and add a term $\theta (\rho_{\rm m} - \rho_{\rm c})$ to the bias function,
where $\theta$ denotes the step function:
\begin{equation}
\theta (x) =
\begin{cases}
0, & x < 0 ; \\
1, & x \ge 0 .
\end{cases}
\end{equation}

Apart from the density threshold, \citet[][]{Chuang2015EZ} applies also a density saturation, i.e., regions with densities above the saturation $\rho_{\rm sat}$ are treated equally for the stochastic generation of haloes. This saturation is responsible for the amplitude of the power spectrum of the resulting tracers.
Besides, \citet[][]{Neyrinck2014} finds an exponential cut-off of the halo bias relation:
\begin{equation}
\rho_{\rm h} \propto \rho_{\rm m}^\alpha \exp{( \rho_{\rm m} / \rho_{\rm exp} )^{-\epsilon}} .
\end{equation}
For simplicity, we account for both effects with the following form \citep[][]{Baumgarten2018}:
\begin{equation}
\rho_{\rm t} = \theta (\rho_{\rm m} - \rho_{\rm c})\, \rho_{\rm sat} \, [ 1 - \exp{(-\rho_{\rm m} / \rho_{\rm exp})} ] \, B_{\rm s},
\label{eq:det_bias}
\end{equation}
where $B_{\rm s}$ denotes the stochastic bias term, which serves as a random rescaling factor of the deterministic biased density field.
Moreover, since there are strong degeneracies between $\rho_{\rm sat}$ and other parameters, we fix $\rho_{\rm sat}=10$ in practice.

\subsubsection{Stochastic bias relations}

We introduce a scatter to the bias relation to account for the stochasticity of tracers, i.e. \citep[][]{Chuang2015EZ}
\begin{equation}
B_{\rm s} =
\begin{cases}
1 + G (\lambda), &  G (\lambda) \ge 0 ; \\
\exp ( G (\lambda) ) , &  G (\lambda) < 0 .
\end{cases}
\label{eq:sto_bias}
\end{equation}
Here, $G(\lambda)$ indicates a random number drawn from a Gaussian distribution centred at 0, and with the standard deviation $\lambda$. In particular, the exponential function is for avoiding negative bias values.

In general, the stochastic bias of galaxies is non-Poissonian and depends on the environments \citep[][]{Somerville2001, Casas2002}. Nevertheless, it is the order of tracer densities in different cells that matters in this work, rather than the actual functional form for the scatter. This is because the densities are further modified by rank ordering with a PDF mapping scheme detailed in the next subsection.
For a more realistic description of stochastic biases, see the negative binomial distribution proposed by \citet[][]{Kitaura2014}, and further validated in \citet[][]{Vakili2017, Pellejero2020}.

In practice, the value of $\lambda$ in Eq.~\eqref{eq:sto_bias} alters mainly the amplitude of the power spectrum and bispectrum, and the same effect can be achieved by the other parameters, such as $\rho_{\rm c}$ and $\rho_{\rm exp}$, hence we set $\lambda = 10$ throughout this work.

\subsubsection{PDF mapping scheme}
\label{sec:ezmock_pdfmap}

To further correct the tracer number density $\rho_{\rm t}$, and map it to the number of tracers per grid cell $n_{\rm t}$, we model the probability distribution function (PDF) of the tracers by a power-law relation:
\begin{equation}
P(n_{\rm t}) = A b^{n_{\rm t}} ,
\label{eq:pdf_map}
\end{equation}
where $P(n_{\rm t})$ denotes the probability of having a cell with $n_{\rm t}$ tracers, and $b$ and $A$ are two free parameters, with the restrictions $A > 0$ and $0 < b < 1$. This serves as an additional effective bias description.

Moreover, since we aim at generating mock catalogues with desired number densities in the cubic volume (see Eqs~\eqref{eq:lrg_dens} -- \eqref{eq:qso_dens}), the expected total number of tracers $N_{\rm t}^{\rm tot}$ is given, which can also be expressed by
\begin{equation}
N_{\rm t}^{\rm tot}
= \sum_{n_{\rm t} = 1}^{n_{\rm t, max}} n_{\rm c} (n_{\rm t}) \cdot n_{\rm t} \, ,
\end{equation}
with
\begin{align}
&n_{\rm c} (n_{\rm t}) = \left\lfloor N_{\rm cell} P(n_{\rm t}) \right\rceil , \\
&n_{\rm t, max} = \min_{n_{\rm t} > 0} \left\{ n_{\rm t} \,\middle|\, N_{\rm cell} P(n_{\rm t}) < 0.5 \right\}.
\end{align}
Here, $n_{\rm c} (n_{\rm t}) $ indicates the number of cells containing $n_{\rm t}$ tracers, $N_{\rm cell}$ indicates the total number of cells ($1024^3$ in this work), $n_{\rm t, max}$ is the maximum expected number of tracers per grid cell, and the operator $\lfloor \,\cdot\, \rceil$ denotes the nearest integer.
Thus, there is only one degree-of-freedom for the PDF model. So we treat only the base $b$ as a free parameter.

We then map $n_{\rm c} (n_{\rm t})$ to the expected tracer number density $\rho_{\rm t}$, which is estimated by the bias relations described in the previous sections, in descending order. For instance, we rank the cells by $\rho_{\rm t}$, and assign $n_{\rm t, max}$ tracers to $n_{\rm c} (n_{\rm t, max})$ cells with the highest $\rho_{\rm t}$ values, and then $( n_{\rm t, max} - 1 )$ tracers to the next $n_{\rm c} (n_{\rm t, max} - 1)$ cells, etc. Thus, the exact values of $\rho_{\rm t}$ defined in Eq.~\eqref{eq:det_bias} is irrelevant for our purpose, as they are effectively modified based on their orders.

The tracers are then assigned randomly to the dark matter particles in each grid cell, if there are any. For cells without enough number of dark matter particles, we randomly pick a position in the cell for the tracer, which potentially damps the BAO feature.
However, this effect is subdominant in this work, as the fractions of LRGs, ELGs, and QSOs that are randomly placed, are only $\sim$ 6\,\%, 16\,\%, and 4\,\% respectively.
Moreover, the strength of the BAO feature has little impacts on the covariance matrices, compared to the contribution of the broad-band amplitudes.
Thus, the enhancement of the BAO feature in the input power spectrum introduced by \citet[][]{Chuang2015EZ}, for correcting the BAO smearing due to the smooth galaxy distribution inside grid cells, is no longer necessary.

\subsubsection{Redshift space distortions}
\label{sec:ezmock_rsd}

The linear peculiar velocity field in the ZA is \citep[see e.g.][]{Bernardeau2002}
\begin{equation}
\boldsymbol{u}_{\rm ZA} (\boldsymbol{q}, a)
= a f(a) H(a) \boldsymbol{\Psi}_{\rm ZA} (\boldsymbol{q}, a) ,
\end{equation}
where $f(a)$ is the dimensionless linear growth rate:
\begin{equation}
f(a) = \frac{{\rm d}\,\ln{D_1 (a)}}{{\rm d}\,\ln{a}} .
\end{equation}
In the linear regime of gravitational instability, galaxy velocities are unbiased, i.e., they follow faithfully dark matter velocities \citep[][]{Hamilton1998}.
To account for the random motion of individual tracers with respect to the bulk flow of dark matter, we further introduce an isotropic 3D Gaussian motion to the linear coherent velocity field, for the modelling of tracer peculiar velocities $\boldsymbol{u}_{\rm t}$:
\begin{equation}
\boldsymbol{u}_{\rm t} = \boldsymbol{u}_{\rm ZA} + \boldsymbol{G} (\nu) ,
\label{eq:pec_vel}
\end{equation}
where $\boldsymbol{G} (\nu)$ denotes a random vector drawn from an isotropic 3D Gaussian distribution centred at $\boldsymbol{0}$, and with the standard deviation $\nu$ (in km\,s$^{-1}$).
This is essentially a modelling of the Maxwellian peculiar velocity distribution.
Another formula for the local random velocities commonly used is the exponential distribution, but we did not test it since the Gaussian distribution has given already reasonable agreements within the scales we are interested in, in terms of the redshift-space clustering statistics.
In general, the random motion accounts only for small-scale clustering measurements, i.e., 2PCF monopole and quadrupole on scales smaller than $10$ and $50\,h^{-1}\,{\rm Mpc}$ respectively. Furthermore, the effects are more obvious for Fourier-space clustering, at $k \gtrsim 0.1\,h\,{\rm Mpc}^{-1}$.

\subsubsection{Summary of model parameters}

So far we have introduced six \ez{} parameters for the effective modelling of tracer biases, and they are summarised in Table~\ref{tab:ez_par}. Since some of the parameters are highly correlated, and we fix the density saturation $\rho_{\rm sat}$ and the width $\lambda$ of the Gaussian random distribution for the stochastic biasing. Thus, there are four free parameters to be calibrated with the two- and three-point clustering statistics of the reference catalogues, i.e., the BOSS DR12 CMASS and eBOSS DR16 catalogues. In order to take into account the impact of survey geometry on the clustering statistics, we calibrate these free parameters only with the \ez{} light-cone catalogues that mimic the geometry of eBOSS DR16 data.

\begin{table}
    \centering
    \begin{threeparttable}
    \caption{A list of the parameters for the effective bias modelling of \ez{}.}
    \begin{tabular}{cccc}
        \toprule
        Parameter & Equation & Description & Value \\
        \midrule
        $\rho_{\rm c}$ & \eqref{eq:det_bias} & Critical density & Free \\
        $\rho_{\rm sat}$ & \eqref{eq:det_bias} & Density saturation & 10 \\
        $\rho_{\rm exp}$ & \eqref{eq:det_bias} & Density modification & Free \\
        $\lambda$ & \eqref{eq:sto_bias} & Stochastic bias & 10 \\
        $b$ & \eqref{eq:pdf_map} & Base of PDF mapping & Free \\
        $\nu$ & \eqref{eq:pec_vel} & Random local motion & Free \\
        \bottomrule
    \end{tabular}
    \label{tab:ez_par}
    \end{threeparttable}
\end{table}

Since the observational systematics affect mainly scales outside the range of clustering statistics for \ez{} calibrations (see Section~\ref{sec:ezmock_calib} for scales relevant for the calibration, and Section~\ref{sec:result} for the comparison between {\it complete} and {\it realistic} mocks), and it is relatively computational expensive to apply observational effects to the mock catalogues, we calibrate the \ez{} parameters with the {\it complete} set of \ez{} light-cones, rather than the {\it realistic} ones.
In practice, the calibration is done with a single mock realization by manually fine-tuning. The results are then validated using 50 realizations to eliminate impacts of cosmic variances, before the mass production.

\subsection{Complete light-cone catalogue construction}
\label{sec:ez_lightcone}

To construct practical mock catalogues, various geometrical features of the observed data have to be applied to the cubic mocks. To this end, we use the \textsc{make\char`_survey}\footnote{\url{https://github.com/mockFactory/make\_survey}} toolkit \citep[][]{White2014} to rotate the cubic \ez{} catalogues, map the tracers with observational coordinates, and trim the catalogues according to the survey footprints and veto masks defined by \textsc{mangle}\footnote{\url{https://space.mit.edu/~molly/mangle/}} \citep[][]{Swanson2008} polygon files.
This procedure is similar to that of the SUrvey GenerAtoR code \citep[\textsc{sugar};][]{Rodriguez2016} used for BOSS DR12 Patchy mocks \citep[][]{Kitaura2016}.

\subsubsection{Coordinate conversion}
\label{sec:coord_conv}

Taking into account the periodic boundary conditions, we firstly remap the $(5\,h^{-1}\,{\rm Gpc})^3$ \ez{} box into a cuboid, with the side lengths being $5$, $5\sqrt{2}$, and $5/\sqrt{2}\,h^{-1}\,{\rm Gpc}$, respectively. The cuboid is then shifted and rotated without rescaling, such that all eBOSS DR16 tracers can be covered by it.
To this end, the translation and rotation parameters are determined by comparing the cuboid with the eBOSS tracers in comoving coordinates, with the observer placed at the origin.
Here, equatorial coordinates (RA, dec) and redshift $z$ of the eBOSS data are transformed to Cartesian comoving coordinates $(x_{\rm c}, y_{\rm c}, z_{\rm c})$ following 
\begin{align}
r_{\rm c} &= \int_0^{z} \frac{c\,{\rm d} z^\prime}{H_0 \sqrt{\Omega_\Lambda + \Omega_{\rm m} (1+z^\prime)^3}} ,
\label{eq:coord_trans1} \\
x_{\rm c} &= r_{\rm c} \cos{(\rm dec)} \cos{(\rm RA)} ,
\label{eq:coord_trans2} \\
y_{\rm c} &= r_{\rm c} \cos{(\rm dec)} \sin{(\rm RA)},
\label{eq:coord_trans3} \\
z_{\rm c} &= r_{\rm c} \sin{(\rm dec)} ,
\label{eq:coord_trans4}
\end{align}
where $r_{\rm c}$ denotes the radial comoving distance, and $H_0 = 100\,h\,{\rm km}\,{\rm s}^{-1}\,{\rm Mpc}^{-1}$ is the Hubble parameter at $z=0$.

Furthermore, we have ensured that there is enough space between the surface of the cuboid and the boundaries of the survey volume in comoving space, for preserving a complete mock sample inside the survey volume in redshift space, in which the redshifts of tracers are modified by their radial peculiar velocities \citep[][]{Harrison1974}:
\begin{equation}
z_{\rm s} = z_{\rm r} + (\boldsymbol{u}_{\rm t} \cdot \hat{\boldsymbol{r}}_{\rm c}) (1 + z_{\rm r}) / c .
\label{eq:coord_rsd}
\end{equation}
Here, $z_{\rm r}$ and $z_{\rm s}$ are the redshifts of tracers in real and redshift space respectively, $\hat{\boldsymbol{r}}_{\rm c}$ denotes the unit line-of-sight vector in comoving space, and the peculiar velocity $\boldsymbol{u}_{\rm t}$ is described in Section~\ref{sec:ezmock_rsd}. In particular, $z_{\rm r}$ is obtained by applying the inverse transformation of Eqs~\eqref{eq:coord_trans1} -- \eqref{eq:coord_trans4} to all the mock tracers, together with their equatorial coordinates. Note that Eq.~\eqref{eq:coord_rsd} is slightly different from the original implementation of \textsc{make\char`_survey}, which uses a single value of $z_{\rm r}$ for all the mock tracers in the cuboid for the redshift due to peculiar velocities.

\subsubsection{Survey volume trimming}
\label{sec:ezmock_trim}

To mimic the angular area of the BOSS DR12 CMASS and eBOSS DR16 data, we trim the \ez{} catalogues with the BOSS/eBOSS footprints, which are defined by groups of sectors -- regions covered by a unique set of plates \citep[][]{Ross2020} -- in the \textsc{mangle} polygon format.
To have reliable clustering measurements, the sectors are further selected according to the associate $C_{\rm eBOSS}$ and $C_z$ values -- $C_{\rm eBOSS} > 0.5$ and $C_z > 0.5$ for LRGs and QSOs, and $C_{\rm eBOSS} \ge 0.5$ and $C_z \ge 0$ for ELGs -- where $C_{\rm eBOSS}$ denotes the fraction of targets that are assigned fibres, or without fibres only due to fibre-collision, and $C_z$ indicates the proportion of valid tracers with fibres, for which a reliable redshift is obtained \cite[see][for more details]{Ross2020, Raichoor2021}.

\begin{figure*}
\centering
\begin{tabular}{rccccl}
     & CMASS LRG & eBOSS LRG & eBOSS ELG & eBOSS QSO &
    \\[1ex]
    \rotatebox[origin=c]{90}{NGC} &
    \raisebox{-0.5\height}{\includegraphics[width=.2\textwidth]{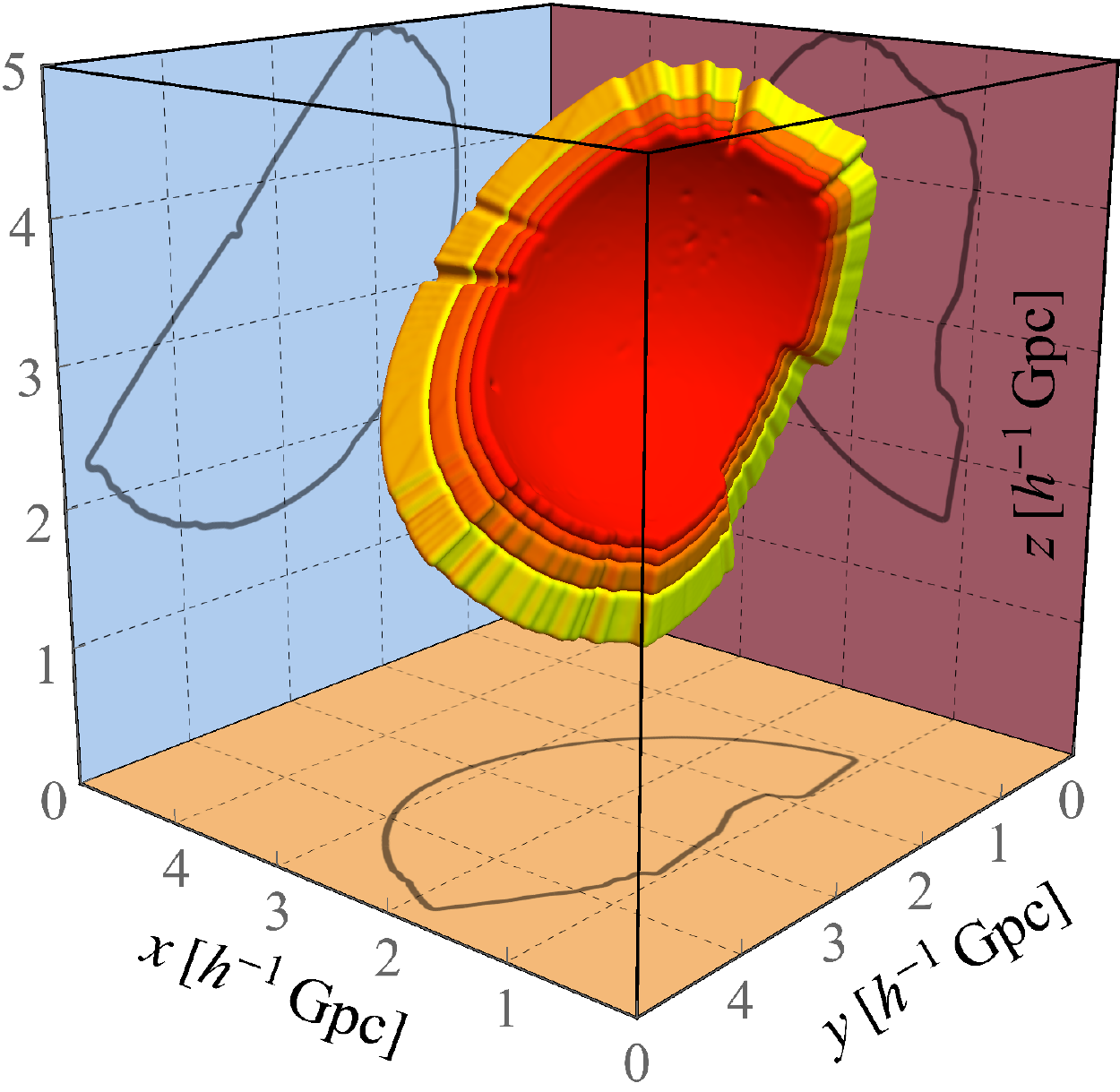}} &
    \raisebox{-0.5\height}{\includegraphics[width=.2\textwidth]{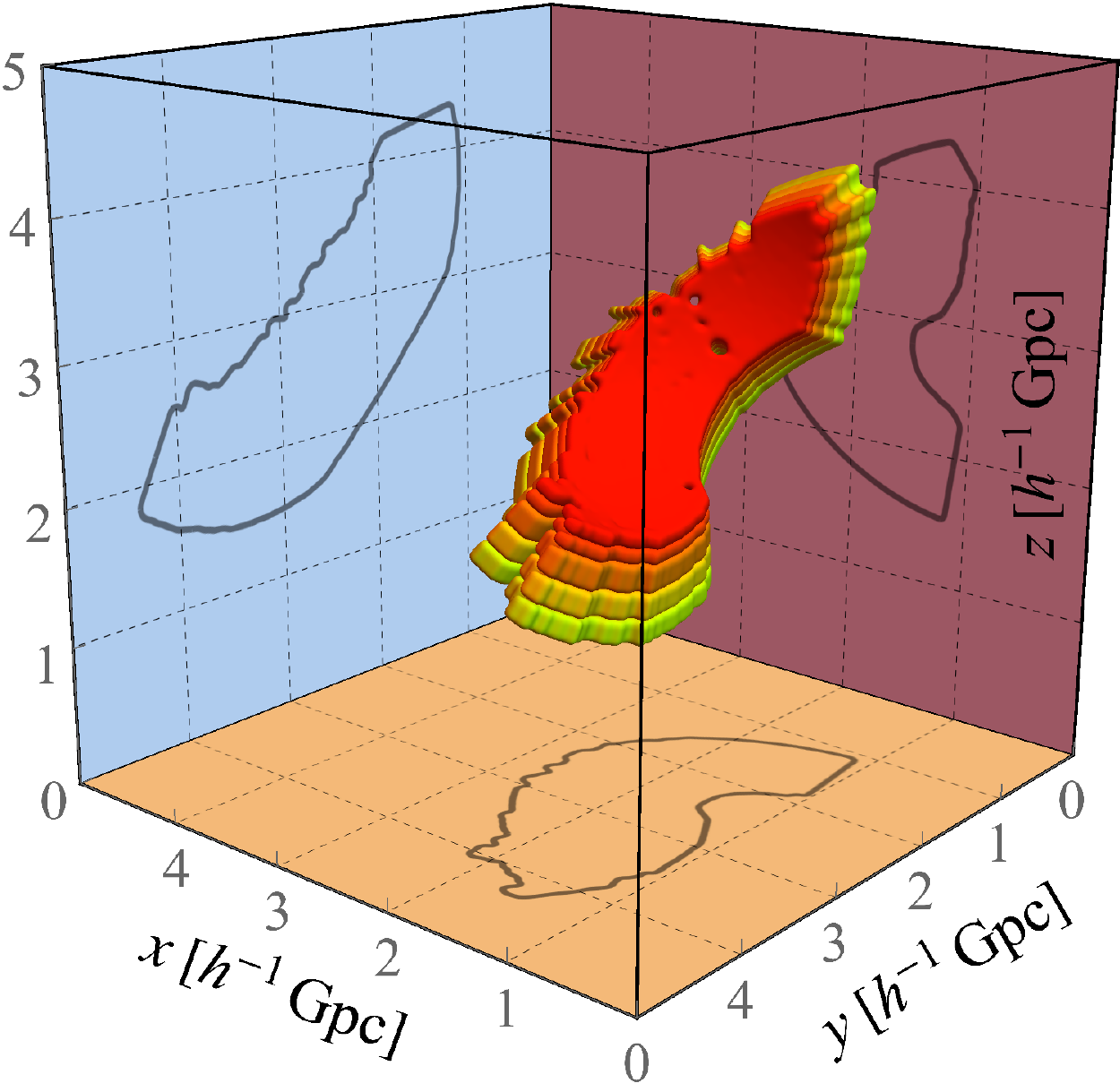}} &
    \raisebox{-0.5\height}{\includegraphics[width=.2\textwidth]{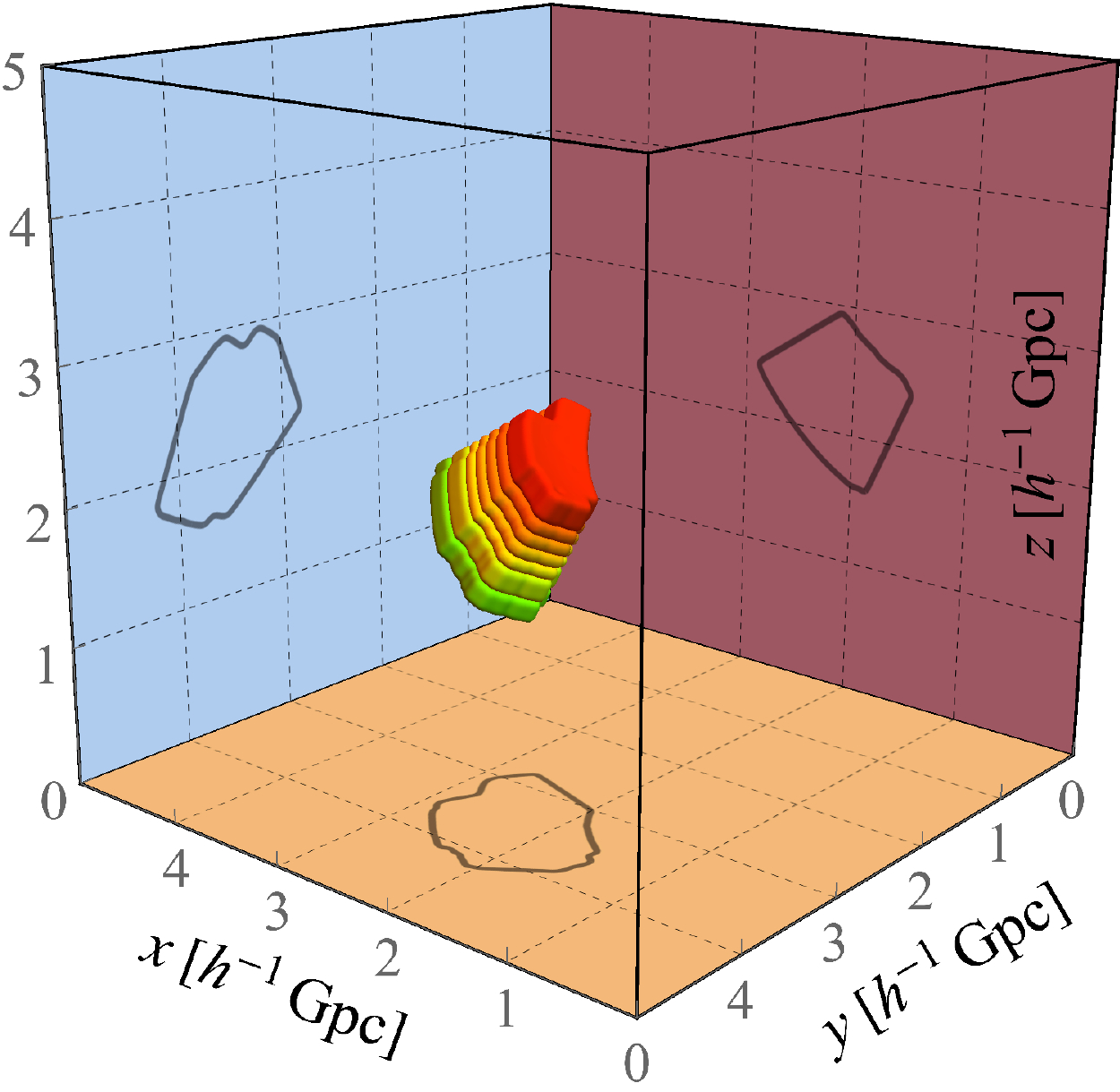}} &
    \raisebox{-0.5\height}{\includegraphics[width=.2\textwidth]{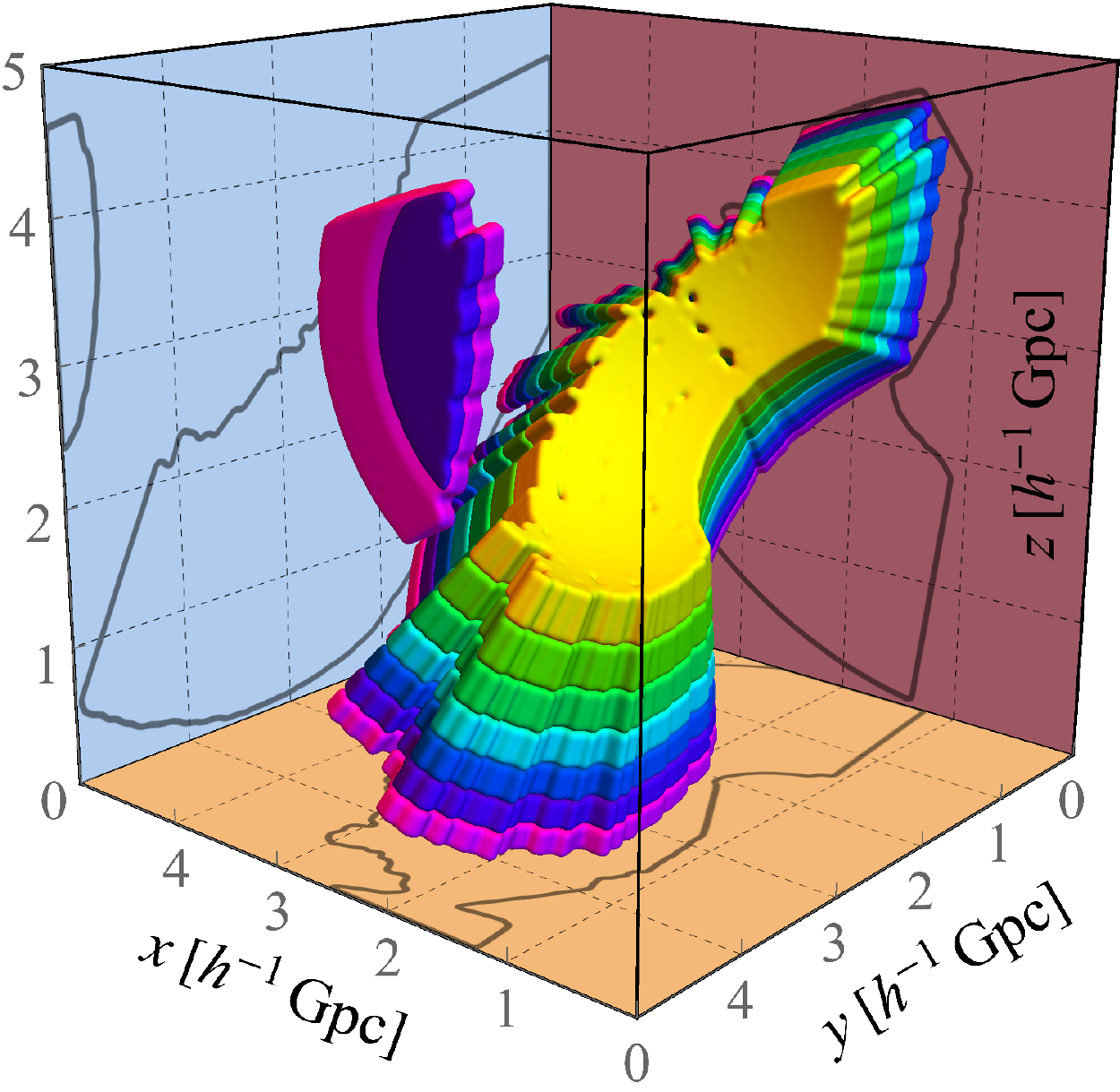}} &
    \multirow{3}{*}[2em]{\includegraphics[width=.05\textwidth]{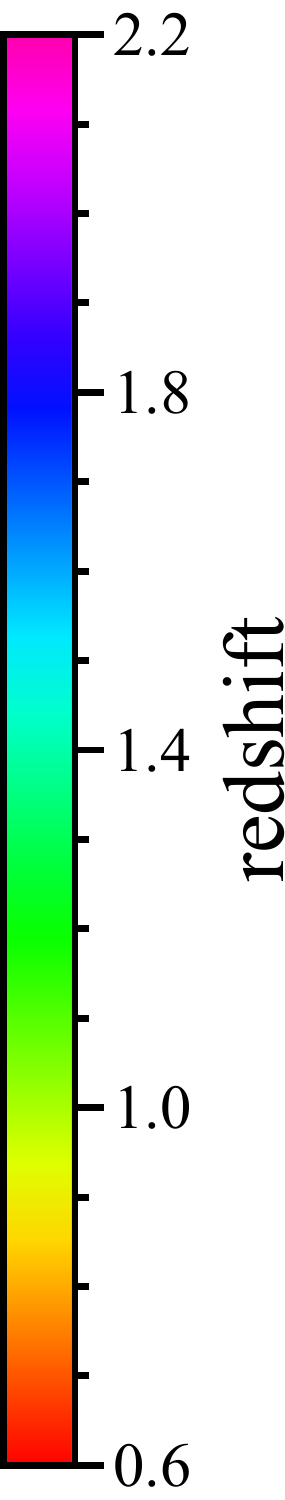}}
     \\
    \rotatebox[origin=c]{90}{SGC} &
    \raisebox{-0.5\height}{\includegraphics[width=.2\textwidth]{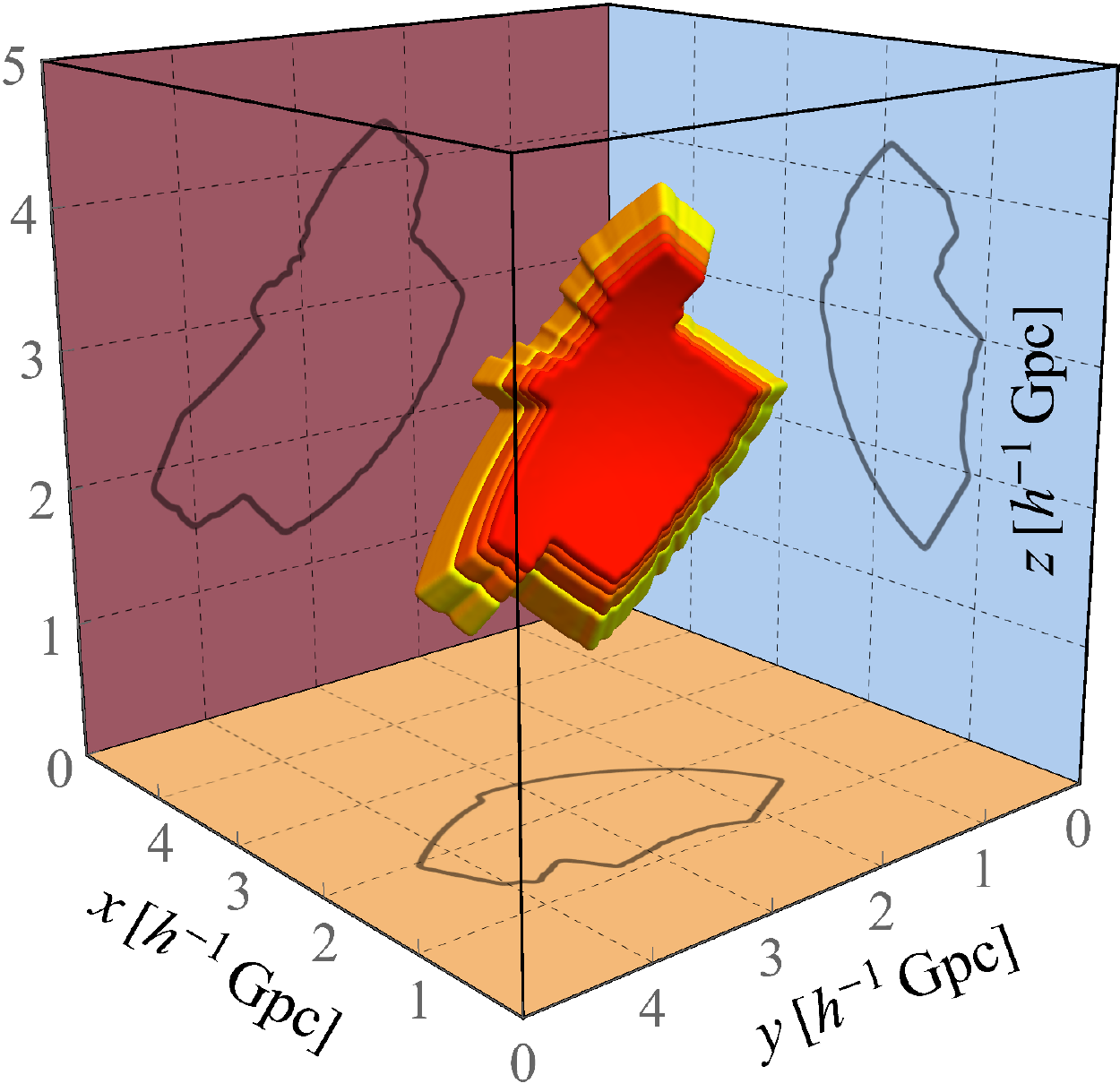}} &
    \raisebox{-0.5\height}{\includegraphics[width=.2\textwidth]{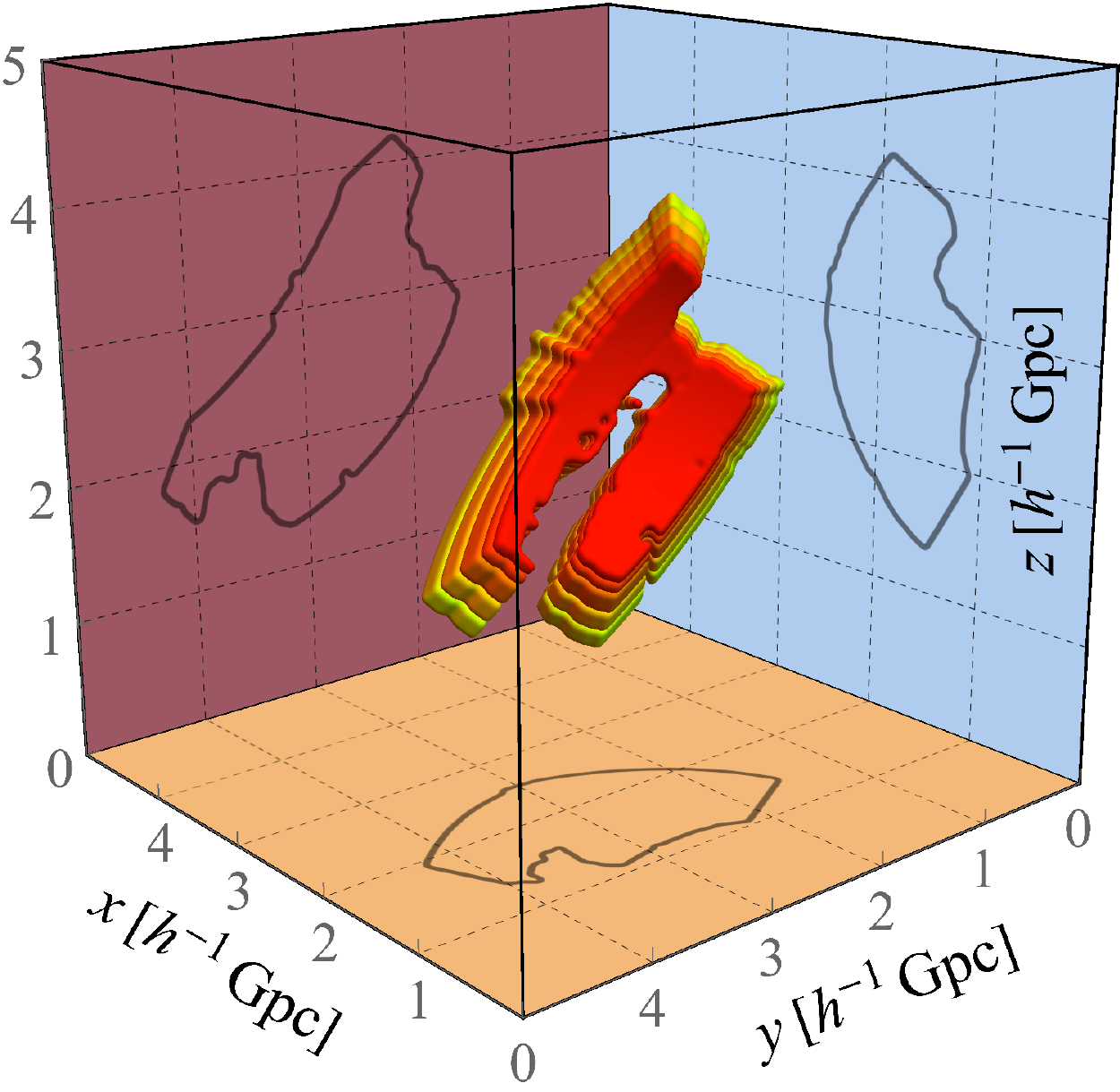}} &
    \raisebox{-0.5\height}{\includegraphics[width=.2\textwidth]{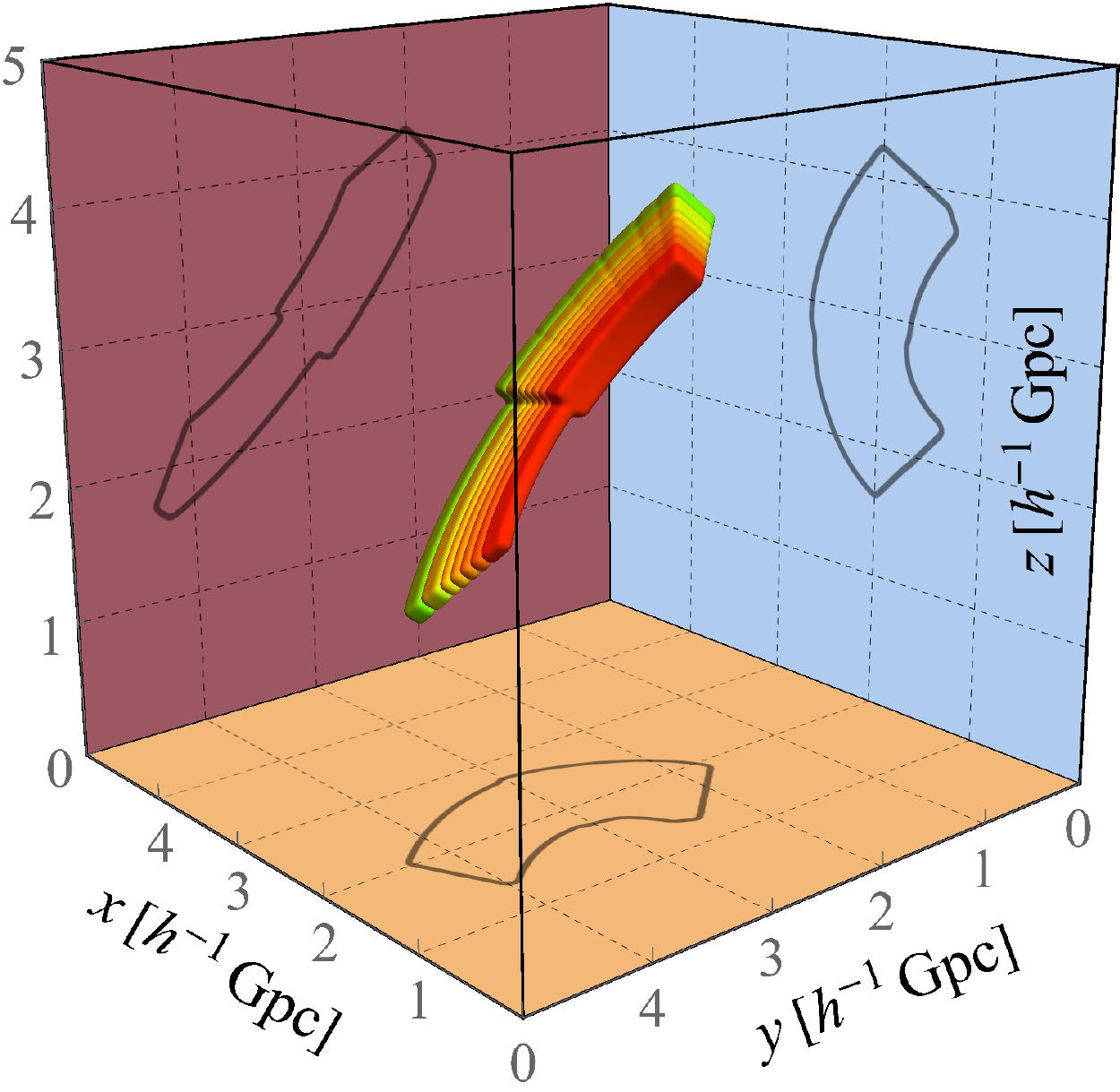}} &
    \raisebox{-0.5\height}{\includegraphics[width=.2\textwidth]{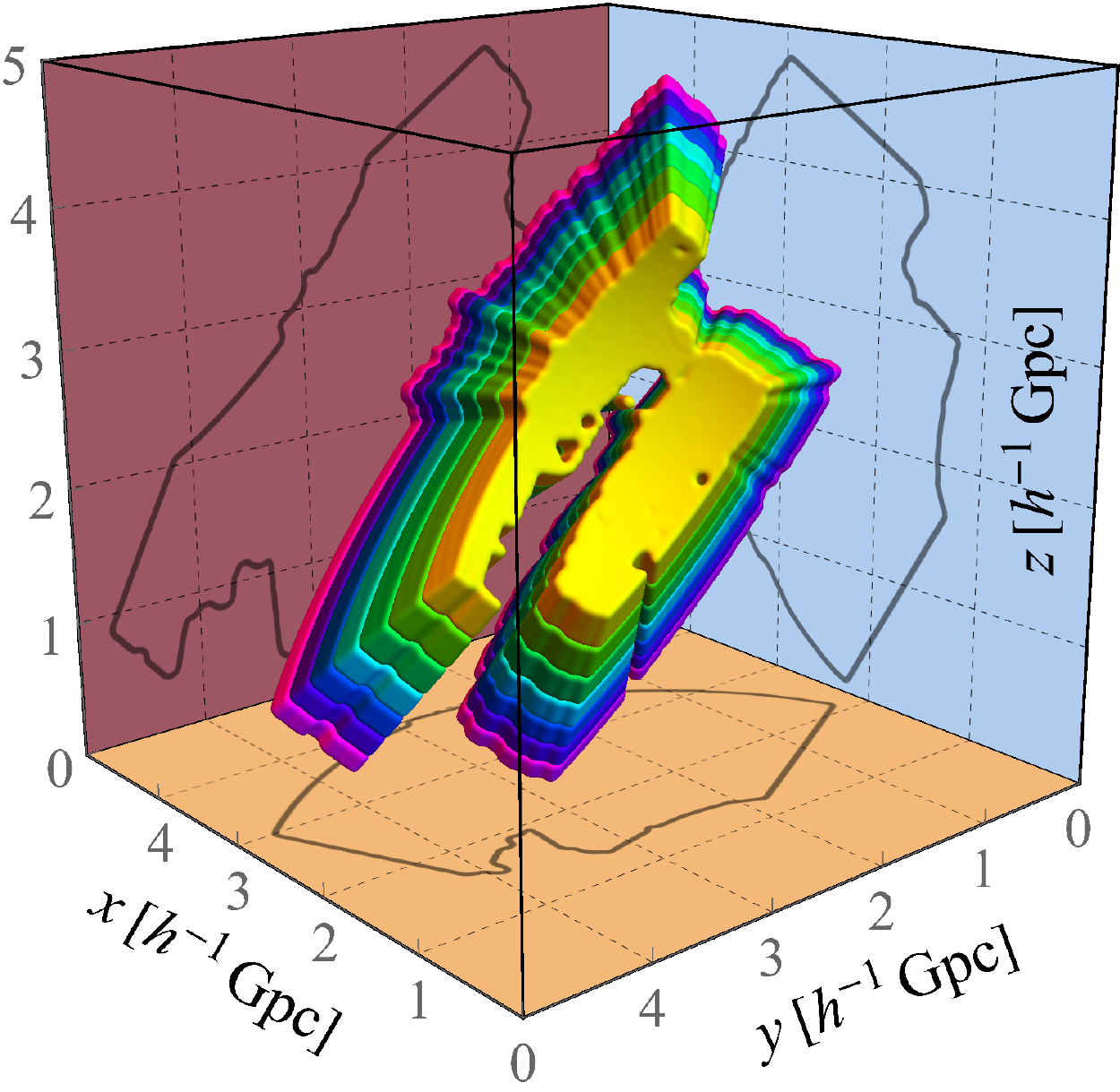}} & \\
\end{tabular}
\caption{The comoving volume of \ez{} catalogues after survey volume trimming, compared with the $(5\,h^{-1}\,{\rm Gpc})^3$ periodic box for constructing the mocks. Regions with different colours indicate the redshift slices used for reproducing the redshift evolution of clustering statistics (see Section~\ref{sec:ezmock_zbin}). The QSO sample in the NGC benefits from the periodic boundary conditions, as its comoving volume is too large to be placed inside the box.}
\label{fig:cutsky_3d}
\end{figure*}

In the radial direction, we simply select \ez{} tracers with redshifts inside the redshift range of the corresponding BOSS/eBOSS data catalogues (see Eqs~\eqref{eq:lrg_zrange} -- \eqref{eq:qso_zrange}).
The comoving volume of the \ez{} catalogues after survey volume trimming, compared to the original cubic periodic boxes, are shown in Fig.~\ref{fig:cutsky_3d}\footnote{The corresponding 3D illustrations are available at \url{https://skfb.ly/6TRz9}}.

\subsubsection{Veto masks}

\begin{figure}
\centering
\includegraphics[width=.95\columnwidth]{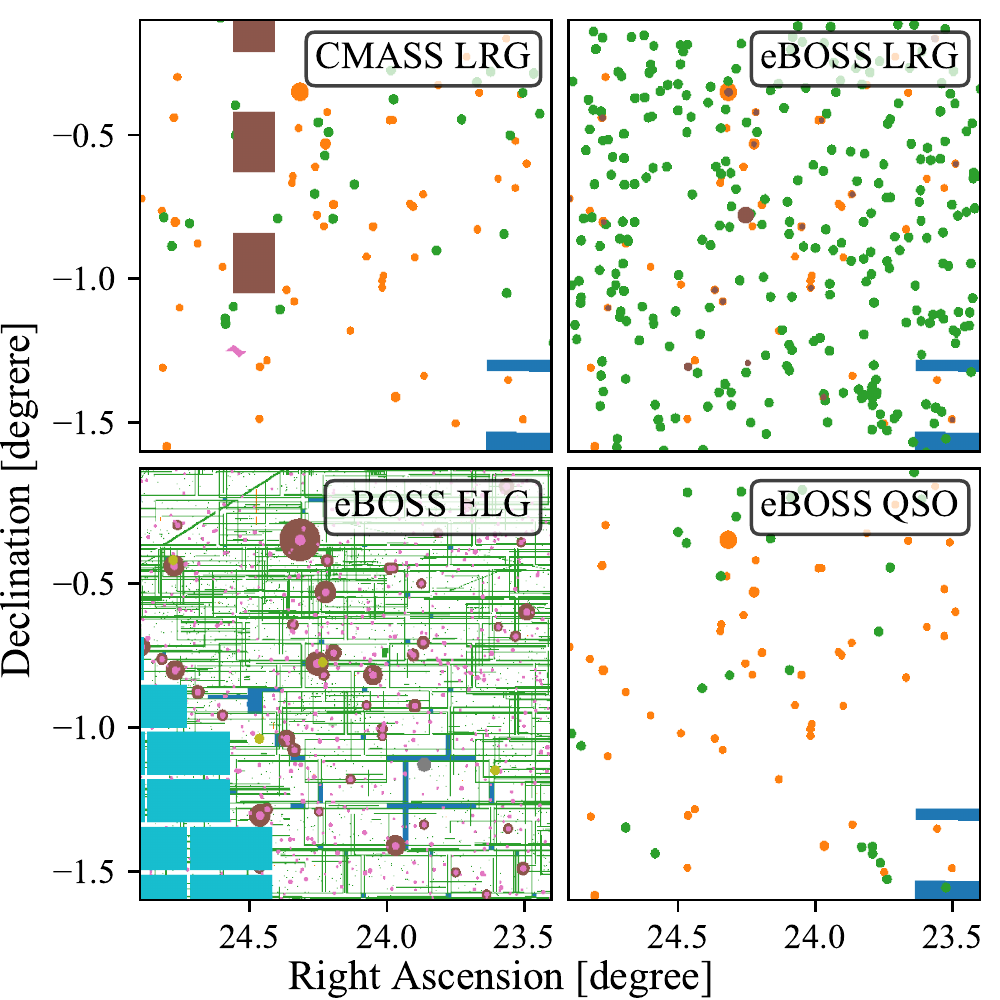}
\caption{Angular veto masks for eBOSS DR16 tracers and BOSS DR12 CMASS LRGs, for the same patch of the sky. Tracers in the coloured regions are removed for various reasons, such as bright source contaminations or unreliable photometric measurements.}
\label{fig:eboss_veto}
\end{figure}

Inside the survey volume there are still angular patches to be removed, such as fields that were not observed, or regions that are too close to bright object to have reliable redshift measurements.
In general, the shapes and distributions of these masks depend on the brightness of the tracers, sources of the images, and the calibration process.
The angular veto masks of the eBOSS DR16 and BOSS DR12 samples are shown in Fig.~\ref{fig:eboss_veto}, where the colours indicate different types of masks, i.e., regions removed for different reasons \citep[see][for more details]{Ross2020, Raichoor2021}.

In practice, the LRG and QSO veto masks are encoded as \textsc{mangle} polygons, and can be simply applied with the \textsc{mply\char`_trim} tool of the \textsc{make\char`_survey} package. However, as can be seen in Fig.~\ref{fig:eboss_veto}, the eBOSS ELG veto masks are much more complicated than those of the other tracers. Thus, it is not practical to translate the ELG masks to simple polygons. Instead, the mask information is associated with each pixel of the DECaLS bricks \citep[][]{Raichoor2021}.
We then use the \textsc{brickmask}\footnote{\url{https://github.com/cheng-zhao/brickmask}} code to apply the ELG masks, which is made publicly available.

\subsubsection{Radial selection and FKP weighting}
\label{sec:ezmock_nbar}

To replicate the radial number densities $n(z)$ of BOSS/eBOSS tracers, we randomly discard mock objects at a given redshift with the probability
\begin{equation}
P_{\rm discard} (z) = 1 - \frac{n_{\rm data} (z)}{n_{\rm mock}^{\rm box}} ,
\label{eq:ezmock_radsel}
\end{equation}
where $n_{\rm data}$ indicates the radial comoving number density distribution of the observed data, which is rescaled to the cosmology for the \ez{} construction ($\Omega_{\rm m} = 0.307115$), and $n_{\rm mock}^{\rm box}$ denotes the number density of mock tracers in periodic boxes (see Eqs~\eqref{eq:lrg_dens} -- \eqref{eq:qso_dens} and Fig.~\ref{fig:eboss_nbar}). In particular, $n_{\rm data} (z)$ and $P_{\rm discard} (z)$ are evaluated for different subsamples separately, i.e., the two Galactic caps. Moreover, since the ELG data is further split into four chunks\footnote{`Chunks' are regions in which the plate and fibre assignments are performed independently.} -- \texttt{eboss23} and \texttt{eboss25} for NGC, \texttt{eboss21} and \texttt{eboss22} for SGC -- due to their different spectroscopic properties \citep[][]{Raichoor2021}, radial selections for \ez{} ELG catalogues are applied for different chunks independently.

Since the radial distribution of the tracers is no longer uniform, to minimize the variance of the clustering measurements, we weight mock objects by the redshift-dependent FKP scheme \citep[][]{Feldman1994}:
\begin{equation}
w_{\rm FKP} (z) = \frac{1}{1 + n_{\rm mock} (z) P_0} ,
\label{eq:weight_fkp}
\end{equation}
where $n_{\rm mock} (z)$ denotes the number densities of light-cone mock tracers, and $P_0$ is the typical power spectrum value of the tracers in the $\boldsymbol{k}$ range that we are interested in. In principle $n_{\rm mock}$ differs for each mock realization. However, for the {\it complete}  \ez{} catalogues, with the radial down-sampling described by Eq.~\eqref{eq:ezmock_radsel}, the difference between $n_{\rm data} (z)$ and $n_{\rm mock} (z)$ is only from the shot noise of the random sampling process, which introduces a small variation (around 2.5 per cent, see Fig.~\ref{fig:ezmock_zhist}) for the number densities of \ez{} catalogues, given the bin size for our $n_{\rm data} (z)$ evaluation: $\Delta z = 0.01$ for LRGs and QSOs, and $\Delta z = 0.005$ for ELGs. Thus, we interpolate $n_{\rm data} (z)$ measured from individual Galactic caps with cubic splines, as an approximation of $n_{\rm mock} (z)$, and apply it to all the mock realizations. Finally, we take the same $P_0$ values as the ones used for the creation of BOSS/eBOSS data catalogues, \citep[][]{Ross2020, Raichoor2021}:
\begin{align}
P_{0, {\rm LRG}} &= 10000 \,h^{-3}\,{\rm Mpc}^3 , \\
P_{0, {\rm ELG}} &= 4000 \,h^{-3}\,{\rm Mpc}^3 , \\
P_{0, {\rm QSO}} &= 6000 \,h^{-3}\,{\rm Mpc}^3 ,
\end{align}
for LRGs (including CMASS), ELGs, and QSOs, respectively.
They broadly correspond to the power spectrum amplitude at $k \sim 0.1\,h\,{\rm Mpc}^{-1}$, for the different tracers.

\subsubsection{Redshift slices}
\label{sec:ezmock_zbin}

With the FKP weights evaluated, the {\it complete} \ez{} catalogues are ready for clustering measurements. However, since the cubic mock catalogues are constructed at a specific redshift, the redshift evolution of structure growth, galaxy bias, and peculiar motion are not taken into account. To be more accurate, we construct the \ez{} catalogues in several redshift slices, with the cubic mocks generated at the effective redshift $z_{\rm eff}$ inside the bins. In particular, the effective redshift is measured from the data catalogues, with the definition \citep[][]{Samushia2014}
\begin{equation}
z_{\rm eff}^{(j)} = \enspace \left. \sum_{\mathclap{\substack{i \\ z_{\rm low} < z_i < z_{\rm high}}}}^{N_{\rm data}} ( w_i^2 z_i^j ) \enspace \middle/ \quad \enspace \sum_{\mathclap{\substack{i \\ z_{\rm low} < z_i < z_{\rm high}}}}^{N_{\rm data}} w_i^2 \right. \, , \quad j = 1, 2 ,
\label{eq:z_eff}
\end{equation}
where $z_{\rm low}$ and $z_{\rm high}$ are respectively the lower and upper boundaries of the redshift bin, and $w_i$ stands for the total weight used for clustering measurements for each object.
This effective redshift definition is different from the ones used for the eBOSS clustering analysis \citep[e.g.][]{Tamone2020, Bautista2021, Hou2021}, which are computed from pairs of tracers with the separation range relevant for the likelihood evaluations, to optimize the cosmological parameter constraints. Nevertheless, the different is small, and the choice of effective redshift should not bias the covariance matrix estimation, as long as the clustering statistics of the mocks are well calibrated.
The effective redshift squared $z_{\rm eff}^{(2)}$ is used later for the \ez{} parameter calibration (see Section~\ref{sec:ezmock_calib}).

The redshift slices used in this work are listed in Table~\ref{tab:ezmock_zbin} (see also Fig.~\ref{fig:cutsky_3d} for the illustration in comoving space).
Catalogues generated at different effective redshifts are trimmed with the corresponding $z_{\rm low}$ and $z_{\rm high}$ values, after performing coordinate conversions (Section~\ref{sec:coord_conv}). These slices are then combined to construct the sample in the full redshift range of the data. Finally, the survey footprint, veto masks, and radial selections (Sections~\ref{sec:ezmock_trim} -- \ref{sec:ezmock_nbar}) are all applied to the combined catalogues.

\begin{table}
\centering
\begin{threeparttable}
\caption{The final redshift slices for the production of \ez{} catalogues, with the corresponding effective redshift (Eq.~\eqref{eq:z_eff}) and (weighted) number of tracers from the observed data. $N_{\rm data}$ denotes the number of objects in each redshift bin, and $w_i$ indicates the total photometric and spectroscopic systematic weights.}
\begin{tabular}{ccccccc}
\toprule
Sample & $z_{\rm low}$ & $z_{\rm high}$ & $z_{\rm eff}$ & $z_{\rm eff}^{(2)}$ & $N_{\rm data}$ & $\sum\limits_{\mathclap{i}}^{\mathclap{N_{\rm data}}} w_i$ \\
\midrule
\multirow{4}{*}{\shortstack{CMASS\\LRG}}
& 0.6 & 0.65 & 0.626 & 0.392 & 114441 & 122385 \\
& 0.65 & 0.7 & 0.675 & 0.455 & 57561 & 61461.9 \\
& 0.7 & 0.8 & 0.737 & 0.545 & 30899 & 33024.2 \\
& 0.8 & 1.0 & 0.847 & 0.719 & 2473 & 2643.8 \\
\midrule
\multirow{5}{*}{\shortstack{eBOSS\\LRG}}
& 0.6 & 0.65 & 0.625 & 0.391 & 28152 & 29983.1 \\
& 0.65 & 0.7 & 0.675 & 0.456 & 33557 & 35828.4 \\
& 0.7 & 0.8 & 0.751 & 0.564 & 64460 & 68592.7 \\
& 0.8 & 0.9 & 0.847 & 0.719 & 37080 & 39099.7 \\
& 0.9 & 1.0 & 0.940 & 0.885 & 11567 & 12130.2 \\
\midrule
\multirow{7}{*}{\shortstack{eBOSS\\ELG}}
& 0.6 & 0.7 & 0.658 & 0.434 & 10046 & 11667.4 \\
& 0.7 & 0.75 & 0.725 & 0.526 & 20275 & 23373.6 \\
& 0.75 & 0.8 & 0.775 & 0.601 & 33487 & 38857.9 \\
& 0.8 & 0.85 & 0.825 & 0.682 & 34631 & 40140.4 \\
& 0.85 & 0.9 & 0.876 & 0.767 & 27831 & 32231.0 \\
& 0.9 & 1.0 & 0.950 & 0.903 & 32721 & 37792.2 \\
& 1.0 & 1.1 & 1.047 & 1.097 & 14745 & 16997.8 \\
\midrule
\multirow{7}{*}{\shortstack{eBOSS\\QSO}}
& 0.8 & 1.0 & 0.907 & 0.826 & 35988 & 38026.4 \\
& 1.0 & 1.2 & 1.104 & 1.223 & 47025 & 50276.2 \\
& 1.2 & 1.4 & 1.301 & 1.697 & 57120 & 61230.1 \\
& 1.4 & 1.6 & 1.500 & 2.252 & 55758 & 59573.2 \\
& 1.6 & 1.8 & 1.700 & 2.894 & 56678 & 60640.0 \\
& 1.8 & 2.0 & 1.898 & 3.606 & 50774 & 54310.4 \\
& 2.0 & 2.2 & 2.094 & 4.389 & 40357 & 42731.2 \\
\bottomrule
\end{tabular}
\label{tab:ezmock_zbin}
\end{threeparttable}
\end{table}

\subsubsection{Sample combination}

Since the BOSS DR12 CMASS and eBOSS DR16 LRG samples overlap widely in both angular and radial directions (see Figs~\ref{fig:eboss_foot} and \ref{fig:eboss_nbar}), and they consist of the same type of galaxies \citep[][]{Prakash2016, Reid2016}, it is reasonable to combine the two datasets directly for joint clustering analyses. Nevertheless, special care has to be taken since the footprint of the two samples are not identical. We follow the combination procedure described in \citet[][]{Ross2020} for the observational data. In brief, we detect eBOSS LRG sectors that contain CMASS LRGs, and add their comoving number densities in redshift bins, with the bin size of $\Delta z = 0.01$, to obtain the number density of the combined sample. The FKP weights are then revised accordingly, following Eq.~\eqref{eq:weight_fkp}. By contrast, eBOSS galaxies in sectors that do not contain CMASS objects, and CMASS galaxies outside eBOSS sectors, are not altered.

Moreover, when combining CMASS and eBOSS \ez{} catalogues, we have ensured that they are constructed with the same initial conditions. This restriction is also applied for the combination of redshift slices, or ELG chunks.

\subsection{Random catalogue creation}
\label{sec:ezmock_rand}

In order to account for the survey window function, including the radial number density of tracers, random catalogues are required for clustering measurements.
One simple way to generate random catalogue for \ez{} catalogues is to apply the light-cone catalogue creation procedure described in Section~\ref{sec:ez_lightcone} (except the redshift division in Section~\ref{sec:ezmock_zbin}, as there is no evolution for a random catalogue), to a uniform random sample in comoving space. In this case, a single random catalogue is necessary for each type of the tracers in individual Galactic caps.

However, the radial selection function of the BOSS DR12 CMASS and eBOSS DR16 catalogues are not directly sampled from the number density of data. Instead, redshifts of the observed data are shuffled, and randomly assigned to the angular random catalogues \citep[][]{Reid2016,Ross2020, Raichoor2021}. This is because the true redshift distribution of data is usually complicated and unknown, since it depends on various imaging and spectroscopic effects. Indeed, the comoving number density shown in Fig.~\ref{fig:eboss_nbar} is only a binned estimation of the true radial selection function, whereas, the shuffled approach ensures the correct radial distribution of random objects automatically. Nevertheless, this method introduces a radial effect that is similar to an additional window function, and bias the clustering measurements on large scales significantly \citep[][]{deMattia2019}. To investigate this problem, we create also random catalogues for the mocks with the shuffled method.

In practice, we generate firstly the random catalogue with redshifts sampled from the spline interpolation of the comoving number density measured from the BOSS/eBOSS data, as is done in Section~\ref{sec:ezmock_nbar}.  And we dub them the `sampled' random catalogues. Then, we keep only the angular positions of these random catalogues, and randomly assign the shuffled redshifts of the \ez{} catalogues to the angular random positions, to create the `shuffled' random catalogues. Note that there is one `shuffled' random catalogue for each of the \ez{} realizations. The consequences of the two sets of random catalogues are shown in Section~\ref{sec:result}. 

\subsection{\ez{} parameter calibration}
\label{sec:ezmock_calib}

We aim at encoding redshift evolution in the \ez{} light-cone catalogues. To this end, besides constructing mocks at different effective redshifts, the effective bias model of \ez{} has to be adjusted for each of the redshift slices. This requires individual calibrations of \ez{} parameters (see Table~\ref{tab:ez_par}) for each bin, with the clustering of observed data catalogues measured in corresponding redshift ranges. However, for many of the redshift bins listed in Table~\ref{tab:ezmock_zbin}, the number of tracers are too low for precise measurements of two- and three-point statistics from the observational data, and the calibration results may be dominated by statistical noise.

To circumvent this problem, we use larger but overlapping redshift bins to determine the \ez{} parameters, as shown in Table~\ref{tab:ezmock_zbin}.
When calibrating \ez{} parameters for each bin, we compare the following clustering statistics measured from the {\it complete} \ez{} light-cone catalogues for both NGC and SGC, with the ones obtained from BOSS/eBOSS data:
\begin{enumerate}
\item
$\xi_0 (s)$, $\xi_2 (s)$: 2PCF monopole and quadrupole, with the galaxy pair separation range of $s \in [10, 50]\,h^{-1}\,{\rm Mpc}$.
\item
$P_0 (k)$, $P_2 (k)$: power spectrum monopole and quadrupole, with the Fourier mode range of $k \in [0.1,0.3]\,h\,{\rm Mpc}^{-1}$ (apart from eBOSS QSO NGC, which is only calibrated on scales up to $k \sim 0.24\,h\,{\rm Mpc}^{-1}$, see Section~\ref{sec:pk}).
\item
$B(k_1, k_2, \theta_{12})$: bispectrum, with $k_1 = 0.1 \pm 0.01\,h\,{\rm Mpc}^{-1}$, $k_2 = 0.05 \pm 0.01\,h\,{\rm Mpc}^{-1}$, and $\theta_{12}$ being the angle between $\boldsymbol{k}_1$ and $\boldsymbol{k}_2$.
\end{enumerate}
In particular, the ranges of the two-point statistics are chosen to be nonsensitive to observational systematic effects, and the same \ez{} parameters are used for the two Galactic caps.

\begin{table}
\centering
\begin{threeparttable}
\caption{The redshift slices used for the calibration of \ez{} catalogues, with the corresponding effective redshift (Eq.~\eqref{eq:z_eff}) and (weighted) number of tracers from the observed data. $N_{\rm data}$ denotes the number of objects in each redshift bin, and $w_i$ indicates the total photometric and spectroscopic systematic weights.}
\begin{tabular}{ccccccc}
\toprule
Sample & $z_{\rm low}$ & $z_{\rm high}$ & $z_{\rm eff}$ & $z_{\rm eff}^{(2)}$ & $N_{\rm data}$ & $\sum\limits_{\mathclap{i}}^{\mathclap{N_{\rm data}}} w_i$ \\
\midrule
\multirow{4}{*}{\shortstack{CMASS\\LRG}}
& 0.6 & 0.7 & 0.652 & 0.426 & 172002 & 183847 \\
& 0.65 & 0.75 & 0.696 & 0.485 & 81206 & 86695.2 \\
& 0.7 & 0.8 & 0.737 & 0.545 & 30899 & 33024.2 \\
& 0.75 & 1.0 & 0.791 & 0.628 & 9727 & 10434.8 \\
\midrule
\multirow{4}{*}{\shortstack{eBOSS\\LRG}}
& 0.6 & 0.7 & 0.652 & 0.426 & 61709 & 65811.5 \\
& 0.65 & 0.8 & 0.727 & 0.531 & 98017 & 104421 \\
& 0.7 & 0.9 & 0.797 & 0.638 & 101540 & 107692 \\
& 0.8 & 1.0 & 0.878 & 0.773 & 48647 & 51229.9 \\
\midrule
\multirow{4}{*}{\shortstack{eBOSS\\ELG}}
& 0.6 & 0.8 & 0.714 & 0.512 & 63808 & 73898.9 \\
& 0.7 & 0.9 & 0.805 & 0.651 & 116224 & 134603 \\
& 0.8 & 1.0 & 0.905 & 0.821 & 95183 & 110164 \\
& 0.9 & 1.1 & 0.994 & 0.991 & 47466 & 54790 \\
\midrule
\multirow{5}{*}{\shortstack{eBOSS\\QSO}}
& 0.8 & 1.3 & 1.077 & 1.181 & 110950 & 118238 \\
& 1.1 & 1.5 & 1.305 & 1.717 & 110326 & 118197 \\
& 1.3 & 1.7 & 1.499 & 2.262 & 113477 & 121358 \\
& 1.5 & 1.9 & 1.699 & 2.899 & 110370 & 117993 \\
& 1.7 & 2.2 & 1.930 & 3.747 & 65077 & 69150.3 \\
\bottomrule
\end{tabular}
\label{tab:ezmock_calib}
\end{threeparttable}
\end{table}

Then, we use a similar way as in \citet[][]{Ata2018}, to model the redshift evolution of the parameters, i.e.
\begin{equation}
p ( z_{\rm eff}, z_{\rm eff}^{(2)} ) = c_{0, p} + c_{1, p} \, z_{\rm eff} + c_{2, p} \, z_{\rm eff}^{(2)} ,
\end{equation}
where the coefficients $c_{0, p}$, $c_{1, p}$, and $c_{2, p}$ are obtained from linear regressions with the redshift bins shown in Table~\ref{tab:ezmock_calib}, for the \ez{} parameter $p$.
This relationship is applied to all the redshift slices listed in Table~\ref{tab:ezmock_zbin}, to infer the \ez{} parameters for the fine bins. For parameters that do not vary much with redshift, we use only a fixed value for all redshift slices. Finally, we examine the fitting results with 50 \ez{} realizations, and fine tune the parameters if necessary. The resulting \ez{} parameters for different redshift slices are shown in Table~\ref{tab:ezmock_par}.

\begin{table}
\centering
\begin{threeparttable}
\caption{The calibrated \ez{} parameters for different redshift slices, that are used for both NGC and SGC.}
\begin{tabular}{ccccccc}
\toprule
Sample & $z_{\rm low}$ & $z_{\rm high}$ & $\rho_{\rm c}$ & $\rho_{\rm exp}$ & $b$ & $\nu$ \\
\midrule
\multirow{4}{*}{\shortstack{CMASS\\LRG}}
& 0.6 & 0.65 & 0.90 & 2.80 & 0.240 & 175 \\
& 0.65 & 0.7 & 1.14 & 3.84 & 0.249 & 175 \\
& 0.7 & 0.8 & 1.37 & 4.19 & 0.252 & 175 \\
& 0.8 & 1.0 & 1.55 & 3.88 & 0.251 & 175 \\
\midrule
\multirow{5}{*}{\shortstack{eBOSS\\LRG}}
& 0.6 & 0.65 & 0.35 & 2.50 & 0.180 & 190 \\
& 0.65 & 0.7 & 0.63 & 3.46 & 0.205 & 190 \\
& 0.7 & 0.8 & 0.80 & 3.00 & 0.220 & 190 \\
& 0.8 & 0.9 & 1.05 & 3.79 & 0.257 & 190 \\
& 0.9 & 1.0 & 0.93 & 4.40 & 0.295 & 190 \\
\midrule
\multirow{7}{*}{\shortstack{eBOSS\\ELG}}
& 0.6 & 0.7 & 0.50 & 1.00 & 0.181 & 150 \\
& 0.7 & 0.75 & 0.50 & 1.00 & 0.180 & 150 \\
& 0.75 & 0.8 & 0.50 & 1.00 & 0.186 & 150 \\
& 0.8 & 0.85 & 0.50 & 1.00 & 0.195 & 150 \\
& 0.85 & 0.9 & 0.50 & 1.00 & 0.211 & 150 \\
& 0.9 & 1.0 & 0.50 & 1.00 & 0.243 & 150 \\
& 1.0 & 1.1 & 0.50 & 1.00 & 0.300 & 150 \\
\midrule
\multirow{7}{*}{\shortstack{eBOSS\\QSO}}
& 0.8 & 1.0 & 1.00 & 0.47 & 0.0100 & 200 \\
& 1.0 & 1.2 & 0.88 & 0.66 & 0.0089 & 217 \\
& 1.2 & 1.4 & 0.57 & 0.81 & 0.0057 & 330 \\
& 1.4 & 1.6 & 0.41 & 0.92 & 0.0033 & 415 \\
& 1.6 & 1.8 & 0.37 & 0.99 & 0.0017 & 474 \\
& 1.8 & 2.0 & 0.49 & 1.02 & 0.0010 & 501 \\
& 2.0 & 2.2 & 0.74 & 1.01 & 0.0011 & 501 \\
\bottomrule
\end{tabular}
\label{tab:ezmock_par}
\end{threeparttable}
\end{table}

\subsection{Observational effects}
\label{sec:ezmock_syst}

The {\it complete} set of \ez{} catalogues do not present the inhomogeneity of the angular distribution of tracers due to various observational effects, e.g. the quality of photometric and spectroscopic data. These effects are typically treated as systematics, and are (partially) corrected by imaging and spectroscopic weights \citep[e.g.][]{Ross2020, Raichoor2021}. To account for their impacts on the covariance matrices for clustering measurements, we generate more  realistic \ez{} catalogues by introducing observational effects to the {\it complete} mocks for eBOSS DR16 tracers.

\subsubsection{Depth dependent radial density}

For the eBOSS ELG \ez{} catalogues, we start from the {\it complete} realizations before applying radial selection (see Section~\ref{sec:ezmock_nbar}). This is because the imaging depth of the DECaLS data used for eBOSS ELG target selection is not homogenous inside eBOSS chunks, especially for \texttt{eboss23}, resulting in an imaging depth dependent number density of ELGs \citep[][]{Raichoor2017}. This effect is migrated to the \ez{} catalogues by the same strategy for generating the random catalogues for the observed data \citep[see][]{Raichoor2021}. Basically the $g$-, $r$-, and $z$-band imaging depths are combined linearly, and the radial number densities of ELGs are evaluated inside three quantiles of the combined depth, for each eBOSS chunk. The \ez{} ELGs are then split into the quantiles, and applied radial selections separately. In this case, the actual radial density of ELGs is anisotropic, and cannot be described by a simple redshift-dependent function. Consequently, the random catalogues for the {\it realistic} ELG \ez{} catalogues are generated using the `shuffled' scheme, by taking redshifts of galaxies in the quantiles separately.

\subsubsection{Angular photometric systematics}

Anisotropic effects that the photometric process carries, such as stellar density, Galactic extinction, seeing, and imaging depth, are correlated with the angular distributions of the samples for large-scale analysis \citep[e.g.][]{Ross2017, Xavier2019}. To mimic these effects in \ez{} catalogues, we extract an angular map of the photometric properties from the imaging sample, and randomly discarding mock tracers with the probability following this map. For LRGs and QSOs, the map is generated by linear regressions for different photometric effects \citep[][]{Ross2020}, while for ELGs we use directly a smoothed angular target density map of the data, with a beam size of 1\,{\rm deg} \citep[][]{deMattia2021}. The corrections are then done by adding photometric weights to the mocks, which are estimated by linear regressions to the angular completeness \citep[see][for details]{Ross2020, Raichoor2021} for each mock realization individually, thus allowing stochasticity for the systematic weights.

\subsubsection{Fibre collision}

Due to the finite size of optical fibres, the spectra of two targets with the angular separation less than 62\,arcsec cannot be both measured with one plate. Thus, one of the targets has to be rejected if they do not reside in sectors covered by different plates.

We use the angular friends-of-friends (FoF) algorithm provided in the \textsc{nbodykit}\footnote{\url{https://github.com/bccp/nbodykit}} \citep[][]{Hand2018} package, to detect groups of \ez{} tracers that are in collision, and mark objects to be removed. Then, the groups are distributed to the sectors of the observational data, and some of the collisions can be resolved when the objects are in sectors belonging to multiple plates.

To correct the clustering statistics with fibre collision, remaining mock tracers in collision groups are up-weighted by the ratio between the original number of targets, and the number of assigned fibres, for each of the groups \citep[cf.][for more investigations on the fibre collision weights]{Hou2021}. The fibre collision effects on the configuration space measurements can be further suppressed by the pairwise-inverse-probability (PIP) weighting scheme, which is an unbiased procedure for all scales \citep[][]{Mohammad2020}.

\subsubsection{Redshift failure}

Reliable redshifts are not always obtained from the spectra in practice. The redshift failure rate $f_{\rm fail}$ for the eBOSS data are modelled by regressions with the signal-to-noise ratio of the spectra, as well as IDs and positions on the focal plane of optical fibres \citep[][]{Ross2020, Raichoor2021}.
This effect is introduced to the \ez{} catalogues with a similar approach for eBOSS DR14 LRG QPM mocks \citep[][]{Bautista2018}.
We associate \ez{} objects with the fibre of the closest valid eBOSS tracer, and randomly down-sample mocks according to the modelled redshift failure rate of the data. We then use the same procedure as with the data to fit our model for $f_{\rm fail}$ for each individual mock, and the remained mock tracers are up-weighted by $1 / (1 - f_{\rm fail})$.

\section{Results: statistical comparison between EZmock catalogues and BOSS/eBOSS data}
\label{sec:result}

We generate 1000 realizations of \ez{} catalogues, for each of the dataset, i.e., BOSS DR12 CMASS LRG, and eBOSS DR16 LRG/ELG/QSO. Thus, 46,000 \ez{} boxes are generated, with the side length of $5\,h^{-1}\,{\rm Gpc}$, for the 23 redshift slices listed in Table~\ref{tab:ezmock_zbin}, and both northern and southern Galactic caps. The number of tracers for each of the LRG, ELG, and QSO boxes are $4\times 10^7$, $8 \times 10^7$, and $3 \times 10^6$, respectively. It takes $\sim 1$ million CPU hours in total, to generate the {\it complete} set of \ez{} mock light-cone catalogues, on the Cori Haswell nodes of the National Energy Research Scientifc Computing Center (NERSC)\footnote{\url{https://docs.nersc.gov/systems/cori}}. The \textsc{ezmock} code is parallelized with OpenMP, and multiple realizations are run simultaneously with the \textsc{jobfork}\footnote{\url{https://github.com/cheng-zhao/jobfork}} tool, which distributes serial or OpenMP based jobs to multiple computing nodes using MPI.

In this section, we present various statistical properties of the \ez{} catalogues and compare them with those from the BOSS/eBOSS data. In particular, results of both the {\it complete} and {\it realistic} mocks are shown.
Moreover, we measure the clustering statistics for the {\it complete} mocks with both the `sampled' and `shuffled' random catalogues, and the results are denoted by `EZmock comp.' and `EZmock R-shuf.', respectively. While results for the {\it realistic} mocks are always obtained using the `shuffled' random catalogues (denoted by `EZmock syst.').
Note however that the {\it realistic} joint BOSS and eBOSS LRG samples (denoted by `COMB BOSS') are constructed with the combination of the {\it complete} CMASS LRG mocks and {\it realistic} eBOSS LRG mocks.
The meanings of these notations are summarised in Table~\ref{tab:ez_notation}.

\begin{table}
\centering
\begin{threeparttable}
\caption{A list of notations for different \ez{} samples.}
\begin{tabularx}{.95\columnwidth}{cX}
\toprule
Notation & Description \\
\midrule
\ez{} comp. & {\it Complete} mocks with `sampled' randoms: no  observational systematics, and redshifts of the random catalogues are sampled from the spline interpolation of the $n(z)$ of observational data. \\
\midrule
\ez{} R-shuf. & {\it Complete} mocks with `shuffled' randoms: no observational systematics, and redshifts of the random catalogues are taken randomly from the corresponding data catalogues. \\
\midrule
\ez{} syst. & {\it Realistic} mocks with `shuffled' randoms: all known observational systematics are applied to the data and random catalogues, and redshifts of the random catalogues are taken randomly from the corresponding data catalogues. \\
\bottomrule
\end{tabularx}
\label{tab:ez_notation}
\end{threeparttable}
\end{table}

Note that the clustering measurements of the {\it complete} mocks are used for \ez{} parameter calibration (see Section~\ref{sec:ezmock_calib}), while the covariance matrices of the {\it realistic} mocks are our final products for the data analyses.
The fiducial cosmological model used for coordinate conversion hereafter, is flat $\Lambda$CDM with $\Omega_{\rm m} = 0.31$ (see Eqs~\eqref{eq:coord_trans1} -- \eqref{eq:coord_trans4}).

\subsection{Spatial distribution}
\label{sec:result_spacial}

\begin{figure}
\centering
\includegraphics[width=.95\columnwidth]{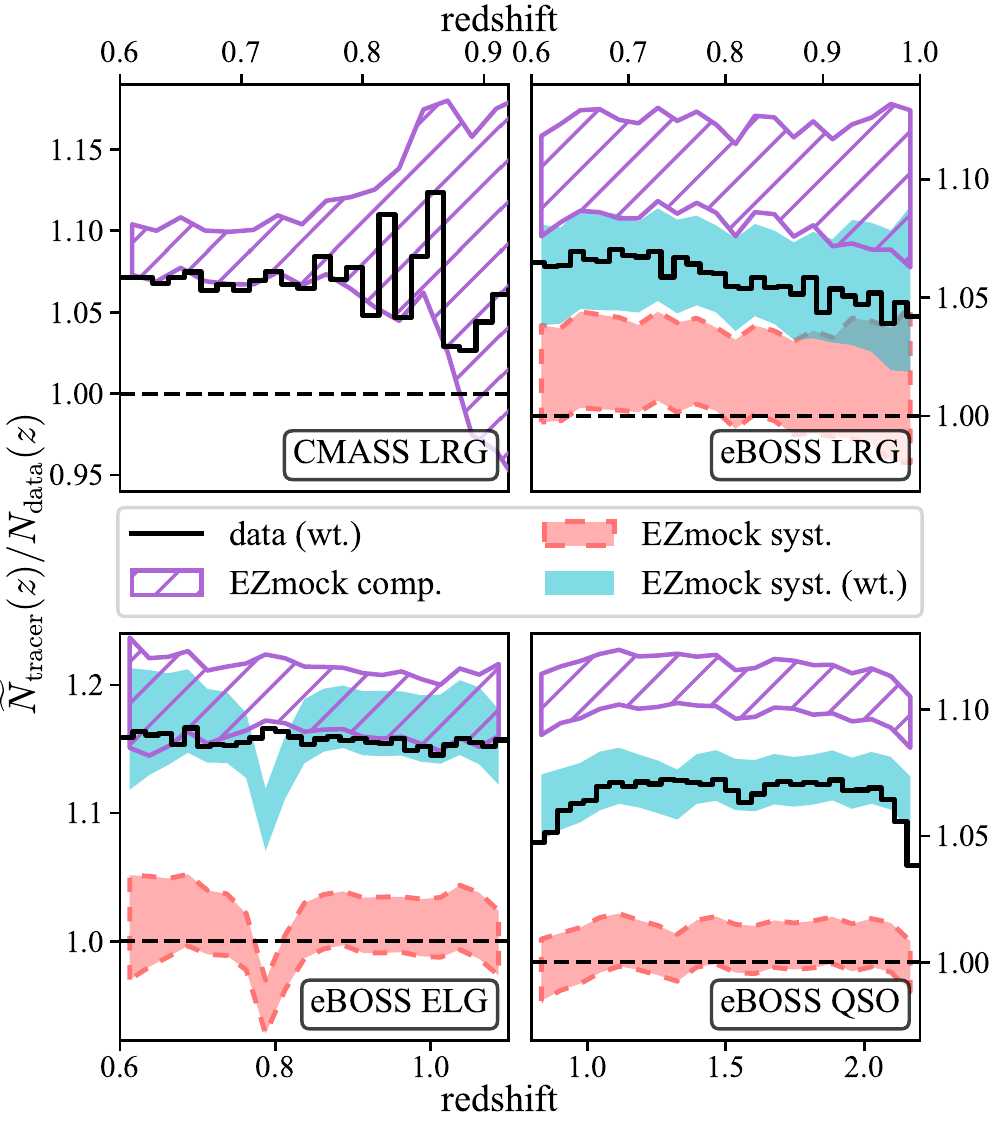}
\caption{(Weighted) tracer distribution of the BOSS/eBOSS data and \ez{} catalogues, normalized by the number of objects in the corresponding data catalogues. `EZmock comp.' and `EZmock syst.' denote the {\it complete} and {\it realistic} \ez{} catalogues respectively, and `wt.' indicates results evaluated with weights, which are the total photometric and spectroscopic weights used for clustering analyses. The upper and lower boundaries of the filled regions show the 1\,$\sigma$ deviation obtained from 1000 realizations of mocks.}
\label{fig:ezmock_zhist}
\end{figure}

The radial distributions of the {\it complete} \ez{} catalogues in comoving space follows those measured from the data with all photometric and systematic weights by construction (see Eq.~\eqref{eq:ezmock_radsel}).
However, the fraction of targets without fibres \citep[$C_{\rm (e)BOSS}$; see][]{Reid2016, Ross2020} are not considered by the weights. Thus, there can be discrepancies on the actual weighted radial counts between data and the corresponding mocks.
This can be seen in Fig.~\ref{fig:ezmock_zhist}, where the comparisons between \ez{} tracers and BOSS/eBOSS data are shown, in terms of the (weighted) number of objects at different redshifts.

For the eBOSS samples, the number of targets without fibres is about 3.4 per cent of the total weighted number of LRGs, and the fractions are 0.9 and 2.3 per cent for ELGs and QSOs respectively. These numbers are consistent with the mismatch between \ez{} catalogues and eBOSS data illustrated in Fig.~\ref{fig:ezmock_zhist}. This effect is due to the definition of the effective area for measuring $n_{\rm data}$, and for sectors with $C_{\rm (e)BOSS} = 1$, the radial comoving number density of tracers from the mocks and data are still consistent (see Appendix~\ref{sec:fourier_norm} for more discussions).

To have accurate estimates of the clustering covariance matrices, it is necessary to reproduce faithfully the sample size of the observational data. Hence, the effect of $C_{\rm eBOSS}$ is considered in the {\it realistic} \ez{} catalogues.
Moreover, after including both photometric and spectroscopic effects (see Section~\ref{sec:ezmock_syst}), a considerable fraction of the mock tracers are removed. Consequently, the number of objects in the mocks and data become more comparable, though they are still not identical, since the small-scale clustering of \ez{} catalogues does not allow precise reproduction of some of the observational systematics, such as fibre collision. Finally, the systematics of the {\it realistic} \ez{} catalogues are corrected by various weights. Thus, the weighted redshift distribution of mocks and data agree well again, as shown in Fig.~\ref{fig:ezmock_zhist}.

Furthermore, since the number density of the cubic \ez{} ELG catalogues ($6.4\times 10^{-4}\,h^3\,{\rm Mpc}^{-3}$) are only slightly larger than the peak density of the eBOSS data in chunk \texttt{eboss22}, after down-sampling with observational systematics, the density of \ez{} ELGs at $z\sim 0.8$ are lower than that of the eBOSS data by at most 5 per cent. We then rescale the radial selection function (see Section~\ref{sec:ezmock_nbar}) of ELGs in chunk \texttt{eboss22}, to obtain the correct number of objects in the full sample. Since this affects only a small number of \ez{} ELGs, the consequences on the covariance matrices are sub-dominant.

\begin{figure*}
\centering
\includegraphics[width=.98\textwidth]{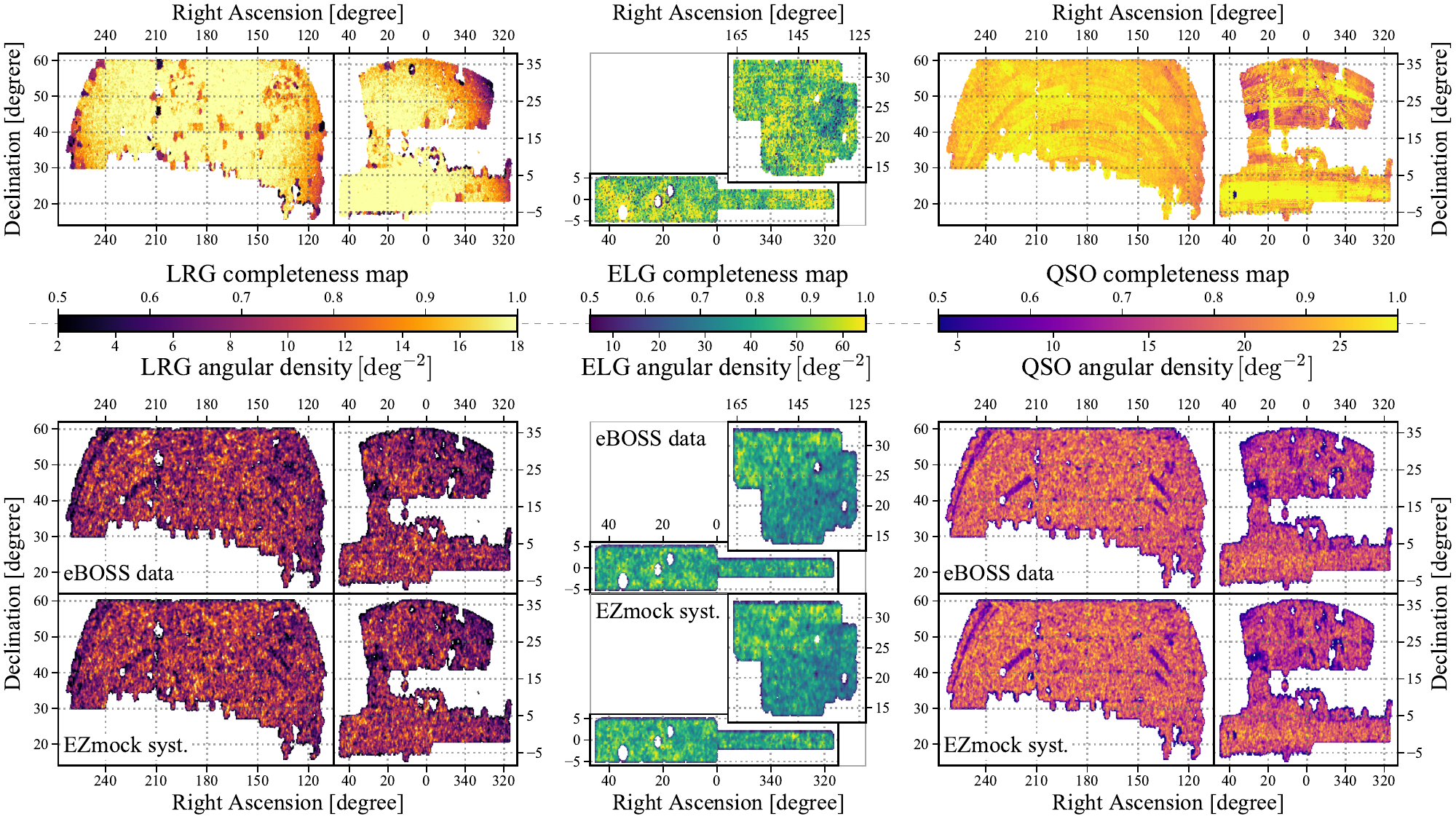}
\caption{\textit{Top} panel: angular completeness map of eBOSS DR16 tracers, modelled with the observational effects discussed in Section~\ref{sec:ezmock_syst}. \textit{Bottom} panel: angular density distribution of tracers in the eBOSS data (first row), and one realization of \ez{} catalogues with observational systematics (second row).}
\label{fig:ezmock_map}
\end{figure*}

Fig.~\ref{fig:ezmock_map} shows that angular systematic map extracted from the eBOSS DR16 data -- including all the effects discussed in Section~\ref{sec:ezmock_syst} -- as well as the comparison of the unweighted angular tracer density between the data and one arbitrary \ez{} realization. Note however that for better illustration, veto masks due to bad photometric calibrations are not shown for ELG SGC \citep[cf.][]{Raichoor2021}. The large-scale angular distribution of both data and \ez{} catalogues agree well with the systematic map: for regions with low completeness, the tracer densities are also low. Moreover, the small-scale clustering pattern of the data and mocks are also similar. We shall compare the clustering statistics quantitatively in the next section.

\subsection{Configuration space clustering}

We express the anisotropic 2PCF in two ways, the 2D 2PCF $\xi (s_{\parallel}, s_{\perp})$, and 2PCF multipoles $\xi_\ell (s)$. Here, $s$ denotes the separation of galaxy pairs, and $s_{\parallel}$ and $s_{\perp}$ indicate the projected separation along and perpendicular to the line-of-sight, respectively. To measure both quantities from the catalogues, we rely on the Landy--Szalay estimator \citep[][]{Landy1993}:
\begin{equation}
\xi = \frac{ {\rm DD} - 2 {\rm DR} + {\rm RR} }{{\rm RR}} ,
\label{eq:xi_LS}
\end{equation}
where DD, DR, and RR stand for the number of data--data, data--random, and random--random pairs, normalized by the total number of pairs, respectively. In practice, we use the Fast Correlation Function Calculator\footnote{\url{https://github.com/cheng-zhao/FCFC}} (\textsc{FCFC}; Zhao et al. in preparation) to count pairs of tracers in the catalogues.

\subsubsection{2D two-point correlation function}
\label{sec:xi2d}

Denoting the positions of two galaxies as $\boldsymbol{s}_1$ and $\boldsymbol{s}_2$, the separation of the pair $\boldsymbol{s} = \boldsymbol{s}_2 - \boldsymbol{s}_1$, and the line-of-sight vector is defined as
\begin{equation}
\boldsymbol{l} = \frac{\boldsymbol{s}_1 + \boldsymbol{s}_2}{2} .
\end{equation}
The two projected separations are then
\begin{align}
s_\parallel &= \frac{ \boldsymbol{s} \cdot \boldsymbol{l} }{|\boldsymbol{l}|} , \\
| s_\perp | &= \sqrt{ |\boldsymbol{s}|^2 - s_\parallel^2 } .
\end{align}
The sign of $s_\perp$ is typically defined by the order of the two galaxies, and pair counts are symmetric about both $s_\parallel = 0$ and $s_\perp = 0$.

\begin{figure*}
\centering
\includegraphics[width=.8\textwidth]{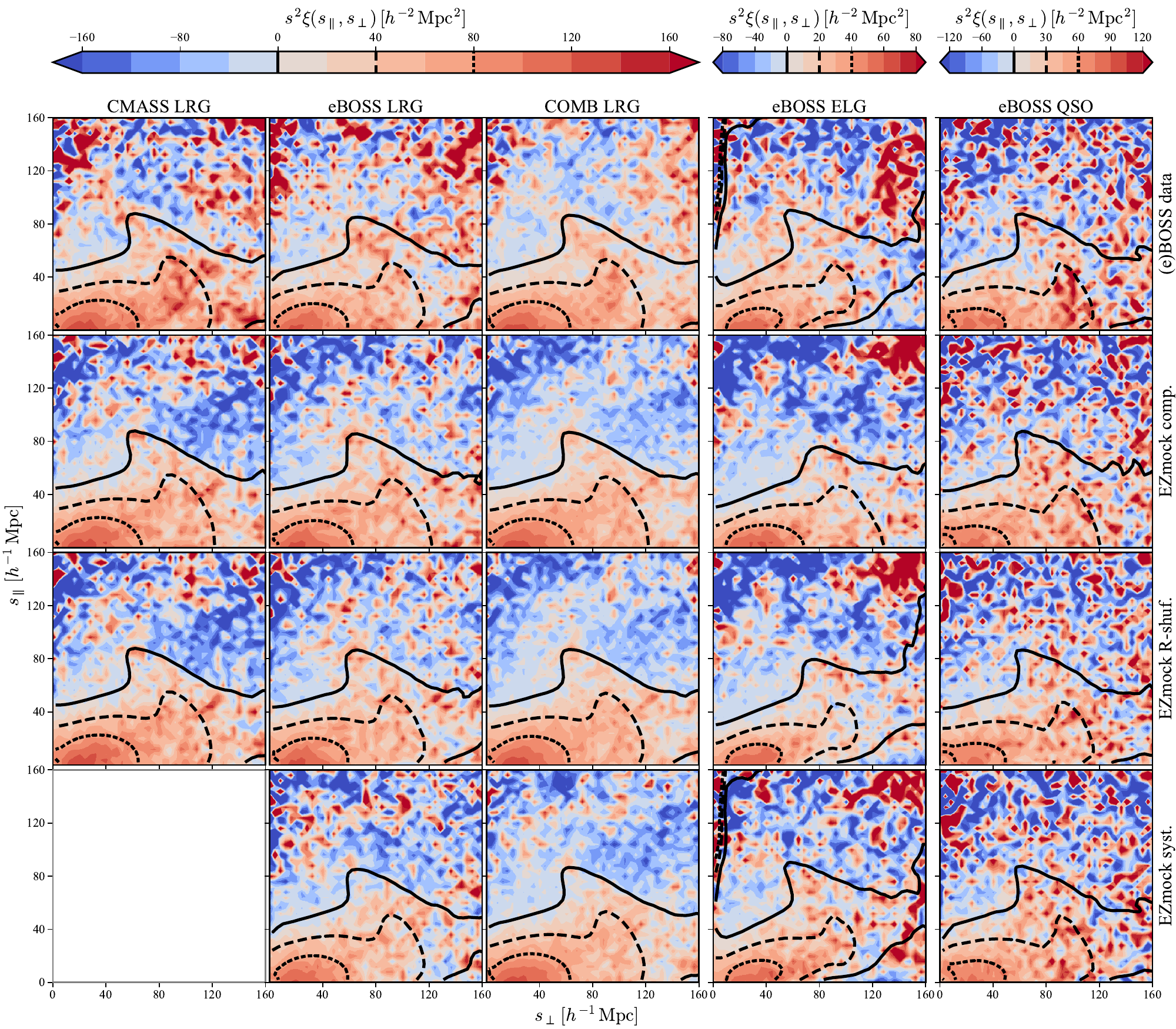}
\caption{2D two-point correlation function $\xi (s_\parallel, s_\perp)$ of the BOSS/eBOSS data ({\it first} row), the {\it complete} ({\it second} and {\it third} rows for the `sampled' and `shuffled' random catalogues respectively) and {\it realistic} ({\it fourth} row) \ez{} catalogues. Only the first quadrant is shown, since $\xi (s_\parallel, s_\perp)$ is symmetric about both $s_\parallel = 0$ and $s_\perp = 0$. Pair counts for NGC and SGC are combined. `COMB LRG' denotes the joint sample with both BOSS and eBOSS LRGs. The colour plots are obtained from single realizations, while the contour lines indicate the averaged results of all the mocks. In particular, in the {\it first} row, the contour lines for the CMASS sample are computed from the {\it complete} mocks with `shuffled' randoms, while for the other samples they are obtained using the {\it realistic} mocks.}
\label{fig:ezmock_xi2d}
\end{figure*}

The 2D 2PCF of the BOSS DR12 and eBOSS DR16 data, as well as the corresponding \ez{} catalogues, are shown in Fig.~\ref{fig:ezmock_xi2d}. In particular, the colour plots show $\xi (s_{\parallel}, s_{\perp})$ for single catalogues, while the contour lines for the mocks indicate levels (see the colour bars) of the mean results obtained from all the 1000 mock realizations, and pair counts for both Galactic caps are combined. On scales smaller than $\sim 120\,h^{-1}\,{\rm Mpc}$, the results from the mocks are generally consistent with those of the data.

By using the `shuffled' random catalogues, the 2PCFs are suppressed when $s_\parallel$ is small, especially for ELGs. The effect is more obvious on large $s_\perp$, as the 2PCFs are rescaled by $s^2$. This is because the data and random have common redshifts, resulting in a higher chance to find data--random pairs with $s_\parallel \sim 0$, compared to the case with `sampled' random catalogues. The 2PCFs are then reduced according to the Landy--Szalay estimator (Eq.~\eqref{eq:xi_LS}). Moreover, since the angular area of the ELG distribution is smaller, this effect starts to be evident from smaller scales.

The impacts of observational effects are also more significant for ELGs, due to the relatively more complicated sources of systematics \citep[see][for details]{Raichoor2021}. Apart from distortions on BAO scale, we also observe excess clustering strength on small angular scales ($s_\perp \sim 0$).

\subsubsection{Two-point correlation function multipoles}

The 2D 2PCF can also be expressed as $\xi (s, \mu)$, where $s=|\boldsymbol{s}|$, and $\mu = s_\parallel / s$. Furthermore, $\mu = \cos \theta$, with $\theta$ being the intersection angle between $\boldsymbol{s}_1$ and $\boldsymbol{s}_2$. The full 2D 2PCF can then be decomposed into a series of 1-D projections, by weighting the angular components with Legendre polynomials $\mathcal{L}_\ell (\mu)$:
\begin{equation}
\xi_\ell (s) = \frac{2 \ell + 1}{2} \int_{-1}^{1} \xi (s, \mu) \mathcal{L}_\ell (\mu) \, {\rm d}\mu .
\label{eq:xi_ell}
\end{equation}
Since the correlation function is symmetric about $\mu = 0$, only the even multipoles ($\ell = 0, 2, 4, \dots$) are relevant.

\begin{figure*}
\centering
\includegraphics[width=.98\textwidth]{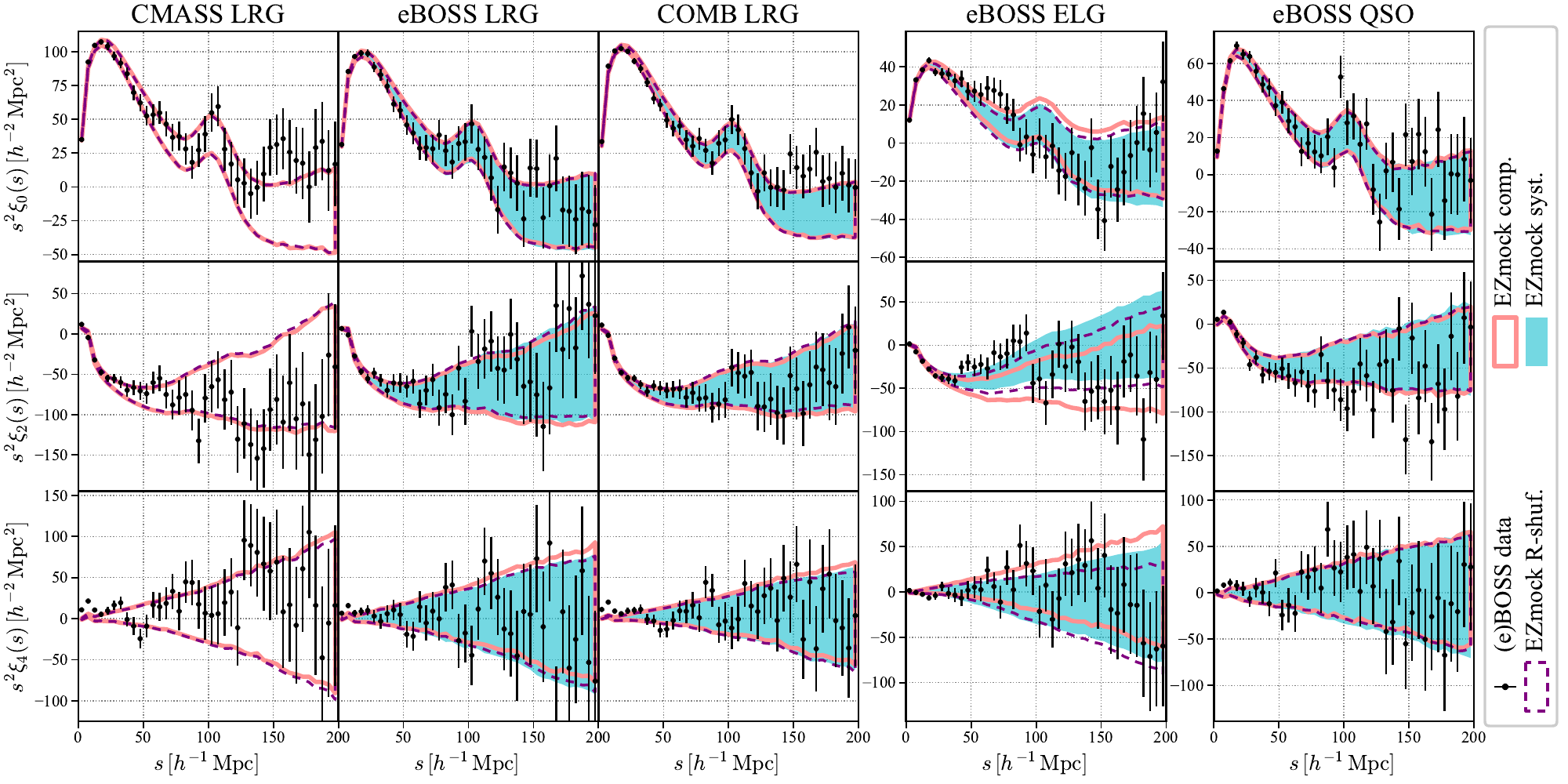}
\caption{Two-point correlation function multipoles of the BOSS/eBOSS data and the corresponding \ez{} catalogues in NGC. The solid/dashed envelopes and shadowed areas indicate the 1\,$\sigma$ regions evaluated from 1000 mock realizations. The error bars for the CMASS LRG sample are obtained from the {\it complete} \ez{} catalogues, while for the other tracers they are taken from the {\it realistic} mocks with systematics.}
\label{fig:ezmock_xi_ngc}
\end{figure*}

\begin{figure*}
\centering
\includegraphics[width=.98\textwidth]{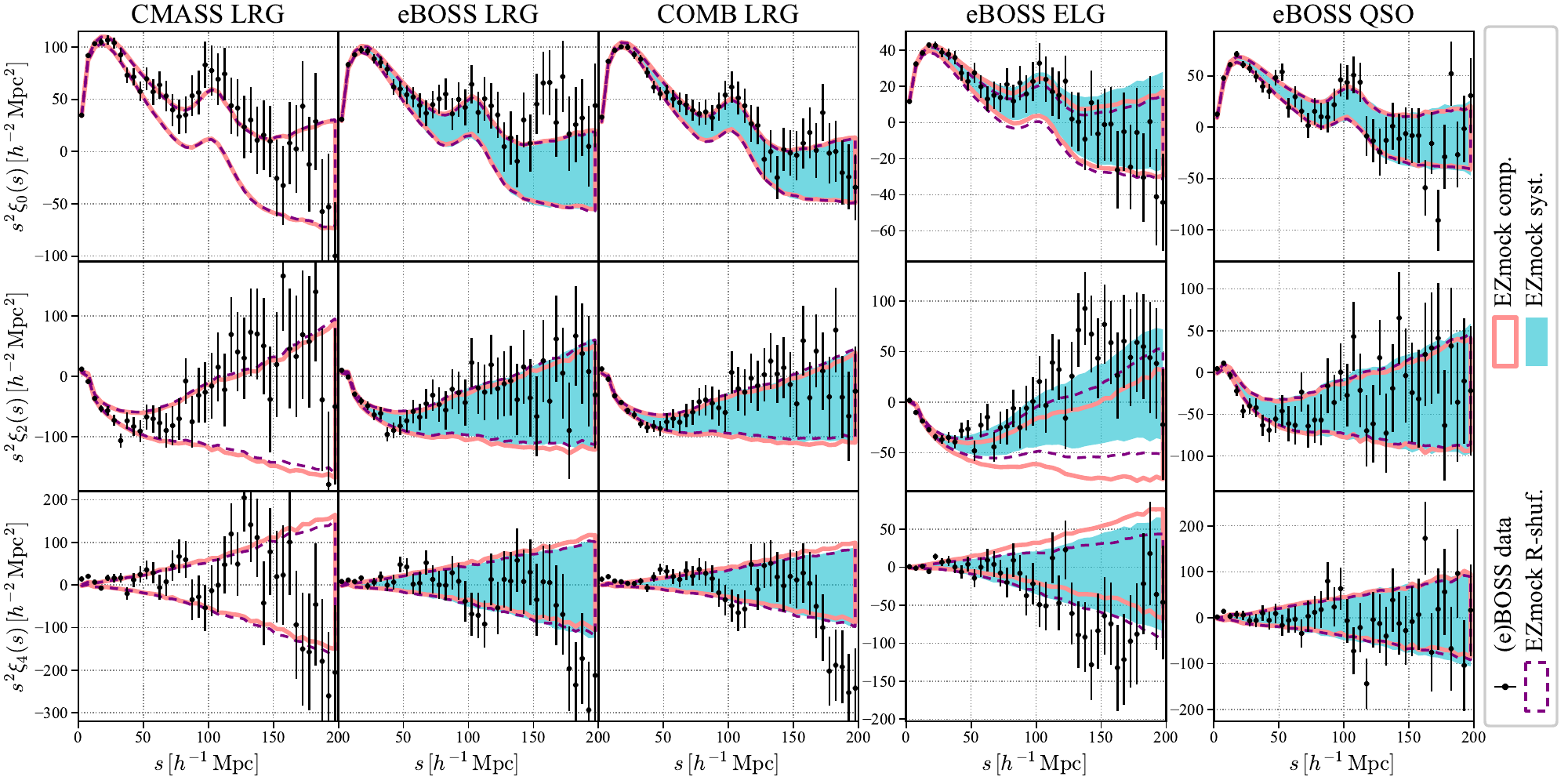}
\caption{Two-point correlation function multipoles of the BOSS/eBOSS data and the corresponding \ez{} catalogues in SGC. The solid/dashed envelopes and shadowed areas indicate the 1\,$\sigma$ regions evaluated from 1000 mock realizations. The error bars for the CMASS LRG sample are obtained from the {\it complete} \ez{} catalogues, while for the other tracers they are taken from the {\it realistic} mocks with systematics.}
\label{fig:ezmock_xi_sgc}
\end{figure*}

For the BOSS/eBOSS data and \ez{} catalogues, we compute the 2PCF monopole ($\ell = 0$), quadrupole ($\ell = 2$), and hexadecapole ($\ell = 4$), with 240 $\mu$ bins from $-1$ to $1$, and 40 $s$ bins from $0$ to $200\,h^{-1}\,{\rm Mpc}$, and the results are shown in Figs~\ref{fig:ezmock_xi_ngc} and \ref{fig:ezmock_xi_sgc}, for NGC and SGC, respectively. On scales down to $\sim 10\,h^{-1}\,{\rm Mpc}$, the 2PCF multipoles of the observational data are well recovered by the corresponding \ez{} catalogues, especially for the {\it realistic} mocks. Indeed, deviations over 1\,$\sigma$ are mainly observed on fairly large scales ($s \gtrsim 150\,h^{-1}\,{\rm Mpc}$), where the impact of observational systematics are relatively more obvious.
A quantitative consistency check between the data and mocks is done in Section~\ref{sec:ezmock_check}.

Furthermore, the 2PCFs measured from the `sampled' and `shuffled' random catalogues differ mainly in the quadrupole and hexadecapole. This is because the differences are only found at fairly small $s_\parallel$. Thus they are more obvious in anisotropic multipole measurements. Besides, observational systematic effects do not play important roles on the 2PCF multipoles for LRGs and QSOs. While for ELGs their impacts are significant.

\begin{figure*}
\centering
\includegraphics[width=.9\textwidth]{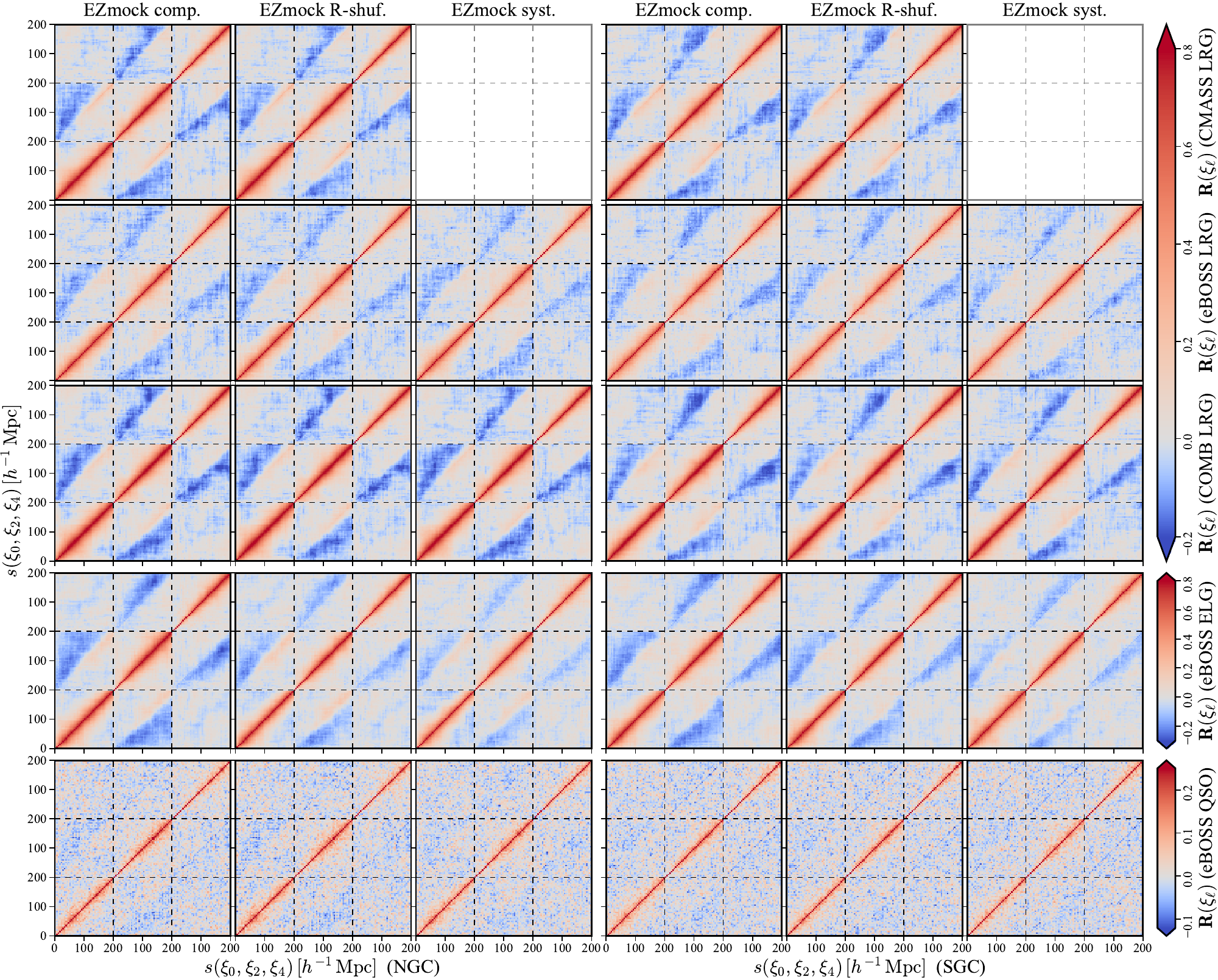}
\caption{Correlation matrices of two-point correlation function multipoles obtained from 1000 \ez{} realizations.}
\label{fig:ezmock_xi_cov}
\end{figure*}

The covariance matrix of the correlation function multipole $\xi_\ell (s)$ can be estimated as
\begin{equation}
\mathbf{C}_{i j} = \frac{1}{N_{\rm m} - 1} \sum_{k}^{N_{\rm m}} \left[ \xi_{\ell, k} (s_i) - \bar{\xi}_\ell (s_i) \right] \left[ \xi_{\ell, k} (s_j) - \bar{\xi}_\ell (s_j) \right] ,
\label{eq:cov_mat}
\end{equation}
where $N_{\rm m}$ is the number of mock realizations, $\xi_{\ell, k} (s)$ denotes the 2PCF multipole of the $k$-th mock with separation $s$, and $\bar{\xi}_\ell $ indicates the mean 2PCF multipole of all the mocks. For illustrative purposes, we further compute the normalized covariance matrices (i.e. correlation matrices) of the 2PCF multipoles:
\begin{equation}
\mathbf{R}_{i j} = \frac{\mathbf{C}_{i j} }{ \sqrt{ \mathbf{C}_{i i} \cdot \mathbf{C}_{j j} } },
\label{eq:corr_mat}
\end{equation}
and the results from different sets of \ez{} catalogues are shown in Fig.~\ref{fig:ezmock_xi_cov}.
The results from the `sampled' and `shuffled' random catalogues are only noticeably different for ELGs, while observational systematics do alter the covariance matrices of LRGs and ELGs, especially for the cross covariance between different multipoles.

\subsection{Fourier space clustering}
\label{sec:result_fourier}

In Fourier space, we measure the two- and three-point statistics, i.e., power spectrum and bispectrum, following the estimators described in \citet[][]{Sefusatti2005}. We start with the weighted tracer density field $F (\boldsymbol{r})$:
\begin{equation}
F (\boldsymbol{r}) = w_{\rm FKP} (\boldsymbol{r}) \left[ n_{\rm t} (\boldsymbol{r}) - \alpha n_{\rm r} (\boldsymbol{r}) \right] ,
\end{equation}
where $n_{\rm t} (\boldsymbol{r})$ and $n_{\rm r} (\boldsymbol{r})$ denote the weighted number density fields of the data and random catalogues, respectively. And $\alpha$ indicates the ratio of the total weighted number of objects in the data catalogue, to that of the random catalogue. In particular, the weights involved for $n_{\rm t}$, $n_{\rm r}$, and $\alpha$ are the total but FKP weights.

The power spectrum and bispectrum are then estimated by
\begin{align}
P (\boldsymbol{k}) &= \left. \left[ \left\langle \hat{F} (\boldsymbol{k}) \hat{F} (-\boldsymbol{k}) \right\rangle - ( 1 + \alpha ) I_{12} \right] \middle/ I_{22} \right. , \label{eq:pk_estimator} \\
B (\boldsymbol{k}_1, \boldsymbol{k}_2, \boldsymbol{k}_3 ) &= \Big\{ \left\langle \hat{F} (\boldsymbol{k}_1) \hat{F} (\boldsymbol{k}_2) \hat{F} (\boldsymbol{k}_3) \right\rangle \nonumber \\
&\qquad - \left[ P(\boldsymbol{k}_1) + P(\boldsymbol{k}_2) + P(\boldsymbol{k}_3) \right] I_{23} \\
&\qquad - \left. ( 1 - \alpha^2 ) I_{13} \Big\} \middle/ I_{33} \right. , \nonumber
\end{align}
where $\hat{F} (\boldsymbol{k})$ denotes the Fourier transform of $F (\boldsymbol{r})$, the angle brackets $\langle \, \cdot \, \rangle$ indicate the average over the full survey volume, and the constant terms are given by
\begin{equation}
I_{a b} = \int {\rm d}^3 r \, n_{\rm t}^a (\boldsymbol{r}) w_{\rm FKP}^b (\boldsymbol{r}) .
\label{eq:fourier_norm}
\end{equation}
Moreover, for bispectrum, $\boldsymbol{k}_1 + \boldsymbol{k}_2 + \boldsymbol{k}_3 = \boldsymbol{0}$.
It is worth noting that the normalization factors of the Fourier space measurements are sensitive to the measured comoving densities of tracers (see Appendix~\ref{sec:fourier_norm} for more discussions).

To obtain the tracer density field, the data and random catalogues are placed into cuboids with adaptive side lengths. Note however that given a specific tracer sample, the size of cuboids for the observational data and the corresponding mocks are identical. Besides, we distribute tracers to 3D regular grids using the triangular shaped cloud \citep[TSC;][]{Hockney1981} scheme, and correct the aliasing effects with the grid interlacing technique \citep[][]{Sefusatti2016}.

\subsubsection{Power spectrum multipoles}
\label{sec:pk}

Similar to the 2PCF multipoles, the anisotropic power spectrum can also be decomposed with Legendre polynomials. In this case, Eq.~\eqref{eq:pk_estimator} can be rewritten as \citep[e.g.][]{Yamamoto2006, Beutler2017, Blake2018}
\begin{equation}
P_\ell (k) = \left. \left[ (2\ell + 1) \left\langle \hat{F}_0 (\boldsymbol{k}) \hat{F}_\ell (-\boldsymbol{k}) \right\rangle - ( 1 + \alpha ) I_{12} \right] \middle/ I_{22} \right. ,
\label{eq:pk_ell}
\end{equation}
where
\begin{equation}
\hat{F}_\ell (\boldsymbol{k}) = \int {\rm d}^3 r \, F (\boldsymbol{r}) \, {\rm e}^{i \boldsymbol{k} \cdot \boldsymbol{r}} \mathcal{L}_\ell \left( \frac{\boldsymbol{k}}{|\boldsymbol{k}|} \cdot \frac{\boldsymbol{r}}{|\boldsymbol{r}|} \right) .
\end{equation}
In practice, we use the \textsc{powspec}\footnote{\url{https://github.com/cheng-zhao/powspec}} code to compute power spectrum multipoles, with the estimator introduced by \citet[][]{Hand2017}. For the clustering measurements hereafter, we choose the grid size of $512^3$ for the LRG and ELG density fields, and $1024^3$ for the QSO sample,\footnote{However, for efficiency considerations, we use $512^3$ grids for the calibration of the \ez{} QSO sample. In this case, the Nyquist frequency for the NGC and SGC samples are $\sim 0.24$ and $0.3\,h\,{\rm Mpc}^{-1}$, respectively.
Consequently, \ez{} QSO catalogues in the NGC are only calibrated with the range $k \in [0.1, 0.24]\,h\,{\rm Mpc}^{-1}$. While for the rest of the mock samples, the calibrations of the power spectra are all performed with $k \in [0.1, 0.3]\,h\,{\rm Mpc}^{-1}$} to ensure that the Nyquist frequency for all the tracers are larger than $0.3\,h\,{\rm Mpc}^{-1}$.

\begin{figure*}
\centering
\includegraphics[width=.98\textwidth]{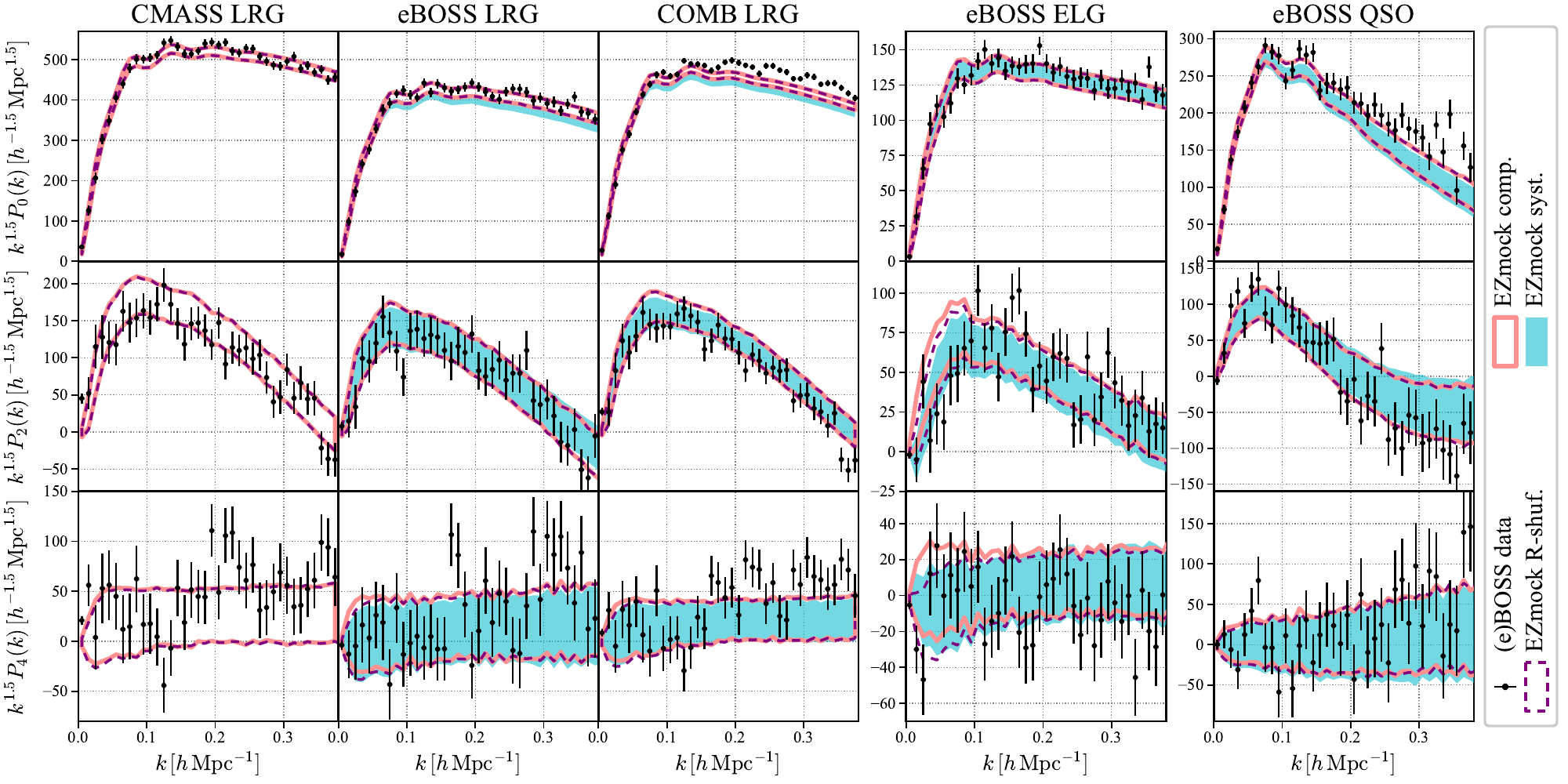}
\caption{Power spectrum multipoles of the BOSS/eBOSS data and the corresponding \ez{} catalogues in NGC. The solid/dashed envelopes and shadowed areas indicate the 1\,$\sigma$ regions evaluated from 1000 mock realizations. The error bars for the CMASS LRG sample are obtained from the {\it complete} \ez{} catalogues, while for the other tracers they are taken from the {\it realistic} mocks with systematics.}
\label{fig:ezmock_pk_ngc}
\end{figure*}

\begin{figure*}
\centering
\includegraphics[width=.98\textwidth]{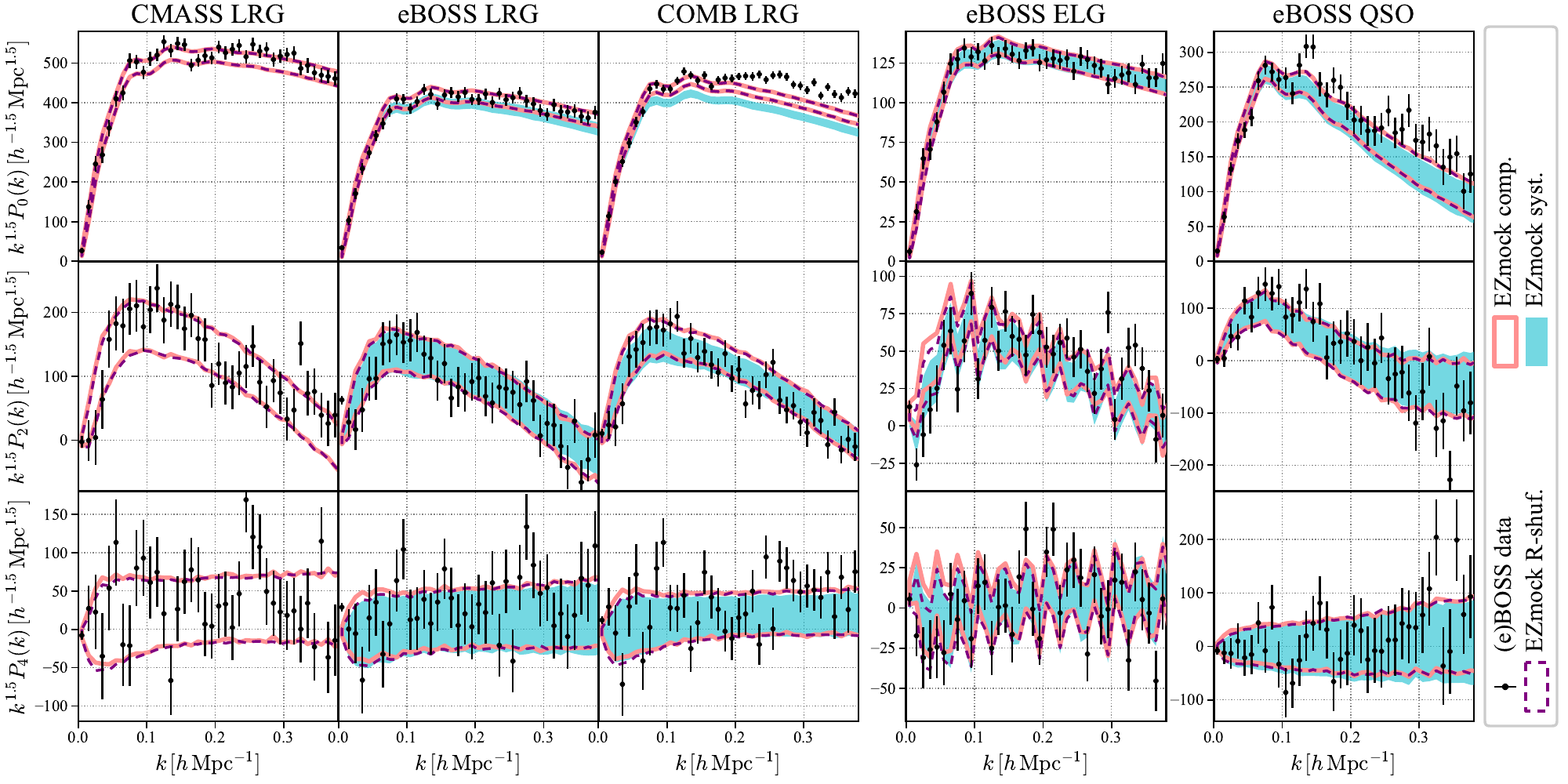}
\caption{Power spectrum multipoles of the BOSS/eBOSS data and the corresponding \ez{} catalogues in SGC. The solid/dashed envelopes and shadowed areas indicate the 1\,$\sigma$ regions evaluated from 1000 mock realizations. The error bars for the CMASS LRG sample are obtained from the {\it complete} \ez{} catalogues, while for the other tracers they are taken from the {\it realistic} mocks with systematics.}
\label{fig:ezmock_pk_sgc}
\end{figure*}

The power spectra monopole ($\ell = 0$), quadrupole ($\ell = 2$), and hexadecapole ($\ell = 4$) for the BOSS/eBOSS data and the corresponding \ez{} catalogues are shown in Figs~\ref{fig:ezmock_pk_ngc} and \ref{fig:ezmock_pk_sgc}, for NGC and SGC respectively, with the bin size of $0.01\,h\,{\rm Mpc}^{-1}$. It can be seen that the differences on the actual number density of tracers between the {\it complete} and {\it realistic} mocks (see Section~\ref{sec:result_spacial}) result in significant biases of the power spectrum amplitude, especially for the monopole of LRGs, which are further enhanced visually due to the small errors. This is because the isotropic number density evaluations are incorrect for the {\it realistic} mocks, resulting in biased normalization factors (see Eq.\eqref{eq:fourier_norm}). Nevertheless, this effect does not alter significantly covariance matrices estimations, provided a constant rescaling (see Section~\ref{sec:fourier_norm} for details).

Apart from the discrepancies on the broad-band amplitude, observational systematics and the `shuffled' random catalogue affects mainly power spectra quadrupole and hexadecapole at $k \lesssim 0.1\,h\,{\rm Mpc}^{-1}$. In general, the measurements from the observational data and mocks are in good agreement. Nevertheless, deviations over 1\,$\sigma$ are seen in the power spectra monopole, at $k \gtrsim 0.25\,h\,{\rm Mpc}^{-1}$ for the eBOSS QSO sample, and $k \gtrsim 0.15\,h\,{\rm Mpc}^{-1}$ for the combined LRG sample.
Since only the eBOSS QSO SGC data is used for the calibration of \ez{} QSO catalogues at $k \gtrsim 0.24\,h\,{\rm Mpc}^{-1}$, it turns out that the data from a single Galactic cap is not enough for optimal \ez{} calibrations at large $k$.

For the joint CMASS and eBOSS LRG sample, there is an additional mismatch at high $k$, this may be due to the fact that small scale cross correlations between the BOSS and eBOSS LRGs are not precisely modelled in \ez{} catalogues. Since the mocks for the two samples are calibrated separately, their cross correlations are only taken into account through the common dark matter density fields. However, both the inaccuracy of ZA on small scales, and the relatively low resolution of the \ez{} density fields ($\sim 5\,h^{-1}\,{\rm Mpc}$) prevent precise reproduction of the cross correlations in Fourier space. Similar effects on the cross power spectra between different types of tracers are also observed in Section~\ref{sec:cross_clustering}. We leave a thorough investigation of this issue to a future study.

Furthermore, in Figure~\ref{fig:ezmock_pk_sgc} we observe some oscillatory patterns in the power spectrum quadrupole and hexadecapole for eBOSS ELGs. They are less significant if placing the catalogues into a large box for FFT, see \citet[][]{deMattia2021}, where a box size of 4\,$h^{-1}\,{\rm Gpc}$ is used. This effect may also be suppressed by the multipole estimations with the regression method \citep[][]{Wilson2016}, but a detailed investigation is outside the scope of this paper.

\begin{figure*}
\centering
\includegraphics[width=.9\textwidth]{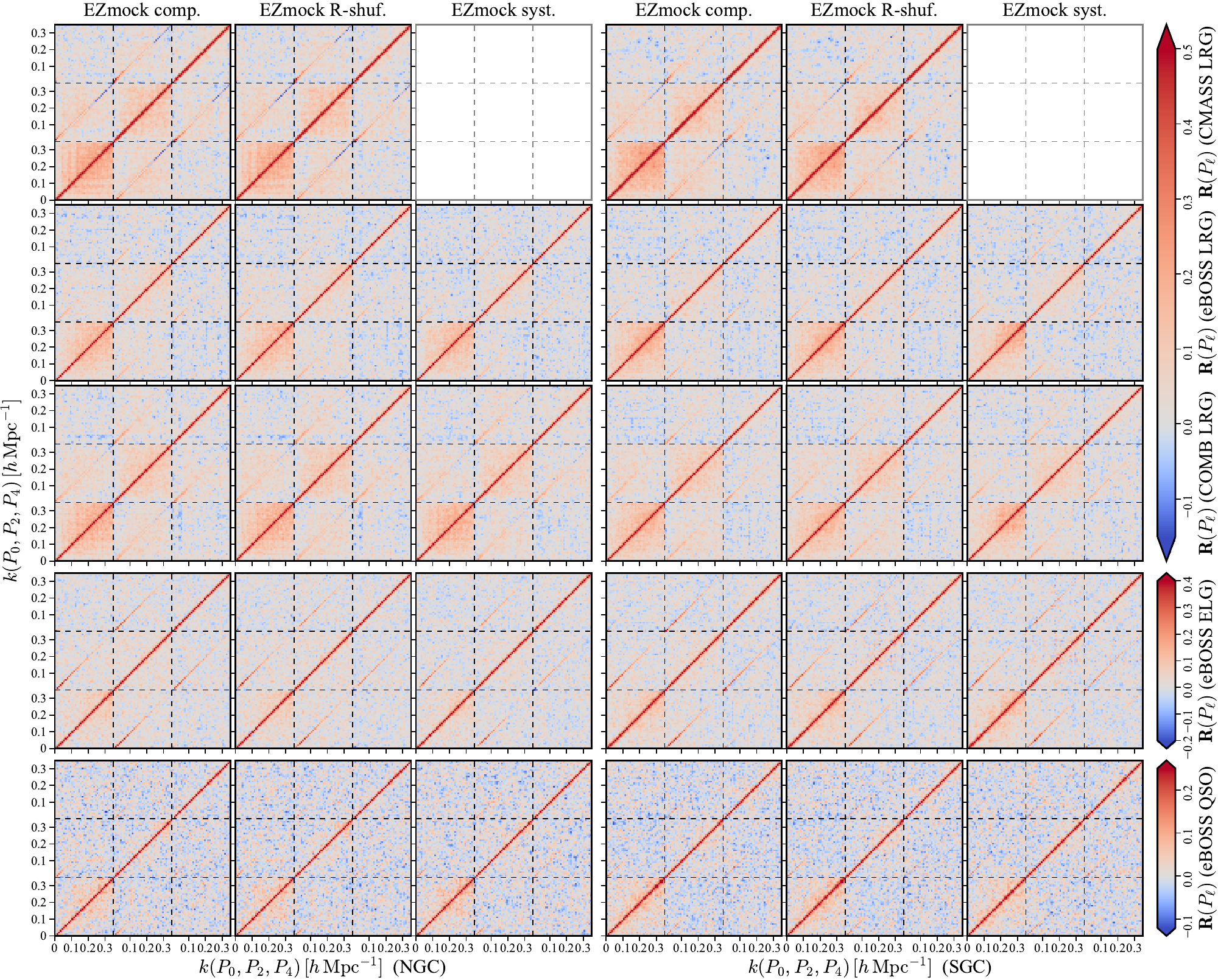}
\caption{Correlation matrices of power spectrum multipoles obtained from 1000 \ez{} realizations.}
\label{fig:ezmock_pk_cov}
\end{figure*}

Finally, we plot the correlation matrices of the power spectrum multipoles for different tracers in Fig.~\ref{fig:ezmock_pk_cov}, with the same definitions as in Eqs~\eqref{eq:cov_mat} and \eqref{eq:corr_mat}, but the data vectors are replaced by power spectrum multipoles. The impacts of observational systematics and random catalogue generation scheme on the correlation matrices appear to be smaller in Fourier space, compared to the results for 2PCF multipoles.

\subsubsection{Bispectrum}

The bispectrum is a function of three Fourier space vectors -- $\boldsymbol{k}_1$, $\boldsymbol{k}_2$, and $\boldsymbol{k}_3$ -- that form a triangle. For simplicity we consider only bispectrum monopole for a special configuration of the triangle: two sides are fixed ($k_1 = 0.1 \pm 0.01\,h\,{\rm Mpc}^{-1}$ and $k_2 = 0.05 \pm 0.01\,h\,{\rm Mpc}^{-1}$), and their intersection angle $\theta_{12}$ is varied from 0 to $\uppi$. The lengths of the sides are chosen to be close to the BAO scale. We use the \textsc{bispec}\footnote{\url{https://github.com/cheng-zhao/bispec}} code to compute bispectra, with the grid size of $512^3$ for the density fields of all tracers.

\begin{figure*}
\centering
\includegraphics[width=.98\textwidth]{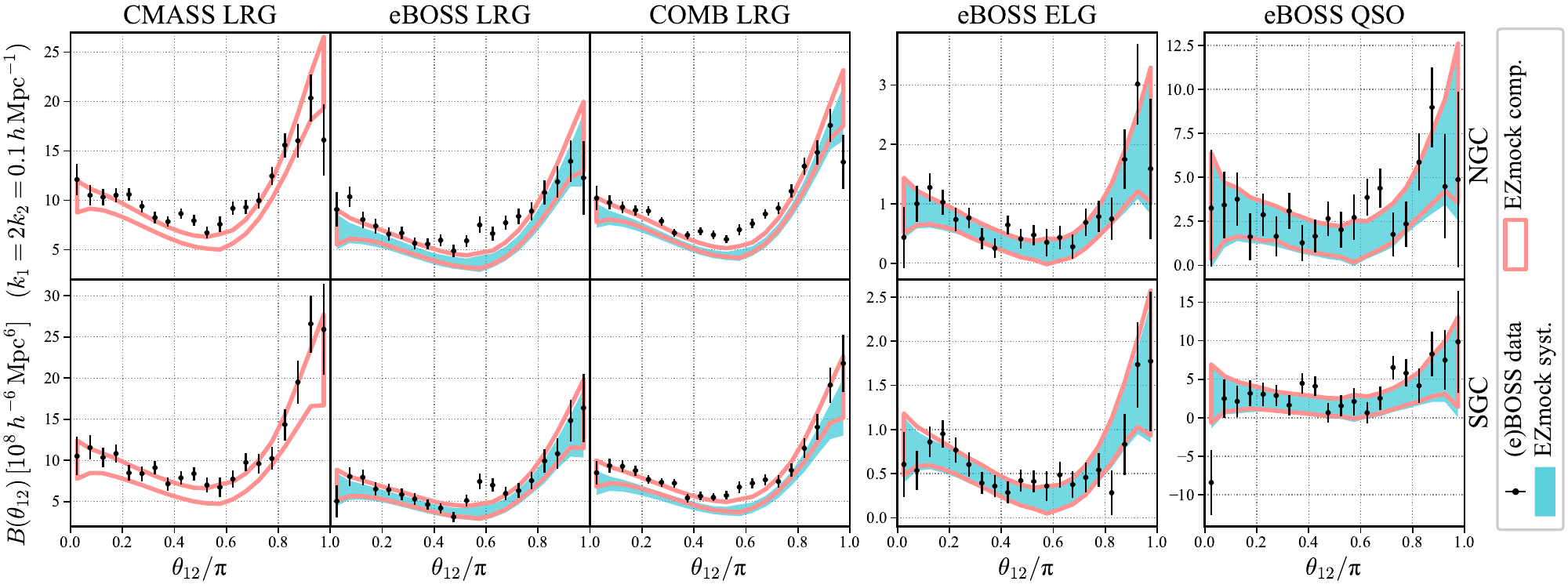}
\caption{Bispectra of the BOSS/eBOSS data and the corresponding \ez{} catalogues, for the two Galactic caps. The solid envelopes and shadowed areas indicate the 1\,$\sigma$ regions evaluated from 1000 mock realizations. The error bars for the CMASS LRG sample are obtained from the {\it complete} \ez{} catalogues, while for the other tracers they are taken from the {\it realistic} mocks with systematics.}
\label{fig:ezmock_bk}
\end{figure*}

Apart from the discrepancies on the amplitude due to the approximation of isotropic number densities (see Section~\ref{sec:fourier_norm}), the agreement between the bispectra of the observational data and \ez{} catalogues are again reasonably well, as shown in Fig.~\ref{fig:ezmock_bk}. For the configuration of the Fourier space triangle we choose, the bispectra are not sensitive to observational systematics and the random catalogue generation method. This ensures that the covariance matrices estimated using \ez{} catalogues for the two-point clustering statistics are robust \citep[][]{Baumgarten2018}.

\subsection{Evolution of clustering}
\label{sec:redshift_evol}

\begin{figure*}
\centering
\includegraphics[width=.95\columnwidth]{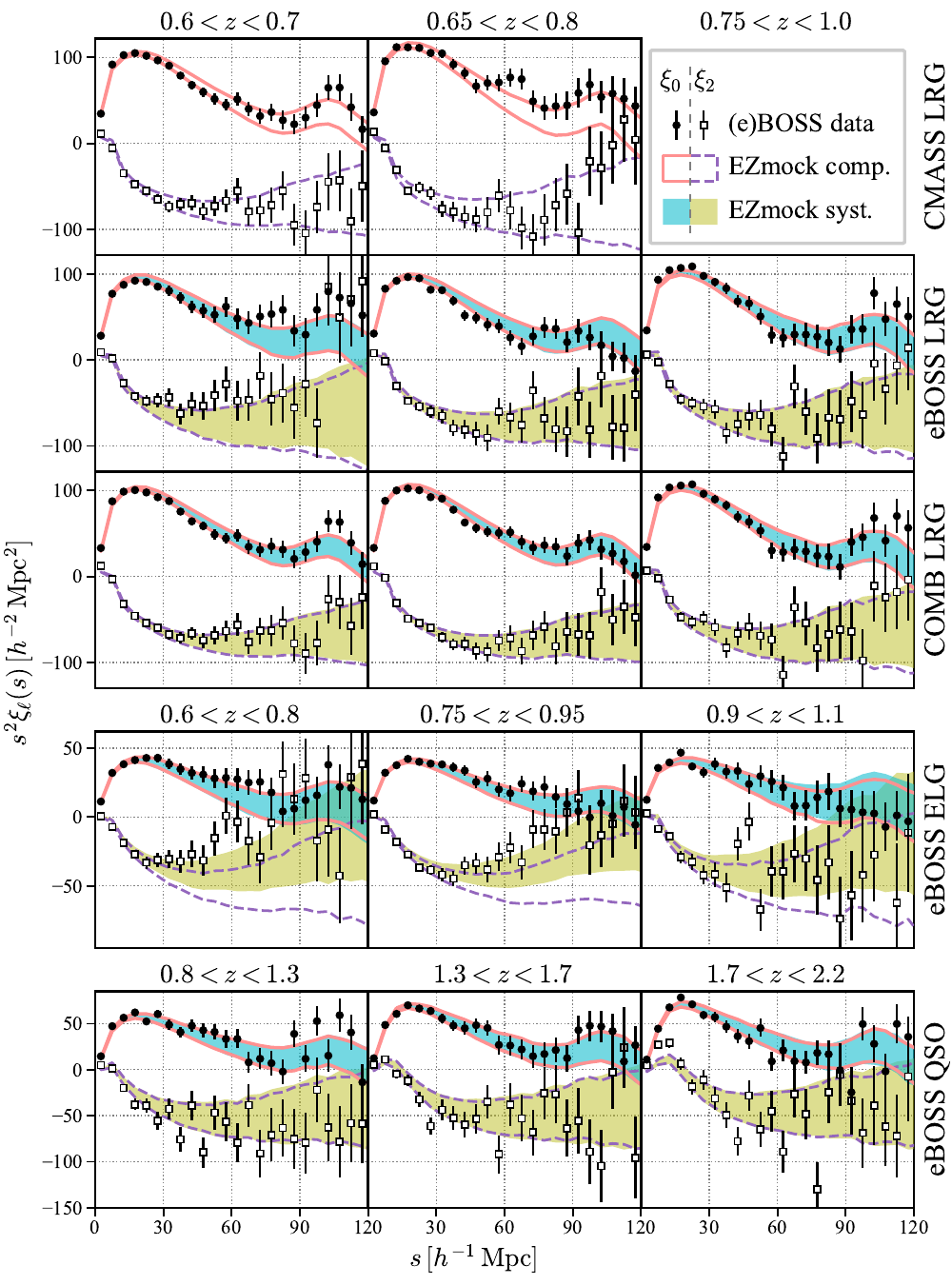}
\hspace{.05\columnwidth}%
\includegraphics[width=.95\columnwidth]{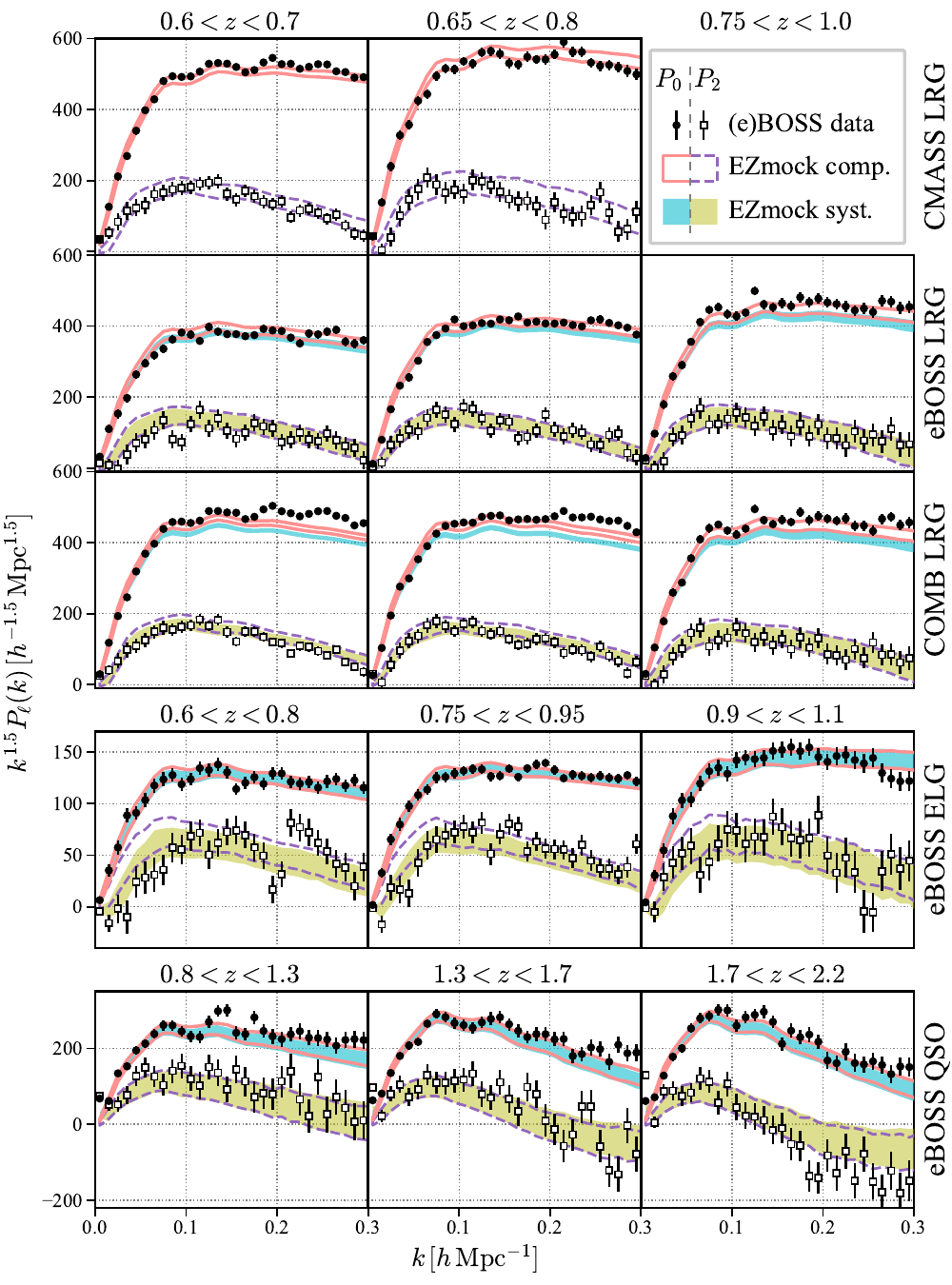}
\caption{2PCF and power spectrum multipoles of the BOSS/eBOSS data and the corresponding \ez{} catalogues in different redshift bins. Measurements from the two Galactic caps are combined. The solid/dashed envelopes and shadowed areas indicate the 1\,$\sigma$ regions evaluated from 500 mock realizations. The error bars for the CMASS LRG sample are obtained from the {\it complete} \ez{} catalogues, while for the other tracers they are taken from the {\it realistic} mocks with systematics.}
\label{fig:ezmock_zbin}
\end{figure*}

The redshift evolution of \ez{} catalogues are modelled by combining snapshots calibrated at several different redshifts (see Section~\ref{sec:ezmock_zbin}). To validate this scheme, we measure the 2PCF and power spectrum multipoles of the BOSS/eBOSS data and 500 realizations of the corresponding \ez{} catalogues in three different redshift bins (apart from CMASS LRGs, for which only two bins are used, due to the low number of galaxies at high redshift). The bins are chosen to contain sufficient data for clustering measurements, as well as close number of tracers in each bin. Besides, we allow overlapping between two adjacent redshift bins. In practice, the redshift bins for the examination of the evolution of \ez{} clustering are listed in Table~\ref{tab:clustering_zbin}. The combined clustering measurements from both Galactic caps are shown in Fig.~\ref{fig:ezmock_zbin}.

\begin{table}
\centering
\begin{threeparttable}
\caption{Redshift bins for the validation of cosmic evolution of \ez{} clustering statistics for different tracers.}
\begin{tabular}{cccc}
\toprule
& bin 1 & bin 2 & bin 3 \\
\midrule
CMASS LRG & $0.6 < z < 0.65$ & $0.65 < z < 0.8$ & -- \\
eBOSS LRG & $0.6 < z < 0.65$ & $0.65 < z < 0.8$ & $0.75 < z < 1.0$ \\
COMB LRG & $0.6 < z < 0.65$ & $0.65 < z < 0.8$ & $0.75 < z < 1.0$ \\
eBOSS ELG & $0.6 < z < 0.8$ & $0.75 < z < 0.95$ & $0.9 < z < 1.1$ \\
eBOSS QSO & $0.8 < z < 1.3$ & $1.3 < z < 1.7$ & $1.7 < z < 2.2$ \\
\bottomrule
\end{tabular}
\label{tab:clustering_zbin}
\end{threeparttable}
\end{table}

For both configuration space and Fourier space measurements, there is a general trend that the amplitudes are larger at higher redshifts. This is because with the same target selection criteria, objects at higher redshift are more luminous, thus having typically higher biases. This selection effect plays a more important role than structure growth. With the density fields and bias models constructed at different redshifts, \ez{} catalogues are able to reproduce both effects.
Fig.~\ref{fig:ezmock_zbin} shows that the cosmic evolution of the clustering statistics of the observational data and \ez{} catalogues are generally in good agreements.

\subsection{Normality check}
\label{sec:ezmock_check}

To further quantify the statistical reliability of the mocks, we measure the chi-squared for the clustering statistics of each \ez{} realization, with respect to the mean results of all mocks:
\begin{equation}
\chi_i^2 = ( \boldsymbol{x}_i - \bar{\boldsymbol{x}} )^T \mathbf{C}^{-1} ( \boldsymbol{x}_i - \bar{\boldsymbol{x}} ) .
\label{eq:ezmock_chi2}
\end{equation}
Here, $\boldsymbol{x}_i$ denotes the data vector (2PCF or power spectrum multipoles) of the $i$-th mock, $\bar{\boldsymbol{x}}$ and $\mathbf{C}$ indicate the corresponding averaged result and covariance matrix evaluated using all the mocks, respectively.

\begin{figure*}
\centering
\includegraphics[width=.98\textwidth]{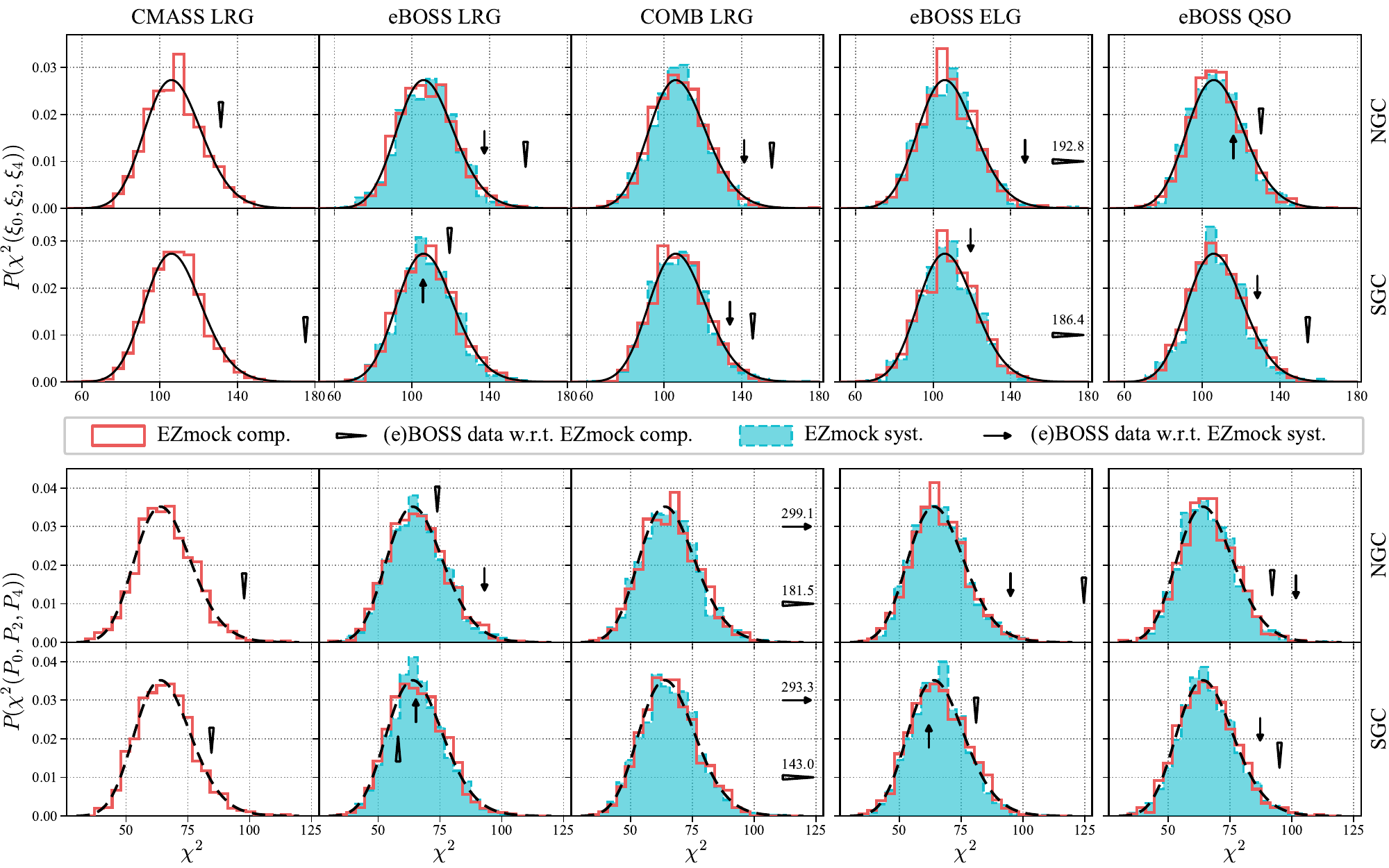}
\caption{The distribution of the chi-squared (Eq.~\eqref{eq:ezmock_chi2}) for the clustering measurements of 1000 individual \ez{} realizations, with respect to the mean results from all the mocks, for 2PCF and power spectrum multipoles (including monopole, quadrupole, and hexadecapole). The ranges of the clustering measurements are $s \in [20, 200]\,h^{-1}\,{\rm Mpc}$ and $k \in [0.03, 0.25]\,h\,{\rm Mpc}^{-1}$, respectively. The solid and dashed lines show the probability density function of the chi-squared distribution, with the degrees of freedom being the number of bins for the corresponding clustering statistics (108 for 2PCF multipoles, and 66 for power spectrum multipoles). Arrows indicate the chi-squared measured with the BOSS/eBOSS data and the mean of the associate mocks.}
\label{fig:ezmock_chi2}
\end{figure*}

The histogram of the chi-squared values for the 2PCF and power spectrum multipoles of all the single mock realizations are shown in Fig.~\ref{fig:ezmock_chi2}. In particular, the monopole, quadrupole, and hexadecapole measurements are all included, for both configuration and Fourier spaces, with the $s$ and $k$ ranges being $[20, 200]\,h^{-1}\,{\rm Mpc}$ and $[0.03, 0.25]\,h\,{\rm Mpc}^{-1}$, respectively. We then compute the probability density function of the chi-squared distribution, with the degrees of freedom being 108 and 66, which are the number of bins for the 2PCF and power spectrum multipole measurements, respectively. Fig.~\ref{fig:ezmock_chi2} shows that the distributions of the chi-squared measured from the mocks follow almost perfectly the analytical probability distribution. Therefore, the variances of the clustering measurements from the mocks are well consistent with Gaussian random variables.

In order to examine the statistical consistency between the BOSS/eBOSS data and \ez{} catalogues, we further compute the chi-squared value of the clustering statistics of the observational data, with respect to both the {\it complete} and {\it realistic} \ez{} catalogues, and the results are marked by arrows in Fig.~\ref{fig:ezmock_chi2}.
It shows that the {\it realistic} mocks are always in better agreements with the observational data in configuration space, compared to the results from the {\it complete} mocks. Indeed, the ELG data and mocks are only consistent with observational systematics applied, and with the `shuffled' random catalogue.
In general, it is possible to regard the observational data as one particular realization of the statistical ensemble of the {\it realistic} mocks, even with the mismatch of the power spectra amplitudes due to the number density evaluation (see Appendix~\ref{sec:fourier_norm}). It is however less representative for the joint CMASS and eBOSS LRG sample, for which the cross correlations between the two data sets may not be well modelled by \ez{} catalogues.

\section{Cross correlations}
\label{sec:cross}

The overlapping volumes between different types of eBOSS tracers -- LRGs, ELGs, and QSOs -- are sufficient for large-scale clustering measurements, which permits multi-tracer analysis with cross correlations. To have reliable estimations on the covariance matrices for cross clustering measurements, for each of the \ez{} realization, all tracers are constructed using the same initial conditions, and applied the same geometric transformation for the light-cone catalogue creation, to ensure the same underlying dark matter density field. Thus, though the mocks for different types of tracers are calibrated separately, their clustering statistics are correlated through the dark matter field.
In this section, we investigate the relationship between different tracers, including their cross clustering statistics.  

\subsection{Spatial relationship}

As the \ez{} catalogues for different types of tracers share the density field, we first examine their spatial distributions, and compare with the dark matter density field from ZA. In practice, \ez{} catalogues for different tracers are populated at different redshifts (see Table~\ref{tab:ezmock_zbin}). Therefore, their dark matter density fields are not identical, but linked through dynamical evolutions. For a direct comparison of tracer distributions, we evaluate the dark matter density field at $z = 0.9$, and interpolate or extrapolate the \ez{} parameters for different eBOSS tracers all at this redshift (see Section~\ref{sec:ezmock_calib}) to construct mock catalogues with exactly the same dark matter distribution.

\begin{figure}
\centering
\includegraphics[width=.98\columnwidth]{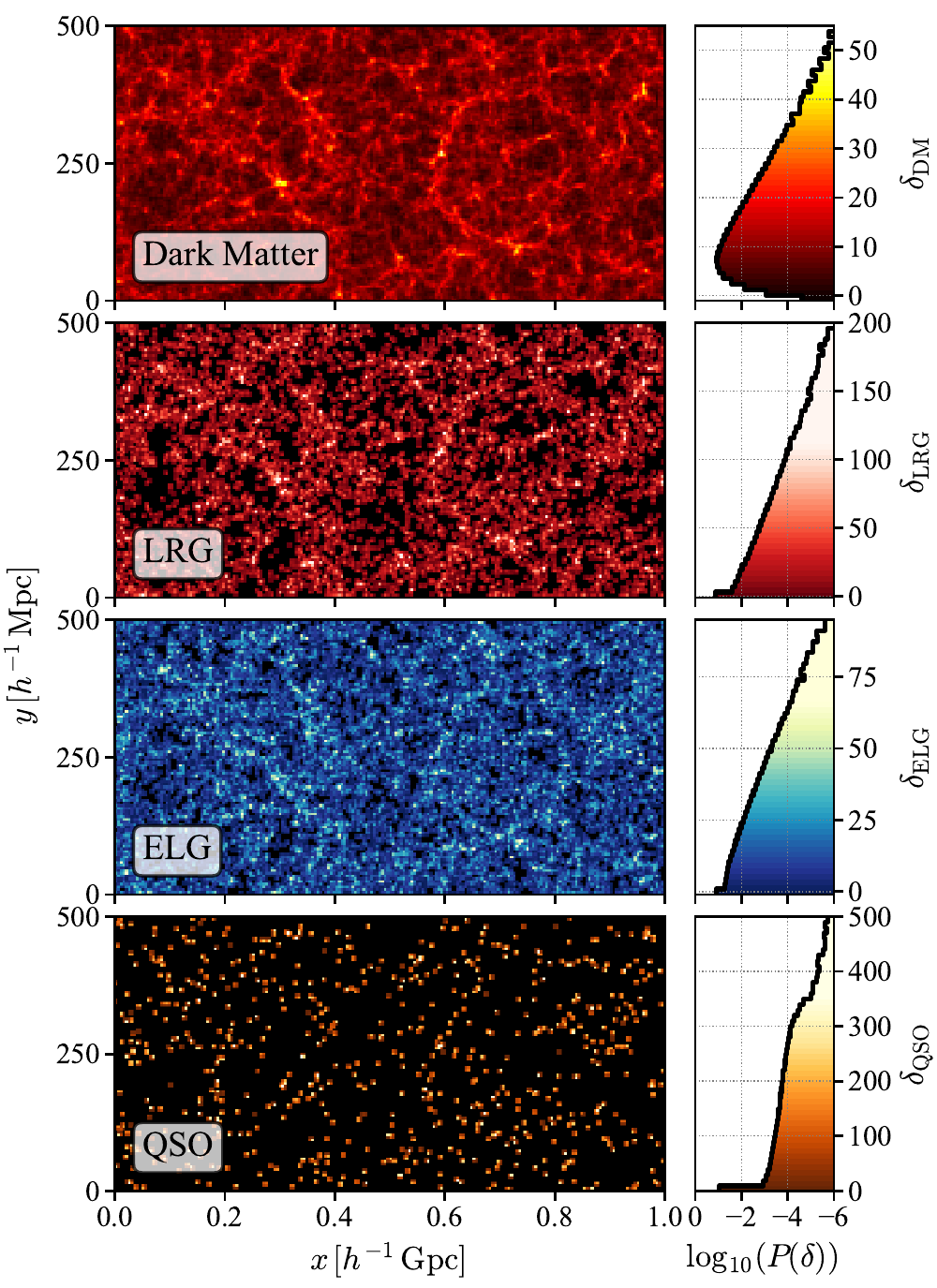}
\caption{Projected overdensity field for different tracers in a $2000 \times 500 \times 50\,h^{-3}\,{\rm Mpc}^3$ sub-volume of an \ez{} box constructed at $z = 0.9$ ({\it left} panel), and the probability distribution function of the tracer densities in the full box ({\it right} panel).}
\label{fig:ezmock_densmap}
\end{figure}

Fig.~\ref{fig:ezmock_densmap} shows the projected dark matter density field, as well as the overdensity distribution of different tracers in the same comoving volume. In particular, the overdensities are defined as 
$\delta_{\rm t} = \rho_{\rm t} / \bar{\rho}_{\rm t} - 1$,
where $\rho_{\rm t}$ indicates the number density of tracers, including dark matter particles, and $\bar{\rho}_{\rm t}$ denotes the mean density in the full comoving volume. Moreover, the density fields are all calculated using the CIC particle assignment scheme.
It can be seen clearly that the large-scale distributions of eBOSS tracers are all in good agreements with the dark matter density field.

Futhermore, the LRG distribution follows closely that of dark matter, while for ELGs the distribution is more diffused, indicating a lower galaxy bias. Thus, the galaxy distributions in \ez{} catalogues are consistent with the mass and environment relationship between passive and star-forming galaxies from observations and simulations \citep[see e.g.][]{Peng2010, Gonzalez2020}.
The overdensity of QSOs appear to be even higher than that of LRGs, but this is mainly due to their low averaged number density. Indeed, the QSO overdensity field does not always match the dark matter distribution, and the densities are generally too low to reveal cosmic web structures. Therefore, the QSO distribution may not be ideal for estimating the density or gravitational field. Consequently, the BAO reconstruction technique \citep[][]{Eisenstein2007} may not work well for QSOs.

\begin{figure}
\centering
\includegraphics[width=.98\columnwidth]{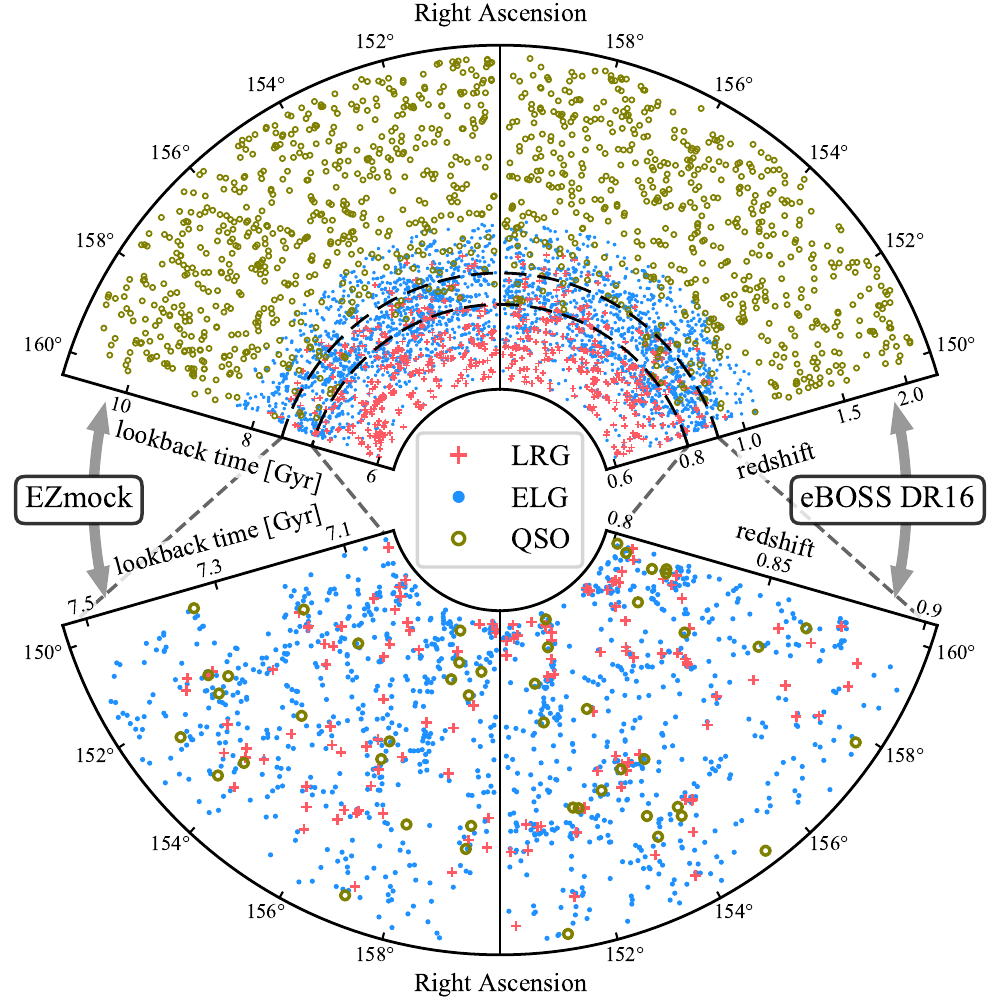}
\caption{Projected distribution of tracers from the eBOSS DR16 data ({\it right}) and one realization of the associated {\it realistic} \ez{} catalogues with observational systematics ({\it left}), in a $10^\circ \times 1^\circ$ region of the sky, with the right ascension between $150^\circ$ and $160^\circ$, and declination between $29^\circ$ and $30^\circ$. The lookback time is evaluated in the flat $\Lambda$CDM cosmology with $\Omega_{\rm m} = 0.31$.}
\label{fig:ezmock_sector}
\end{figure}

In order to illustrate the spatial distribution of different types of tracers in the full \ez{} light-cone catalogues, we further compare the `side view' of tracer distributions from the eBOSS data and one {\it realistic} \ez{} realization in a small angular region ($10^\circ \times 1^\circ$), and the plot is shown in Fig.~\ref{fig:ezmock_sector}. In particular, the {\it upper} panel shows the tracer distribution in the full eBOSS redshift range, while the {\it lower} panel presents a common redshift range ($0.8 < z < 0.9$) for all tracers. Statistically there are no obvious differences between the tracer distributions of the data and \ez{} realization, and similar filamentary and void patterns can be seen in both catalogues. Again, the ELG distribution is more diffused. But thanks to their high number density, ELGs can be used as references for comparing the distributions of tracers in the shared volume. The {\it lower} panel of Fig.~\ref{fig:ezmock_sector} reveals tight links between different tracers: most of the LRGs and QSOs reside with ELGs, and there are typically no tracers inside voids of the ELG distributions.

\subsection{Cross clustering measurements}
\label{sec:cross_clustering}

To quantify the cross correlation between different tracers in the BOSS/eBOSS data and the corresponding \ez{} catalogues, we present in this section the cross clustering measurements between different tracers, in both configuration and Fourier space. Since there are not many LRGs and QSOs in a common volume, we consider only the LRG--ELG and ELG--QSO cross correlations. Furthermore, we use the full catalogues for the cross correlation measurements, rather than taking only tracers in the shared volumes.

In practice, the anisotropic cross correlation functions are measured using the modified Landy--Szalay estimator \citep[][]{Szapudi1997}:
\begin{equation}
\xi^\times (s, \mu) = \frac{{\rm D}_{\rm A} {\rm D}_{\rm B} (s, \mu) - {\rm D}_{\rm A} {\rm R}_{\rm B} (s, \mu) - {\rm R}_{\rm A} {\rm D}_{\rm B} (s, \mu) }{{\rm R}_{\rm A} {\rm R}_{\rm B} (s, \mu)} + 1,
\end{equation}
where the subscripts `A' and `B' indicate the catalogues of the two different tracers to be cross correlated. Thus, we always count pairs based on two catalogues from different tracers.

Similarly, the cross power spectrum estimator is based on the modified auto power spectrum estimator (Eq.~\eqref{eq:pk_estimator}):
\begin{equation}
P^\times (\boldsymbol{k}) = I_{22, {\rm A}}^{-1/2} I_{22, {\rm B}}^{-1/2} \left\langle \hat{F}_{\rm A} (\boldsymbol{k}) \hat{F}_{\rm B} (-\boldsymbol{k}) \right\rangle ,
\end{equation}
but without the shot noise term $I_{12}$. This is because the shot noise of the cross correlation is generally negligible, since objects from the two samples cannot be at the same positions \citep[e.g.][]{Smith2009}.

\begin{figure}
\centering
\includegraphics[width=.98\columnwidth]{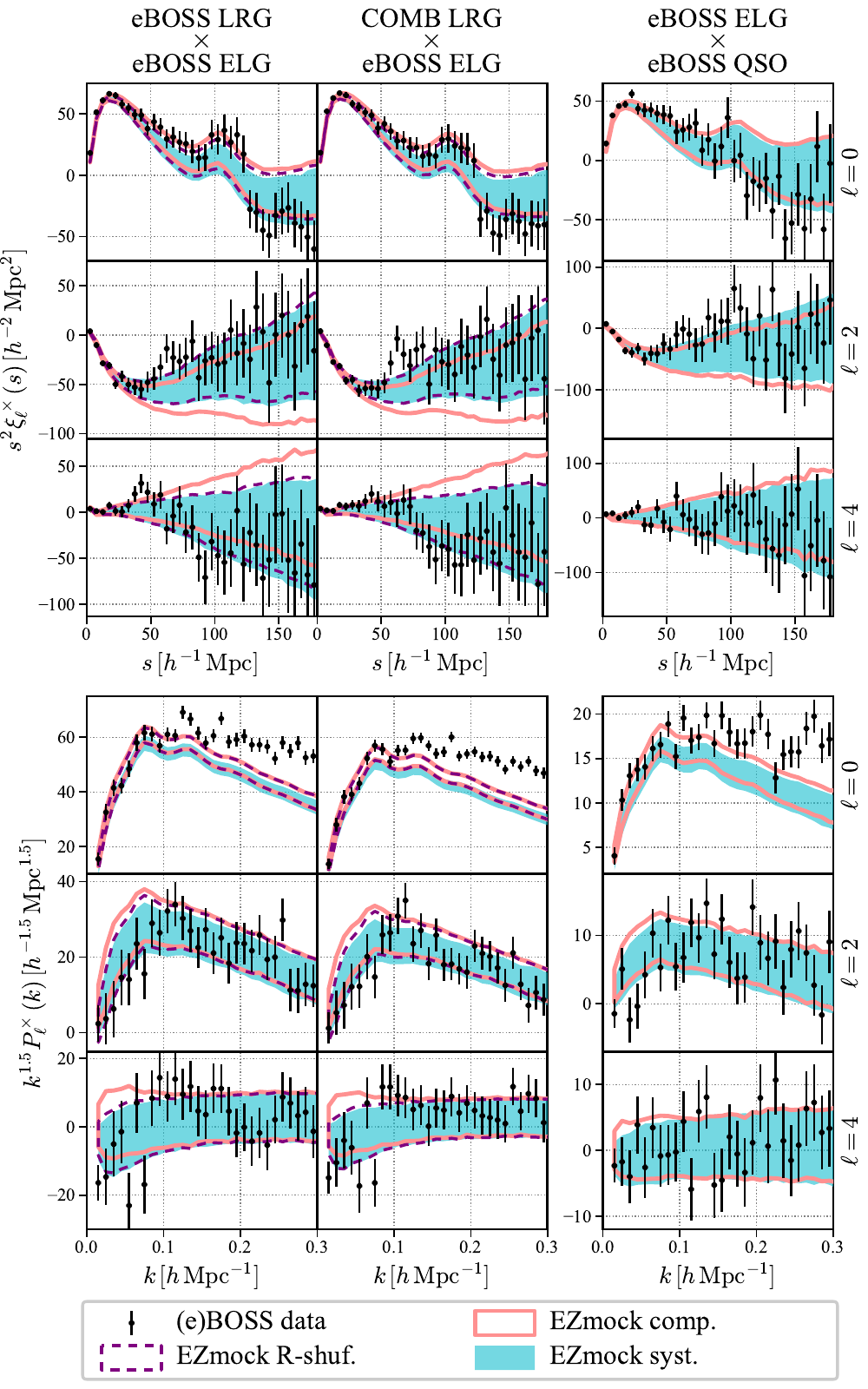}
\caption{The cross clustering measurements of the BOSS/eBOSS data and the corresponding \ez{} catalogues. Results from the two Galactic caps are combined. The solid/dashed envelopes and shadowed areas indicate the 1\,$\sigma$ regions evaluated from 1000 mock realizations.}
\label{fig:ezmock_cross}
\end{figure}

The cross correlation function and cross power spectrum can be decomposed with Legendre polynomials as well, to obtain the multipole measurements. The formulae are similar to those of the auto correlations, i.e. Eqs~\eqref{eq:xi_ell} and \eqref{eq:pk_ell}. As the results, the cross clustering multipoles between different BOSS/eBOSS samples and the associated \ez{} catalogues are shown in Fig.~\ref{fig:ezmock_cross}.

In general, all the configuration space cross correlation measured from the mocks are in good agreements with those of the observational data, especially for the results computed with the `shuffled' random catalogues. However, for the cross power spectrum multipoles, apart from the discrepancies on the normalizations for the {\it realistic} mocks (see Section~\ref{sec:fourier_norm}), there are also significant mismatches between the data and mocks at high $k$. As has been discussed in Section~\ref{sec:result_fourier}, the small scale cross correlations between different tracers may not be modelled correctly in the \ez{} catalogues, by performing the \ez{} calibrations separately for different types of tracers. In reality, different tracers may reside in the same galaxy cluster, and are strongly linked with each other. However, this effect are not considered for the \ez{} catalogues, in which different tracers are populated in the density field independently.
Thus, the cross clustering statistics of the \ez{} tracers should be underestimated on small scales. To correct for this effect, further small-scale connections between different tracers should be included \citep[see][for a multi-tracer HOD approach]{Alam2019}, and we leave the detailed studies for \ez{} catalogues to a future paper.

Though the observational systematics affect the auto correlations of ELGs dramatically, We do not see significant difference comparing the cross-clustering measurements from the {\it realistic} mocks with those from {\it complete} mocks. The `shuffled' random catalogues are used to compute these measurements. This is because the  observational systematics for different tracers are only through foregrounds, e.g. stellar density or galactic extinction, and are not obvious for the cross clustering measurements. For a thorough analysis of the cross correlation function between LRGs and ELGs, see \citet[][]{Wang2020}.

\section{Conclusions}
\label{sec:conclusion}

We have described the construction of 1000 realizations of \ez{} catalogues, for each type of the eBOSS DR16 tracers for LSS analysis, including LRGs ($0.6 < z < 1.0$), ELGs ($0.6 < z < 1.1$), and QSOs ($0.8 < z < 2.2$), as well as the BOSS DR12 CMASS LRGs in the redshift range $0.6 < z < 1.0$ for the joint LRG studies, taking into account the cross correlations between different tracers. To this end, 46\,000 realizations of simulation boxes are generated, with the side length of $5\,h^{-1}\,{\rm Gpc}$. The final mock catalogues are composed of four redshift slices for CMASS LRGs, five slices for eBOSS LRGs, seven slices for eBOSS ELGs, and seven slices for eBOSS QSOs, to account for cosmic evolution of the clustering statistics and sample selection biases at different redshifts.

These mock catalogues encode effective structure formation and tracer bias models, based on the Zel'dovich approximation, and bias descriptions including both deterministic and stochastic effects. Moreover, various geometrical survey features are applied to the mocks, including survey footprints, veto masks, and radial distributions. In addition, both the photometric and spectroscopic systematic effects of the observational data are migrated to the \ez{} catalogues, to have robust estimates of the covariance matrices for BAO and RSD analysis.

The \ez{} catalogues have shown good agreements with the observational data, in terms of two- and three-point auto-clustering statistics, as well as two point cross correlations. The consistencies are generally within $1\,\sigma$ for scales down to a few $h^{-1}\,{\rm Mpc}$ in configuration space, and up to $0.3\,h\,{\rm Mpc}^{-1}$ in Fourier space, apart from offsets on the normalizations of power spectra due to the definition of isotropic radial selection functions (see Appendix~\ref{sec:fourier_norm}), and discrepancies at $k \gtrsim 0.15\,h\,{\rm Mpc}^{-1}$ for cross correlations in Fourier space.
And the covariance matrices obtained from these mock catalogues are used for the BAO and RSD measurements of the LRG samples \citep[][]{Bautista2021, GilMarin2020}, ELG samples \citep[][]{deMattia2021, Raichoor2021, Tamone2020}, and QSO samples \citep[][]{Neveux2020, Hou2021} for the final eBOSS analysis, as well as the cross correlation studies with LRGs and ELGs \citep[][]{Wang2020}, and the cosmological constraints \citep[][]{eBOSS2020}.

The final \ez{} catalogues presented in this paper will be made available to the public\footnote{\url{https://data.sdss.org/sas/dr16/eboss/lss/catalogs/EZmocks}}.

\section*{Acknowledgements}

We thank the anonymous referee for the valuable comments and suggestions.
CZ, AR, and AT acknowledge support from the SNF grant 200020\_175751.
AR and JPK acknowledge support from the ERC advanced grant LIDA.
AdM acknowledges support from the P2IO LabEx (ANR-10-LABX-0038) in the framework ``Investissements d'Avenir'' (ANR-11-IDEX-0003-01) managed by the Agence Nationale de la Recherche (ANR, France).
AJR is grateful for support from the Ohio State University Center for Cosmology and Particle Physics.
RN acknowledges support from ANR-17-CE31-0024-01, NILAC.
Authors acknowledge support from the ANR eBOSS project (ANR-16-CE31-0021) of the French National Research Agency.
GR acknowledges support from the National Research Foundation of Korea (NRF) through Grants No. 2017R1E1A1A01077508 and No. 2020R1A2C1005655 funded by the Korean Ministry of Education, Science and Technology (MoEST), and from the faculty research fund of Sejong University. SA is supported by the European Research Council through the COSFORM Research Grant (\#670193).

The massive production of \ez{} catalogues was performed at the National Energy Research Scientific Computing Center (NERSC)\footnote{\url{https://ror.org/05v3mvq14}}, a U.S. Department of Energy Office of Science User Facility operated under Contract No. DE-AC02-05CH11231. We also made use of the Beiluo cluster at Tsinghua University, and Sciama High Performance Computing cluster supported by the ICG, SEPNet and the University of Portsmouth.

Funding for the Sloan Digital Sky Survey IV has been provided by the Alfred P. Sloan Foundation, the U.S. Department of Energy Office of Science, and the Participating Institutions. SDSS-IV acknowledges support and resources from the Center for High-Performance Computing at the University of Utah. The SDSS web site is \url{www.sdss.org}.

SDSS-IV is managed by the Astrophysical Research Consortium for the Participating Institutions of the SDSS Collaboration including the 
Brazilian Participation Group, the Carnegie Institution for Science,
Carnegie Mellon University, the Chilean Participation Group,
the French Participation Group, Harvard-Smithsonian Center for Astrophysics, 
Instituto de Astrof\'isica de Canarias, The Johns Hopkins University,
Kavli Institute for the Physics and Mathematics of the Universe (IPMU) / University of Tokyo,
the Korean Participation Group, Lawrence Berkeley National Laboratory, 
Leibniz Institut f\"ur Astrophysik Potsdam (AIP),  
Max-Planck-Institut f\"ur Astronomie (MPIA Heidelberg), 
Max-Planck-Institut f\"ur Astrophysik (MPA Garching), 
Max-Planck-Institut f\"ur Extraterrestrische Physik (MPE), 
National Astronomical Observatories of China, New Mexico State University, 
New York University, University of Notre Dame, 
Observat\'ario Nacional / MCTI, The Ohio State University, 
Pennsylvania State University, Shanghai Astronomical Observatory, 
United Kingdom Participation Group,
Universidad Nacional Aut\'onoma de M\'exico, University of Arizona, 
University of Colorado Boulder, University of Oxford, University of Portsmouth, 
University of Utah, University of Virginia, University of Washington, University of Wisconsin, 
Vanderbilt University, and Yale University.

\section*{Data availability}
A python interface for \ez{} generating is available at \url{https://github.com/cheng-zhao/pyEZmock}. Furthermore, the catalogues described in this paper will be made public\footnote{Before the page is brought online, the catalogues can be obtained on request to the corresponding author.} at \url{https://data.sdss.org/sas/dr16/eboss/lss/catalogs/EZmocks}.




\bibliographystyle{mnras}
\bibliography{eBOSS_EZmock} 




\appendix

\section{Normalization of Fourier space clustering measurements with angular incompleteness}
\label{sec:fourier_norm}

For an arbitrary spatial function $f(\boldsymbol{r})$, which can be evaluated at the positions of all tracers and random points with the values $f_{\rm t}$ and $f_{\rm r}$, the following integration can be discretized:
\begin{equation}
\int {\rm d}^3 r \, n_{\rm t} (\boldsymbol{r}) f(\boldsymbol{r})
\approx \sum_i^{N_{\rm t}} w_{{\rm t}, i} \, f_{{\rm t}, i}
\approx \alpha \sum_j^{N_{\rm r}} w_{{\rm r}, j} \, f_{{\rm r}, j} ,
\label{eq:norm_integrate}
\end{equation}
where $n_{\rm t} (\boldsymbol{r})$ indicates the number density field of the tracers, and
\begin{equation}
\alpha = \left. \sum_i^{N_{\rm t}} w_{{\rm t}, i} \, \middle/ \enspace \sum_j^{N_{\rm r}} w_{{\rm r}, j} \right. .
\end{equation}
Here, $w$ denotes the total photometric and spectroscopic weights, $N$ denotes the total number of objects, and the subscripts t and r indicate the quantities for the data and random samples, respectively.

The $n_{\rm t} (\boldsymbol{r})$ term in Eq.~\eqref{eq:norm_integrate} expresses the intrinsic number density of the sample, which does not depend on the estimated comoving density $\tilde{n}_{\rm t}$ from a tracer catalogue. Therefore, the number density estimation only affects the constant factors $I_{a b}$ for the Fourier space clustering statistics (see Eq.~\eqref{eq:fourier_norm}) with $a > 1$. For instance, the normalization factor $I_{22}$ of power spectrum is usually evaluated by \citep[e.g.][]{Beutler2017}
\begin{equation}
I_{22} \approx \alpha \sum_j^{N_{\rm r}} w_{{\rm r}, j} \, \tilde{n}_{{\rm t}, j} \, w_{{\rm FKP}, {\rm t}, j}^2 .
\label{eq:pk_i22}
\end{equation}
Note that $\tilde{n}$ and $w_{\rm FKP}$ are both quantities of the tracer field, so they do not represent the actual number density of random points. The sum is taken over the random catalogue to reduce Poisson noise, given their larger sample size compared to the data catalogue.

\begin{figure}
\centering
\includegraphics[width=.98\columnwidth]{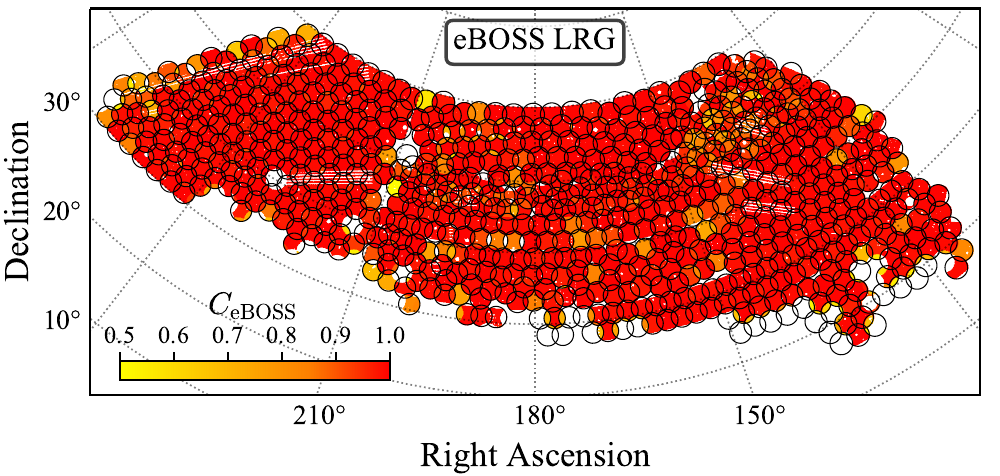}
\caption{The $C_{\rm eBOSS}$ (fraction of targets without fibres) map of eBOSS LRGs in NGC. Circles indicate the plates for the final LRG data.}
\label{fig:LRG_comp}
\end{figure}

In practice, $\tilde{n}_{\rm t}$ is often estimated as an isotropic function of the weighted tracer distribution. However, this is not always true, due to the angular incompleteness caused by missing fibres ($C_{\rm (e)BOSS}$), which are not corrected by weights \citep[][]{Reid2016, Ross2020}. The $C_{\rm eBOSS}$ map of eBOSS LRGs in the NGC is shown in Fig.~\ref{fig:LRG_comp}, for which the total completeness is 96.5 per cent. This anisotropic effect is taken into account for the evaluation of the effective survey area $A_{\rm eff}$, by counting only the corresponding fraction of area of different sectors. So the effective comoving survey volume in a given redshift bin $(z_{\rm low}, z_{\rm high})$ is
\begin{equation}
V_{\rm eff} (z) = \frac{4\uppi}{3} \left[ r_c^3 (z_{\rm high}) - r_c^3 (z_{\rm low}) \right] \cdot \frac{A_{\rm eff}}{A_{\rm sky}} ,
\end{equation}
where $r_c (z)$ denotes the radial comoving distance at redshift $z$, and $A_{\rm sky}$ indicates the full sky area. Then, for this redshift bin, the comoving number density is computed with
\begin{equation}
\tilde{n}_{\rm t} (z) = \enspace \left. \sum_{\mathclap{\substack{i \\ z_{\rm low} < z_i < z_{\rm high}}}}^{N_{\rm t}} w_{{\rm t}, i} \enspace \middle/ \enspace V_{\rm eff} (z) \right. .
\end{equation}
Therefore, the number densities are in general over-estimated, as $\tilde{n}_{\rm t}$ represents the actual number density only when $C_{\rm (e)BOSS} = 1$. Consequently, the normalisation factors of the Fourier space clustering measurements are over-estimated.

\begin{figure}
\centering
\includegraphics[width=.7\columnwidth]{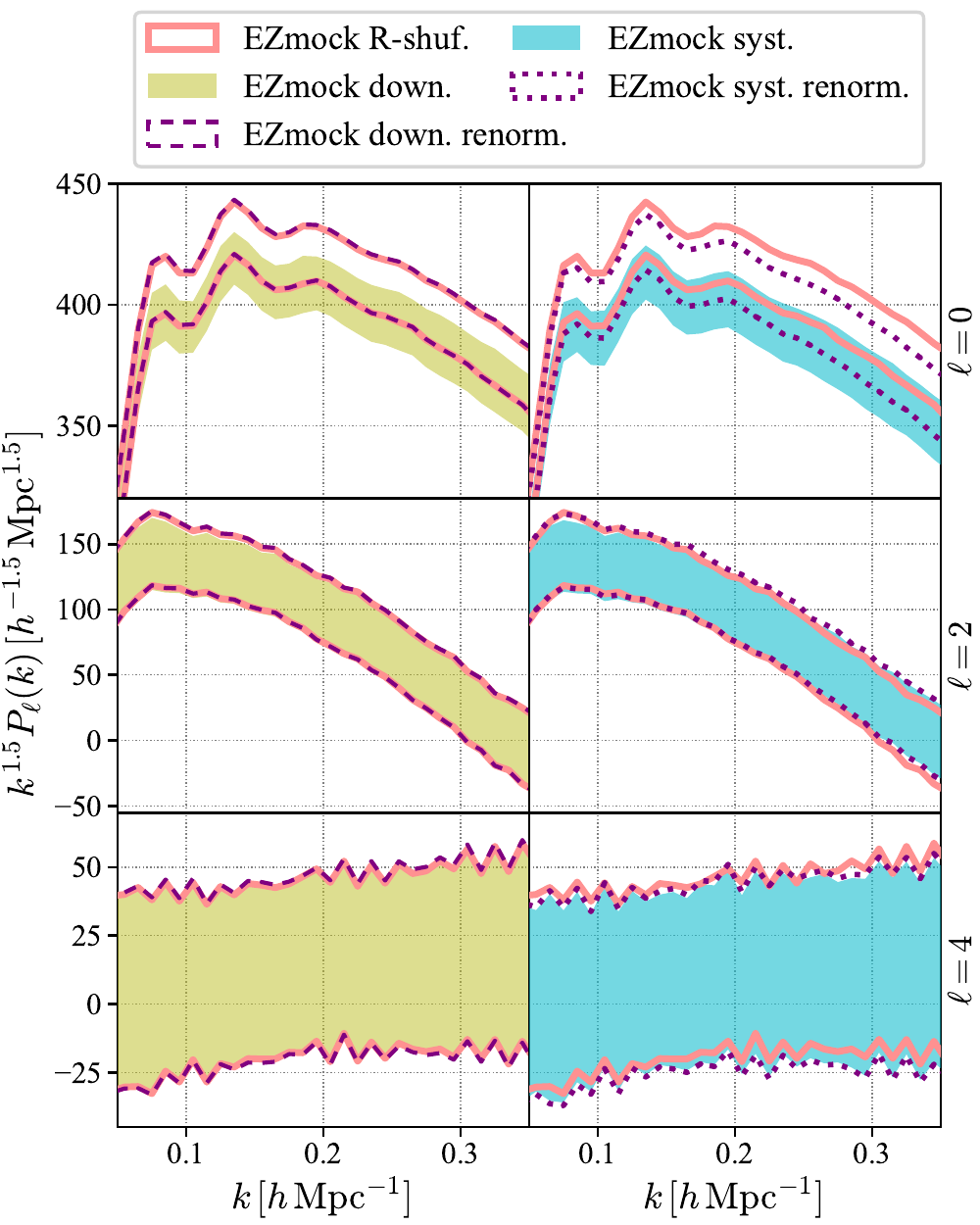}
\caption{Power spectrum multipoles of the {\it complete}, {\it realistic} and `down-sampled' \ez{} catalogues, with normalization factors expressed by Eqs~\eqref{eq:pk_i22} and \eqref{eq:pk_i22_correct} respectively. The solid/dashed envelopes and shadowed areas indicate the 1\,$\sigma$ regions evaluated from 1000 mock realizations.}
\label{fig:pk_renorm}
\end{figure}

To demonstrate this effect, we apply only the incompleteness indicated by the $C_{\rm eBOSS}$ map of the eBOSS LRGs in NGC, to the {\it complete} set of \ez{} catalogues together with the `shuffled' randoms, and compute the power spectrum multipoles with the $\tilde{n}_{\rm t}$ estimation described above. The results for these `down-sampled' mocks (denoted by `EZmock down.') are shown in Fig.~\ref{fig:pk_renorm}. As expected from the definition of $\tilde{n}_{\rm t}$, the amplitude of the power spectrum multipoles from the `down-sampled' mocks are $\sim 3$ per cent lower compared to those of the original catalogues, which is visually particularly obvious for the monopole due to its high amplitude.

This reveals that the power spectrum normalization with Eq.~\eqref{eq:pk_i22} is inappropriate, albeit the cosmological analysis can be unbiased with the same normalization factor for the estimation of the survey window function \citep[][]{deMattia2019}. To solve this problem, one has to take into account the anisotropy of the comoving number densities, which can be simply expressed by
\begin{equation}
n_{\rm t} = \tilde{n}_{\rm t} \, C_{\rm (e)BOSS} .
\end{equation}
Thus, the corrected normalization factor of power spectrum is
\begin{equation}
I_{22} \approx \alpha \sum_j^{N_{\rm r}} w_{{\rm r}, j} \, \tilde{n}_{{\rm t}, j} \, C_{\rm (e)BOSS} \, w_{{\rm FKP}, {\rm t}, j}^2 .
\label{eq:pk_i22_correct}
\end{equation}
And similar corrections should be applied to the other factors for Fourier space clustering measurements, such as $I_{23}$ and $I_{33}$ for the evaluation of bispectrum.

The power spectrum multipoles of the re-normalized `down-sampled' mocks with Eq.~\eqref{eq:pk_i22_correct} (denoted by `EZmock down. renorm.') are shown in the left-hand panel of Fig.~\ref{fig:pk_renorm}, and they are in good agreement with the measurements from the {\it complete} mocks. We then apply the $n_{\rm t}$ correction to the {\it realistic} \ez{} sample, and generate a new set of mocks denoted by `EZmock syst. renorm.'. The clustering results are illustrated in the right-hand panel of Fig.~\ref{fig:pk_renorm}. One can see that the observational systematics applied to the {\it realistic} mocks do not really alter significantly the amplitude of power spectrum multipoles, especially for small wave numbers. However, the discrepancies between the {\it realistic} \ez{} catalogues and the eBOSS data still exist, as the same re-normalization should be applied to the measurements from the observational data too. Actually, the calibration performed with the {\it complete} mocks has already been biased by the inappropriate normalization factor.

\begin{figure}
\centering
\includegraphics[width=.95\columnwidth]{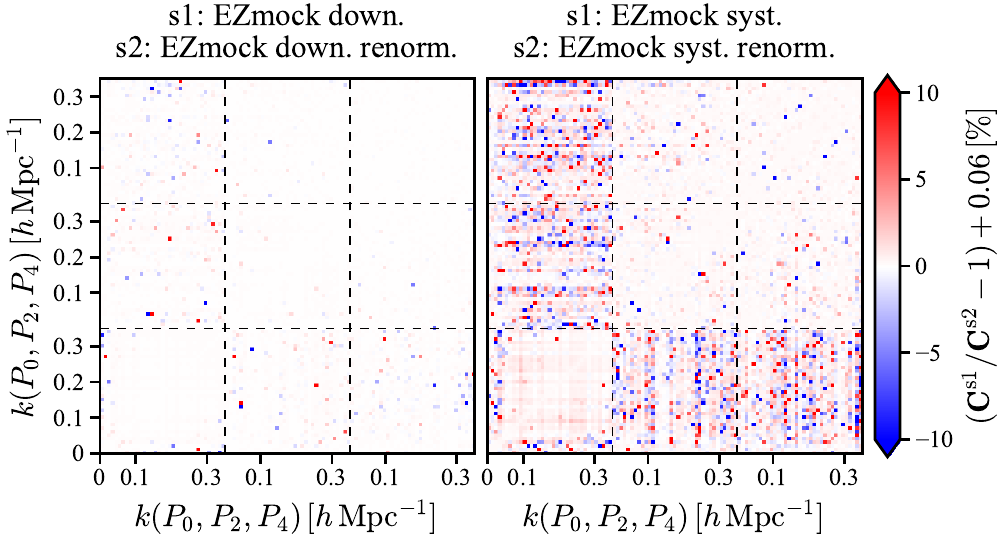}
\caption{Relative difference of the covariance matrices obtained from 1000 \ez{} realizations, with the power spectrum normalizations expressed by Eqs~\eqref{eq:pk_i22} and \eqref{eq:pk_i22_correct} respectively.}
\label{fig:cov_renorm}
\end{figure}

To further quantify the impact of this issue on the covariance matrices estimated using \ez{} catalogues, we compute the covariance matrices of the power spectrum multipoles with the normalization factors expressed by Eqs~\eqref{eq:pk_i22} and \eqref{eq:pk_i22_correct} respectively, and the comparisons for the {\it complete} and {\it realistic} \ez{} samples are shown in Fig.~\ref{fig:cov_renorm}. As expected, the covariance matrices are generally biased by 6 per cent, which is consistent with the 3 per cent difference on the amplitudes. However, there are also fluctuations on the differences of the covariance matrices, especially for the cross covariances between monopole and the other multipoles. This is because there are separate random catalogues for different realizations, and the re-normalization factors suffer Poisson noises. The fluctuations are more significant for the {\it realistic} mocks, since the systematic effects are also different for each random realization.
Nevertheless, the diagonal terms are insensitive to the variations of random catalogues, and in general, a rescaling of the covariance matrices works well for correcting the normalization issue.

\bsp	
\label{lastpage}
\end{document}